%% file: fossacs24.tex
\documentclass{llncs}
\input{usepackages.tex}
\makeatletter
   \def\@citecolor{blue}%
   \def\@urlcolor{blue}%
   \def\@linkcolor{blue}%

\def\orcidID#1{\smash{\href{http://orcid.org/#1}{\protect\raisebox{-1.25pt}{\protect\includegraphics{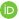}}}}}
\makeatother

\begin{document}
\title{
  Succinctness of Cosafety Fragments of LTL\\
  via Combinatorial Proof Systems \\ (extended version)%
  \thanks{
    To cite this paper, please consider the conference
    version~\cite{DBLP:conf/fossacs/GeattiMM24}
  }
}
%
%
\author{%
  Luca Geatti\inst{1}\orcidID{0000-0002-7125-787X} \and
  Alessio Mansutti\inst{2}\orcidID{0000-0002-1104-7299} \and
  Angelo Montanari\inst{1}\orcidID{0000-0002-4322-769X}}
\authorrunning{L. Geatti, A. Mansutti, A. Montanari}
%
\institute{University of Udine, Italy
\email{[luca.geatti,angelo.montanari]@uniud.it}\\
\and
IMDEA Software Institute, Spain
\email{alessio.mansutti@imdea.org}}
\maketitle              

\begin{abstract}
  This paper focuses on succinctness results for fragments of
  Linear Temporal Logic with Past (\LTL) devoid of binary temporal operators like \emph{until}, and provides methods to establish them. 
  We prove that there is a family of \emph{cosafety} languages $(\lang_n)_{n \geq 1}$ such that $\lang_n$ can be expressed
  with a \emph{pure future formula} of size $\O(n)$, but it requires formulae of size $2^{\Omega(n)}$ to be captured 
  with \emph{past formulae}.
  As a by-product, such a succinctness result shows the optimality of 
  the \emph{pastification algorithm} proposed in \mbox{\emph{[Artale et al., KR, 2023]}}. 
  
%
  We show that, in the considered case, succinctness
  cannot be proven by relying on
  the classical automata-based method introduced 
  in~\emph{[Markey, Bull.~EATCS, 2003]}. 
  In place of this method, we devise and apply a \emph{combinatorial proof system} 
  whose deduction trees represent \LTL formulae. 
  The system can be seen as a proof-centric (one-player) view on 
  the games used by Adler and Immerman to study the 
  succinctness~of~\CTL.

  \keywords{Temporal logics \and LTL \and Succinctness \and Proof systems.}
\end{abstract}

\input{sections/introduction}

\input{sections/background}
\input{sections/difficult-property}
\input{sections/proof-system}

\input{sections/lowerbound-monolitico}

\input{sections/pastification-lowerbounds}

\input{sections/automata-struggles}
\input{sections/related-works}

\subsubsection*{Acknowledgements}
Luca Geatti and Angelo Montanari acknowledge the support from the 2023
Italian INdAM-GNCS project \emph{``Analisi simbolica e numerica di sistemi
ciberfisici''}, ref.~no.~\texttt{CUP\_E53C22001930001}. Angelo Montanari
acknowledges the support of the MUR PNRR project FAIR - Future AI Research
(\texttt{PE00000013}) funded by the NextGenerationEU.
Alessio Mansutti is supported 
by the C\'esar Nombela grant \texttt{2023-T1/COM-29001}, 
funded by the Madrid~Regional~Government, 
and by the
grant \texttt{PID2022-138072OB-I00}, funded by \texttt{MCIN/AEI/10.13039/501100011033} (FEDER, EU).

\bibliographystyle{abbrv}
\bibliography{biblio}

\newpage
\appendix
\input{sections/appendix}

\end{document}

%% file: usepackages.tex
\usepackage{paralist} 
\usepackage[inline]{enumitem}
\usepackage{amsmath,amssymb}

\usepackage{amsthm}
\usepackage{thm-restate}
\usepackage{subfigure}
\usepackage{hyperref}
\usepackage[capitalise]{cleveref}
\usepackage{xspace}
\usepackage{mathpartir}
\usepackage[all]{foreign}
\usepackage{xcolor}
\usepackage{marginnote}
\usepackage{tikz}
\usepackage{todonotes}
\usepackage{packages/commands} 
\usepackage{packages/graphics} 
\usepackage{packages/logic}    
\usepackage{packages/symbols}  

\spnewtheorem{clm}{Claim}{\itshape}{}
\crefname{clm}{Claim}{Claims}

%% file: sections/introduction.tex
\section{Introduction}
\label{sec:introduction}

Linear Temporal Logic with Past
(\LTL~\cite{lichtenstein1985glory,pnueli1977temporal}) is the
\emph{de-facto} standard language for the specification, verification,
and synthesis of reactive systems~\cite{DBLP:books/daglib/0080029}. 
Concerning these reasoning tasks, 
two fundamental subsets of \LTL-definable languages
come into play, namely,
\emph{safety} and \emph{cosafety} languages.
Safety languages express properties 
stating that ``something bad never
happens''; cosafety languages, instead,  express the fact that ``something good will eventually happen''. 
The crucial feature of cosafety (resp., safety) languages is
that checking a \emph{finite prefix} of an infinite trace suffices to establish whether the
entire trace belongs (resp., does not belong) to the language.
Such an ability of reducing reasoning over infinite words to the finite case
plays a fundamental role 
in lowering the
complexity of reasoning tasks~\cite{kupferman2001model}.
Because of this, while \LTL was commonly interpreted over infinite traces,
recent work mainly considers its finite trace
semantics~\cite{DeGiacomoV13,DBLP:conf/aaai/MaggiMP20,DBLP:conf/bpm/PesicA06}.


In what follows, given a set of temporal operators $S$, we write $\LTL[S]$ for the set of all $\LTL$ formulae in \emph{negation normal form} whose temporal operators are restricted to those in $S$. 
Similarly, we denote with $\eventually(\LTL[S])$ the set of formulae of the form $\eventually(\alpha)$, with $\alpha \in \LTL[S]$. Here, $\eventually$ is the \emph{future} modality (a.k.a.~\emph{eventually}).

There are two notable syntactic characterizations of the cosafety languages of~\LTL. The first one is a \emph{pure future} characterization given by the logic $\LTL[\tomorrow,\until]$ featuring modalities \emph{next} $\tomorrow$ and \emph{until} $\until$. The second one is an \emph{eventually pure past}\footnote{``Eventually pure past'' refers to formulae of the form $\eventually(\alpha)$, with $\alpha$ pure past formula.} characterisation given by the logic~\Falpha, where~$\pLTL$ is the \emph{pure past} fragment of $\LTL$, that is, the restriction of $\LTL$ to past modalities. Analogous characterizations have been provided for safety languages.

As for applications, \Falpha is considered to be much more convenient than $\LTL[\ltl{X,U}]$, because, starting from an (eventually) pure past formula of size $n$, it is possible to build an equivalent deterministic finite automaton of singly exponential size in $n$~\cite{de2021pure}. In the case of $\LTL[\ltl{X,U}]$, such an automaton may have size doubly exponential in $n$~\cite{kupferman2001model}. This computational advantage of pure past formulae originated a recent line of research that focuses on the \emph{pastification problem}, i.e., the problem of translating an input \emph{pure future} formula for a cosafety (or safety) language into an equivalent \emph{pure past} (equivalently, \emph{eventually pure past}) formula. While the best known algorithm for $\LTL[\ltl{X,U}]$ 
is triply exponential~\cite{de2021pure}, a singly exponential pastification algorithm to transform \LTLxf formulae into $\ltl{F}(\LTL[\ltl{Y,wY,O}])$ ones has been recently developed in~\cite{ArtaleGGMM23}. Here, modalities \emph{yesterday}~$\ltl{Y}$ and \emph{once}~$\ltl{O}$ are the ``temporal reverses'' of modalities~$\ltl{X}$ and $\ltl{F}$, respectively, whereas the \emph{weak yesterday} operator $\ltl{wY}$ is the dual of~$\ltl{Y}$ (we formally define the semantics of all these modalities in~\Cref{sec:back}). No super-polynomial lower bounds for these pastification problems are known.

While the above two characterisations of cosafety languages have been thoroughly studied in the last decades in terms of expressiveness~\cite{ChangMP92} and complexity~\cite{DBLP:conf/aaai/ArtaleGGMM23}, their \emph{succinctness} is still poorly understood. To the best of our knowledge, the only known result is the one in~\cite{DBLP:conf/time/ArtaleGGMM23} showing that \Falpha \emph{can} be exponentially more succinct than~$\LTL[\ltl{X,U}]$ --- note that lower bounds to pastification problems require the opposite direction.\footnote{A logic $\mathbb{L}$ \emph{can be exponentially more succinct} than a logic $\mathbb{L}'$ whenever there is a family of languages $(\lang_n)_{n \geq 1}$ such that $\lang_n$ can be expressed in $\mathbb{L}$ with a formula of size polynomial in $n$, whereas expressing $\lang_n$ in $\mathbb{L}'$ requires formulae of size $2^{\Omega(n)}$.}

In this paper, we study the succinctness of $\LTLonlyF$ against $\eventually(\LTL[\ltl{Y,wY,O,H}])$, where $\ltl{H}$ is the dual of $\ltl{O}$, as well as the succinctness of their \emph{reverse logics}~\cite{DBLP:conf/time/ArtaleGGMM23}, that is, the succinctness of $\ltl{F}(\LTLPFinv)$ against $\LTLnountil$. For these fragments of~\LTL, we establish the following two results.
\begin{restatable}{theorem}{theoremLTLOsuccinct}
  \label{theorem:LTLO-more-succint}
  $\ltl{F}(\LTLPFinv)$ can be exponentially more succinct than
  $\LTLnountil$.
\end{restatable}
\vspace{-.25cm}%
\begin{restatable}{theorem}{theoremLTLFsuccinct}
  \label{theorem:LTLF-more-succint}
  \LTLonlyF can be exponentially more succinct than $\eventually(\LTLPnountil)$.
\end{restatable}

The two theorems prove an \emph{incomparability result} about the succinctness of the characterizations of cosafety languages in the pure future and eventually pure past fragments of \LTL. \Cref{theorem:LTLO-more-succint}  and~\Cref{theorem:LTLF-more-succint} hold for both the finite and infinite trace semantics of~\LTL (however, due to lack of space, we report the proof of~\Cref{theorem:LTLO-more-succint} only in the case of finite traces).
As a corollary,~\cref{theorem:LTLF-more-succint}
implies that the~pastification algorithm 
proposed in~\cite{ArtaleGGMM23} is optimal.
\begin{restatable}{corollary}{CorrOptimalPastification}
  \label{corollary:optimal-pastification}
  The pastification of~\LTLxf into $\eventually(\LTLPnountil)$ is in $2^{\Theta(n)}$.
\end{restatable}

To prove~\cref{theorem:LTLO-more-succint}, we devise and apply a \emph{combinatorial proof system}.\footnote{%
  We use the term ``combinatorial'' for our proof system to conform with the
  terminology from  the Workshop ``Combinatorial Games in Finite Model
  Theory'', LICS`23.}
Given two sets of finite traces $A$ and $B$, with the proof system one can establish whether there is a formula $\phi$ in \LTLnountil that \emph{separates} $A$ from $B$, 
that is, $\phi$ is satisfied by all traces in $A$
(written $A \models \phi$) and violated by all traces in $B$ (written $B
\pperp \phi$). A proof obtained by applying $k$ rules of the proof system corresponds to the existence of one such separating formula $\phi$ of size $k$.

The proposed combinatorial proof system can be seen as a reformulation in terms of proofs of the games introduced by Adler and Immerman to show that \CTLplus is $\Theta(n)!$ more succinct than \CTL~\cite{adler2003n}. They are two-player games that extend Ehrenfeucht--Fra\"iss\'e games for quantifier depth in a way that captures the notion of formula size instead. However, unlike Ehrenfeucht--Fra\"iss\'e ones, in Adler--Immerman games one of the two players (the duplicator) has always a trivial strategy. With our proof system, we show that removing the duplicator from the game yields a natural one-player game based on building proofs.

To prove~\cref{theorem:LTLO-more-succint} by applying the proposed proof system, we provide, for every $n \geq 1$, a formula $\Phi_n$ in $\ltl{F}(\LTLPFinv)$ of size linear in $n$ and two sets of traces $\AA_n$ and $\BB_n$ such that $\AA_n \models \Phi_n$ and $\BB_n \pperp \Phi_n$, and then we show that the smallest deduction tree that separates $\AA_n$ from $\BB_n$ has size at least $2^n$. This implies that all formulas of $\LTLnountil$ capturing $\Phi_n$ are of size at least $2^n$.

Once~\cref{theorem:LTLO-more-succint} is established, one can prove~\cref{theorem:LTLF-more-succint} by ``reversing'' the direction of time, building 
correspondences between formulae of $\LTL[\ltl{F}]$ and $\eventually{\LTL[\ltl{O}]}$, 
and between formulae of $\eventually(\LTL[\ltl{Y,wY,O,H}])$ and $\LTL[\ltl{X,wX,F,G}]$.

In the context of \LTL, the main technique to prove ``future
against past'' succinctness discrepancies is arguably the
automata method introduced by Markey in~\cite{markey2003temporal}. At
its core, such a method exploits the fact that pure
future formulae of \LTL can be translated into
nondeterministic B\"uchi automata of exponential size, and
thus no property requiring a doubly exponential size automaton
can be represented succinctly. The introduction of our proof
system raises the question of whether Markey's method can be
applied to establish our succinctness results. We prove that
it cannot be used in our context. In order to obtain such a result,
the key observation is that, given a cosafety formula $\eventually\psi$, 
a \emph{deterministic} B\"uchi automaton (\DBA) for $\eventually\psi$ 
of size $\ell$, and a prefix~$\Pi$ consisting of~$k$ temporal operators among $\tomorrow$,
$\eventually$, and $\always$, the minimal \DBA for the formula $\Pi \eventually\psi$
has size polynomial in $k$ and $\ell$.

\paragraph*{Synopsis.}
\cref{sec:back} introduces the necessary background.
\Cref{sec:difficult-property} discusses the languages we use
to prove~\cref{theorem:LTLO-more-succint}.
\cref{sec:psltlxf} introduces the combinatorial proof
system. In~\cref{sec:lbltlxf} we
prove~\cref{theorem:LTLO-more-succint}.
In~\cref{sec:past-lb} we
prove~\cref{theorem:LTLF-more-succint,corollary:optimal-pastification}.
The limits of the automata-based method to prove succinctness lower
bounds are discussed in~\cref{sec:automata-struggles}.
Related and future work are discussed
in~\cref{sec:related-and-future-works}. 
An extended version of the paper, complete of all proofs, can be found
in~\cite{geatti2024succinctness}.

%% file: sections/background.tex
\section{Preliminaries}
\label{sec:back}

In this section, we introduce  background knowledge on~\LTL focusing on finite traces. All definitions admit a
natural extension to the setting of infinite traces.

Let $\Sigma$ be a finite alphabet. We denote by $\Sigma^*$ the set of all finite words over~$\Sigma$ and 
by $\Sigma^+$ the subset of finite non-empty words. We use the term \emph{trace} as a synonym of word. 
A \emph{language} $\lang$ over $\Sigma$ is a subset of~$\Sigma^*$.  Let $\sigma = \seq{w_0,w_1,\dots,w_n}$ be a word in
$\Sigma^*$.  We denote by $|\sigma|$ the \emph{length} of $\sigma$, that is, $n+1$.  A \emph{position} in $\sigma$ is an element in the set ${\position(\sigma) \coloneqq [0,n]}$. For every $i \in \position(\sigma)$, we denote by $\sigma[i]\in \Sigma$ the letter $w_i$, and by $\sigma\sffx{i}$ the word $\seq{w_i,\dots,w_n}$.   We say
that  position $j$ of $\sigma$ has \emph{type} $\tau \in \Sigma$ whenever $\sigma[j] = \tau$. Given two traces
$\sigma_1$ and $\sigma_2$, we write $\sigma_1 \sqsubseteq
\sigma_2$ whenever $\sigma_1$ is a \emph{suffix} of
$\sigma_2$, that is, there is $j \in \position(\sigma_2)$
such that $\sigma_1 = \sigma_2\sffx{j}$.  Given a word
$\sigma'\in \Sigma^*$, we denote the \emph{concatenation of
$\sigma'$ to $\sigma$} as $\sigma\cdot\sigma'$, or simply
$\sigma\sigma'$. Given two languages $\lang$ and $\lang'$,
we define $\lang\cdot\lang' \coloneqq
\set{\sigma\cdot\sigma' \suchthat \sigma\in\lang ,
\sigma'\in\lang'}$. We sometimes apply the concatenation to
a word and a language; in these cases the word is implicitly
converted into a singleton language, e.g., $\sigma \cdot
\lang \coloneqq \{\sigma\} \cdot \lang$. 
With $A \subsetfin B$ we denote the fact that $A$ is a \emph{finite}
subset of the set $B$.

\paragraph{Linear Temporal Logic with Past.}
In the following, we introduce syntax and semantics of Linear
Temporal Logic with Past (\LTL) restricted to those operators that we
are going to use throughout the paper.  In particular, we omit the future operators
\emph{until} and \emph{release}, and their past counterparts
(\emph{since} and \emph{triggers}).  Let $\AP$ be a finite set of atomic
propositions. The syntax of the formulae over $\AP$ is generated by the
following grammar:
\begin{align*}
  \phi \coloneqq p & \choice \neg p \choice \phi\lor\phi \choice 
                           \phi\land\phi  & \text{Boolean connectives} \\
                   & \choice \ltl{X\phi} \choice \ltl{wX\phi} 
                     \choice \ltl{F \phi}
                     \choice \ltl{G \phi} & \text{future operators} \\
                   & \choice \ltl{Y\phi} \choice \ltl{wY\phi} 
                     \choice \ltl{O \phi}
                     \choice \ltl{H \phi} & \text{past operators}
\end{align*}
where $p\in\AP$. 
The temporal operators are respectively called:
$\ltl{X}$, \emph{next};
$\ltl{wX}$, \emph{weak next};
$\ltl{F}$, \emph{future};
$\ltl{G}$, \emph{globally};
$\ltl{Y}$, \emph{yesterday};
$\ltl{wY}$, \emph{weak yesterday};
$\ltl{O}$, \emph{once};
$\ltl{H}$, \emph{historically}.
For the rest of the paper, we let $\operators \coloneqq \set{\ltl{X,wX,F,G,Y,wY,O,H}}$.

For every formula $\phi$, we define the \emph{size of $\phi$}, denoted by
$\size(\phi)$, inductively defined as follows:
\begin{enumerate*}[label=(\roman*)]
  \item $\size(p) \coloneqq 1$ and $\size(\lnot p) \coloneqq 1$,
  \item $\size(\otimes \phi) \coloneqq \size(\phi) + 1$, for $\otimes \in \operators$, and 
  \item $\size(\phi_1 \oplus \phi_2) \coloneqq \size(\phi_1) +$ $\size(\phi_2) + 1$ for $\oplus \in \{\lor,\land\}$.
\end{enumerate*}

We focus on the interpretation of $\LTL$
formulae over \emph{finite non-empty traces} over
the alphabet $2^{\AP}$. From now on, we set the alphabet $\Sigma$ to be
$2^{\AP}$.  Given a word $\sigma\in\Sigma^+$, the \emph{satisfaction} of
a formula $\phi$ by $\sigma$ at time point / position $i \in \position(\sigma)$, 
denoted by $\sigma,i\models\phi$, is defined as follows:
\begin{conditions}
  \item $\sigma,i \models p$                 & $p\in\sigma[i]$;
  \item $\sigma,i \models \ltl{\neg p}$      & $p\not\in\sigma[i]$;
  \item $\sigma,i \models \ltl{\phi_1 || \phi_2}$  &
          $\sigma,i \models \phi_1$ or $\sigma,i \models \phi_2$;
  \item $\sigma,i \models \ltl{\phi_1 && \phi_2}$ &
          $\sigma,i \models \phi_1$ and $\sigma,i \models \phi_2$;
  \item $\sigma,i \models \ltl{X\phi}$     & 
          $i+1<|\sigma|$ and  $\sigma,i+1\models \phi$;
  \item $\sigma,i \models \ltl{wX\phi}$     & 
          either $i+1=|\sigma|$ or $\sigma,i+1\models \phi$;
  \item $\sigma,i \models \ltl{F \phi}$  &
          there exists $i\le j<|\sigma|$ such that $\sigma,j\models\phi$;
  \item $\sigma,i \models \ltl{G \phi}$  &
          for all $i\le j<|\sigma|$, it holds $\sigma,j\models\phi$;
  \item $\sigma,i \models \ltl{Y\phi}$    &
          $i > 0$ and $\sigma,i-1\models \phi$;
  \item $\sigma,i \models \ltl{Z\phi}$    &
          either $i = 0$ or $\sigma,i-1\models \phi$;
  \item $\sigma,i \models \ltl{O \phi}$  &
          there exists $0\le j\le i$ such that $\sigma,j\models\phi$;
  \item $\sigma,i \models \ltl{H \phi}$  &
          for all $0\le j \le i$, it holds $\sigma,j\models\phi$.
\end{conditions}
For every formula $\phi$, we say that a trace $\sigma$ satisfies
$\phi$, written $\sigma\models\phi$, if $\sigma,0\models\phi$. The
\emph{language} of $\phi$, denoted by $\lang(\phi)$, is the set of words
$\sigma\in\Sigma^+$ such that $\sigma\models\phi$. Given two formulae
$\phi$ and $\psi$, we say that $\phi$ is \emph{equivalent} to $\psi$,
written $\phi \equiv \psi$, whenever $\lang(\phi) = \lang(\psi)$.

\paragraph{Fragments of \LTL.}
Given a set of operators $S \subseteq \operators$, we denote by
$\LTL[S]$ the set of formulae only using temporal operators from~$S$. 
When dealing with
a concrete~$S$, we omit the curly brackets and write, e.g.,
\LTLxf instead of $\LTL[\set{\ltl{X,F}}]$.
Whenever $S$
contains only future operators (resp., past operators), the logic $\LTL[S]$
is called a \emph{pure future} (resp., \emph{pure past}) fragment of \LTL.
Finally, we denote by $\ltl{F}(\LTL[S])$ (resp., $\ltl{G}(\LTL[S])$) the
set of formulae of the form $\ltl{F(\alpha)}$ (resp., $\ltl{G(\alpha)}$), where
$\alpha$ is a formula of $\LTL[S]$.
A language $\lang \subseteq \Sigma^*$ is a \emph{cosafety language}
whenever $\lang = K \cdot \Sigma^*$, for some $K \subseteq \Sigma^*$.
A language $\lang$ is a \emph{safety language} whenever its complement
$\overline{\lang}$ is a cosafety language.  For every formula~$\phi$ in the
fragments $\LTL[\ltl{X,F}]$ and $\ltl{F}(\LTL[\ltl{Y,wY,O,H}])$, 
it holds that $\lang(\phi)$ is a cosafety language. Similarly,
for every formula $\phi$ in the fragments $\LTL[\ltl{wX,G}]$ and
$\ltl{G}(\LTL[\ltl{Y,wY,O,H}])$, it holds
that $\lang(\phi)$ is a safety language.

\paragraph{The pastification problem.}
Given two sets $S \subseteq \set{\ltl{X,wX,F,G}}$ and $S' \subseteq
\set{\ltl{Y,wY,O,H}}$, the \emph{pastification problem for $\LTL[S]$ into $\ltl{F}(\LTL[S'])$} asks, given an input
formula $\phi \in \LTL[S]$, to return a formula $\psi$ from $\ltl{F}(\LTL[S'])$ such that $\phi \equiv \psi$.
An algorithm for the
pastification problem is said to be of \emph{$k$-exponential size} (for $k\in\N$ fixed) whenever the output formula $\psi$ is such that $\size(\psi)
\in \exp_2^k(\poly(\size(\phi)))$, where $\exp^k(.)$ is the $k$-th
iteration of the base-$2$ tetration function given by $\exp^0(n) = n$
and $\exp^{i+1}(n) = 2^{\exp^i(n)}$. 
In~\cite{ArtaleGGMM23}, an exponential time, \mbox{$1$-exponential size}, pastification algorithm for $\LTL[\ltl{X,F}]$ into $\ltl{F}(\LTL[\ltl{Y,wY,O}])$ is presented.

\paragraph{Succinctness.}
Given two sets $S,S' \subseteq \operators$, we say that $\LTL[S]$
\emph{can be exponentially more succinct than $\LTL[S']$} if there is a family of languages
$(\lang_n)_{n \geq 1}$ such that, for every $n \geq 1$, $\lang_n \subseteq \Sigma_n^+$, for some alphabet $\Sigma_n$, and:
\begin{itemize}
  \item there is $\phi \in \LTL[S]$ 
  such that $\lang(\phi) = \lang_n$ 
    and $\size(\phi) \in \poly(n)$, and
  \item for every $\psi \in \LTL[S']$, if $\lang(\psi) = \lang_n$ then $\size(\psi) \in 2^{\Omega(n)}$.
\end{itemize}
It is worth noticing that the above-given syntax for \LTL is already in \emph{negation normal form}, that is,
negation may only appear in front of atomic propositions.  Allowing negations to occur freely in the formula neither increase expressiveness nor succinctness, as the grammar above is already closed under dual operators, e.g., $\always \phi \equiv \lnot \eventually \lnot \phi$, and the size of a formula does not depend on the number of
negations occurring in literals. Because of this, all results given in the paper continue to
hold when negation is added to the language.%

%% file: sections/difficult-property.tex

\section{A problematic cosafety language for $\LTL[\tomorrow,\weaktomorrow,\eventually,\always]$}
\label{sec:difficult-property}

We now describe the property that we will exploit to prove that $\ltl{F}(\LTL[\ltl{O}])$ can be exponentially more succinct than $\LTL[\ltl{X,wX,F,G}]$
(\Cref{theorem:LTLO-more-succint}). 
More precisely, we define a family of $\eventually(\LTLPFinv)$ formulae $(\Phi_n)_{n \ge 1}$ 
such that, for every $n \geq 1$, $\Phi_n$ has size in $\O(n)$ and captures a property requiring a formula of size at least~$2^n$ to be expressed in $\LTLnountil$ (as we will see in~\Cref{sec:lbltlxf}).

Let $n \geq 1$. We consider the alphabet of $2n+2$ distinct atomic propositions ${\AP \coloneqq \{\widetilde{p},\widetilde{q}\} \cup P \cup Q}$, with $P \coloneqq \{p_1,\dots,p_n\}$ and $Q \coloneqq
\{q_1,\dots,q_n\}$. For all $n \geq 1$, the formula~$\Phi_n$ of $\eventually(\LTLPFinv)$ is defined as follows:
\[ 
  \textstyle
  \Phi_n \coloneqq \eventually \left( \widetilde{q} \land \bigwedge_{i=1}^n 
  \Big( 
    \big(q_i \land \once( \widetilde{p} \land p_i)\big) 
    \lor
    \big(\lnot q_i \land \once (\widetilde{p} \land \lnot p_i)\big)
  \Big)
  \right).
\]
Observe that, 
for every $n \geq 1$, $\size(\Phi_n)$ belongs to $\mathcal{O}(n)$.
The formula~$\Phi_n$ is satisfied by those traces~$\sigma \in \Sigma^+$ where there is a position $j
\in \position(\sigma)$ such that
\begin{enumerate*}[label=(\roman*)]
  \item $\widetilde{q} \in \sigma[j]$ and 
  \item for every $i \in [1,n]$ there is a position $k_i \in [0,j]$ such that
        $\widetilde{p} \in \sigma[k_i]$ and $q_i \in \sigma[j]$ \emph{if and only if} $p_i \in \sigma[k_i]$.
\end{enumerate*}
Notice that each $k_i \in [0,j]$ depends on an index $i \in [1,n]$. Therefore, for distinct $i,j \in [1,n]$ the positions $k_i$ and $k_j$ might differ. This feature is crucial to get a language which has 
a compact definition in $\eventually(\LTLPFinv)$, but is hard to capture for $\LTL[\ltl{X,wX,F,G}]$.  

As a matter of fact, requiring the various $k_i$ to coincide yields a formula $\Psi_n$ characterising
the property: ``the trace $\sigma$ has two positions $j \geq k$ such that $\widetilde{p} \in \sigma[k]$, $\widetilde{q} \in \sigma[j]$ and, for every~$i \in [1,n]$, $q_i \in \sigma[j]$ if and only if $p_i \in \sigma[k]$''. 
This formula is known to require exponential size in~\LTL~\cite{markey2003temporal}, and therefore in $\eventually(\LTLPFinv)$ as well. 
In a sense, the asymmetry obtained by relaxing the uniqueness of the position $k$ above is what makes $\Phi_n$ being easily expressible in $\eventually(\LTLPFinv)$, but difficult to characterise in $\LTLnountil$. The same trick, applied to position $j$ instead of position $k$, can be used to obtain a family of formulae that can be represented in an exponentially more succinct way in \LTLonlyF than in $\eventually(\LTLPnountil)$. 
This form of ``temporal duality'' is what we will ultimately exploit in~\Cref{sec:past-lb} to prove~\Cref{theorem:LTLF-more-succint}. 


The following lemma shows that $\Phi_n$ can be expressed in $\LTLonlyF$
(and thus in $\LTLnountil$ as well) with a formula of exponential size.

\begin{restatable}{lemma}{LemmaAFormulaForPhiN}
  \label{lemma:a-formula-for-Phi-n}
  For every $n \geq 1$, there is a formula $\Phi_n'$ in $\LTLonlyF$ such that~$\Phi_n' \equiv \Phi_n$ and~$\size(\Phi_n') < 2^{n+1} (n+2)^2$.
\end{restatable}
\begin{proof}[Proof sketch]
  Given $\tau \in 2^P$, we write $\overline{\tau}$ for the
  element of $2^Q$ such that $p_i \in \tau$ if and only if $q_i \in
  \overline{\tau}$, for every $i \in [1,n]$.  
  Then, the formula $\Phi_n'$ is defined as follows:
  \[ 
    \textstyle 
    \Phi_n' \coloneqq \bigvee_{\tau \,\in\, 2^{P}}
      \Big( 
        \bigwedge_{p \,\in\, \tau} \eventually \big(\widetilde{p} \land p \land \eventually (\widetilde{q} \land \psi_{\overline{\tau}})\big)
        \land 
        \bigwedge_{p \,\in\, P \setminus \tau} \eventually \big(\widetilde{p} \land \lnot p \land \eventually (\widetilde{q} \land \psi_{\overline{\tau}})\big)
      \Big),
  \]
  where $\psi_{\overline{\tau}} \coloneqq (\bigwedge_{q \,\in\,
  \overline{\tau}} q \land \bigwedge_{q \,\in\, Q \setminus
  \overline{\tau}} \lnot q)$. 
\end{proof}

%% file: sections/proof-system.tex

\section{A combinatorial proof system for~$\LTL[\tomorrow,\weaktomorrow,\eventually,\always]$}
\label{sec:psltlxf}

In this section, we introduce the proof system that we  will later
employ to prove~\Cref{theorem:LTLO-more-succint}, and discuss its connection
with Adler--Immerman games~\cite{adler2003n}.

\paragraph{Further notation.}
Let $A \subseteq \Sigma^+$, with $\Sigma \coloneqq 2^{\AP}$ for some set of
propositions~$\AP$.  We define $A^{\ltl{X}} \coloneqq \set{\sigma\sffx{1}
: \sigma \in A \text{ s.t. } \abs{\sigma} \geq 2}$, i.e., 
the set of non-empty traces obtained from $A$ by stepping each trace one position to the right. 
We define $A^{\ltl{G}}
\coloneqq {\{ \sigma\sffx{j} : \sigma \in A \text{ and } j \in
\position(\sigma)\}}$, i.e.,
the set of all suffixes of the traces in~$A$.
We say that a map $f : A \to \N$ is a \emph{future
point} for $A$ whenever $f(\sigma) \in \position(\sigma)$ for every $\sigma
\in A$. We write $\FP_A$ for the set of all maps that are future points
for~$A$. Given a future point $f$ for $A$ and $\sigma \in A$ with
$f(\sigma) = i$, we define $\sigma^f \coloneqq \sigma\sffx{i}$ and ${A^f
\coloneqq \set{ \sigma^f : \sigma \in A }}$.  Note that, by definition, $A^{\ltl{G}} = \bigcup_{f \in \FP_A}
A^f$.

For a formula $\phi$ of~\LTL, we write $A \models \phi$ whenever
$(\sigma,0) \models \phi$ for every $\sigma \in A$, and $A \pperp \phi$
whenever $(\sigma,0) \not\models \phi$ for every $\sigma \in A$. Given two
sets of traces $A,B \subseteq \Sigma^+$ we say that $\phi$ \emph{separates}
$A$ from $B$ whenever $A \models \phi$ and $B \pperp \phi$. We write
$\separ{\cdot,\cdot}_S \subseteq \Sigma^+ \times \Sigma^+$ for the
\emph{separable relation on} $S \subseteq \operators$, i.e., the binary
relation holding on pairs $(A,B)$ whenever there is some formula from
$\LTL[S]$ that separates $A$ from~$B$. Note that, when $A$ and $B$ are
finite sets and $\ltl{X} \in S$, deciding whether $\separ{A,B}_{S}$ holds is
trivial. 

\begin{restatable}{lemma}{lemmaTrivialX}
  \label{lemma:unclosable-tree-condition}
  Let $A,B \subseteq \Sigma^+$ and $S \subseteq \operators$. Then, $\separ{A,B}_S$ implies $A \cap B = \emptyset$.
  Moreover, if $A$ and $B$ are finite sets and $\tomorrow \in S$,
  $A \cap B = \emptyset$ implies $\separ{A,B}_S$.
\end{restatable}

\begin{proof}[Proof sketch]
  For the first statement, clearly if $A \cap B \neq \emptyset$ then it is not possible to separate $A$ from $B$. 
  To prove the second statement, one defines a disjunction $\phi$ of formulae, each characterising an element in $A$. For instance, for $\AP = \{p,q\}$, the trace $\{p\}\{q\}$ can be characterised 
  with the formula $(p \land \lnot q) \land \tomorrow (q \land \lnot p \land \weaktomorrow \bot)$, where $\bot \coloneqq p \land \lnot p$.
  Then, $\phi$ 
  separates $A$ from $B$.
\end{proof}

We mainly consider the relation $\separ{\cdot,\cdot}_S$ with $S$ being the
set $\{\tomorrow,\weaktomorrow,\eventually,\always\}$, and thus from now on
simply write $\separ{\cdot,\cdot}$ when considering this concrete choice of $S$.

\subsection{The proof system}
The combinatorial proof system that we define is a
natural-deduction-style proof system. It is made of several
inference rules of the form $\frac{H_1 \ H_2 \ \dots \
H_n}{C}$, to be read as ``if the hypotheses $H_1,\dots,H_n$
hold, then the consequence $C$ holds''.  As usual, proofs
within the proof system have a tree-like presentation. An
example of such a \emph{deduction tree} is given
in~\Cref{figure:deduction-example}, where $a \coloneqq
\set{p}$ and $b \coloneqq \emptyset$, with ${p \in \AP}$.
This is a deduction tree for the \emph{term}
$\separ{\{abaa,aaaa\},\{aaab\}}$, which we call the
\emph{root} of the deduction tree.
In~\Cref{figure:deduction-example}, to the root it is
\emph{applied} the rule~\ruleOr, with hypotheses
$\separ{\{abaa\},\{aaab\}}$ and~$\separ{\{aaaa\},\{aaab\}}$.
In turn, these two hypotheses are derived in the deduction
tree by eventually reaching applications to the
rule~\ruleAtomic.  A deduction tree is always \emph{closed}:
all maximal paths from the root ends with an application of
the rule~\ruleAtomic.  This means that a rule of the proof
system must be applied to each term $\separ{A,B}$ appearing
in the tree.  We call a tree a \emph{partial deduction tree}
if this property is not enforced, namely when there might be
unproven terms $\separ{A,B}$. The \emph{size} of a deduction
tree is the number of rules in it. For instance, the tree in~\Cref{figure:deduction-example} has size $5$. 

\begin{figure}[t]
  \flushright
  \scalebox{0.95}{%
    \begin{mathpar}%
      \ \
      \and
      \inferrule*[Left=\ruleAtomic,Right={$\alpha {\normalfont\text{ literal}}$}]{A \models \alpha \quad B \pperp \alpha}{\separ{A,B}}
      \hspace{2.4cm}
      \inferrule*[Left=\ruleOr]{\separ{A_1,B} \quad \separ{A_2,B}}{\separ{A_1 \uplus A_2, B}}
      \hspace{1.3cm}
      \inferrule*[Left=\ruleAnd]{\separ{A,B_1} \quad \separ{A,B_2}}{\separ{A, B_1
      \uplus B_2}}\\
      \inferrule*[Left=\ruleNext]{\separ{A^{\ltl{X}},B^{\ltl{X}}} \and |A^{\ltl{X}}| = |A|}{\separ{A,B}}
      \hspace{2.4cm} 
      \inferrule*[Left=\ruleWeakNext]{\separ{A^{\ltl{X}},B^{\ltl{X}}} \and \abs{B^{\ltl{X}}} = \abs{B}}{\separ{A,B}}\\
      \inferrule*[Left=\ruleFuture,Right={$f \in \FP_A$}]{\separ{A^f,B^{\ltl{G}}}}{\separ{A,B}}
      \hspace{3.5cm} 
      \inferrule*[Left=\ruleGlobally, Right={$f \in \FP_B$}]{\separ{A^{\ltl{G}},B^f}}{\separ{A,B}}
      \hspace{1.1cm}
      \,
    \end{mathpar}%
  }%
  \caption{The combinatorial proof system. Here, $A,B \subseteq \Sigma^+$.}
  \label{figure:combinatorial-proof-system}
\end{figure}%
\begin{figure}[t]
    \centering
    \scalebox{0.9}{%
      {\begin{mathpar}%
        \inferrule*[Left = \ruleOr]{ \inferrule*[Left =
          \ruleNext]{ \inferrule*[Left = \ruleAtomic]{ \{baa\} \models \lnot p \and \{aab\} \pperp \lnot p }{\separ{\{baa\},\{aab\}}}
              }{\separ{\{abaa\},\{aaab\}}}
          \and 
          \inferrule*[Right =
            \ruleGlobally]{ 
              \inferrule*[Right =
                \ruleAtomic]{\{aaaa,aaa,aa,a\} \models p  \and \{b\} \pperp p}{
                  \separ{\{aaaa,aaa,aa,a\},\{b\}}
              }
            }{\separ{\{aaaa\},\{aaab\}}}
        }{\separ{\{abaa,aaaa\},\{aaab\}}}
      \end{mathpar}}%
    }%
    \hspace{0.2cm}\,
  \caption{A deduction tree proving $\separ{\{abaa,aaaa\},\{aaab\}}$. Here, $a \coloneqq \{p\}$ and~${b \coloneqq \emptyset}$.}\label{figure:deduction-example}
  \vspace{-10pt}
\end{figure}

We define the inference rules of the proof system
in~\Cref{figure:combinatorial-proof-system}. Let us briefly describe these rules.  
The \ruleAtomic rule allows deriving $\separ{A,B}$ if every
trace in $A$ satisfies some literal $\alpha$ and every trace in $B$
violates $\alpha$. The \ruleOr rule 
corresponds the case of $A$ being separable from $B$ via a formula of the form  $\phi_1 \lor \phi_2$. In this and the rule~\ruleAnd, $\uplus$ stands for the union of disjoint sets. Intuitively, \ruleOr can be applied by proving
that $\phi_1$ separates a set~$A_1 \subseteq A$ from $B$
\emph{and} that $\phi_2$ separates the set $A \setminus A_1$ from $B$.  The \ruleNext rule allows separating $A$ from~$B$ with
a formula of the form $\ltl{X\phi}$, by checking whether the sets obtained by
stepping all traces in $A$ and $B$ to next time point are separable by
$\phi$. The condition $|A^{\ltl{X}}| = |A|$ is necessary to ensure that all
traces in $A$ have a next time point. The \ruleFuture rule separates $A$
from $B$ by following this principle: if the set obtained by choosing
one suffix for every trace in $A$ is separable from the set of all suffixes of the
traces in $B$, then there is a formula of the form $\eventually \phi$ separating $A$ from~$B$.
The rules~\ruleAnd, \ruleWeakNext and \ruleGlobally are 
designed to be duals of the rules~\ruleOr, \ruleNext and \ruleFuture, respectively.

By using the proof system one can derive whether a pair of (finite or
infinite) sets of traces $(A,B)$ is in the separable relation
$\separ{\cdot,\cdot}$. Because of~\Cref{lemma:unclosable-tree-condition},
this is not, however, a particularly useful application. Instead, the proof
system is to be used to derive non-trivial lower (or upper) bounds on the
size of the minimal formula that separates $A$ from $B$. This is done by
studying the sizes of the possible deduction trees of $\separ{A,B}$ in the
proof system.

For instance, the deduction tree of~\Cref{figure:deduction-example} shows that there 
is a formula~$\phi$ having $\size(\phi) = 5$ and separating $\{abaa,aaaa\}$ from $\{aaab\}$.
This formula is found by simply reading bottom-up, starting from the root,
the rules in the deduction tree, associating to each rule the homonymous
operator of \LTL.  In the case of the tree
in~\Cref{figure:deduction-example} we have $\phi \coloneqq {(\tomorrow
\lnot p) \lor \always p}$.  Note that the formula $\phi$ is not the
smallest separating formula, because the formula $\ltl{X X G p}$ also
separates $\{abaa,aaaa\}$ from $\{aaab\}$ and corresponds to a tree of
size $4$.

The correspondence between deduction trees and formulae is
formalised in the next theorem (we remark that $A$ and $B$
below do not need to be finite sets).

\begin{restatable}{theorem}{MainTheoremProofSystem}
\label{th:ps}
  Consider $A,B \,\subseteq \,\Sigma^+$. Then, the term $\separ{A,B}$ has
  a deduction tree of size~$k$ if and only if there is a formula~$\phi$ of
  \LTLnountil separating $A$ from $B$ and such that $\size(\phi) = k$.
\end{restatable}

\begin{proof}[Proof sketch]
  We leave to the reader the proof of the left to right direction of the theorem (shown by induction on $k$), as it is not required to establish lower bounds on the sizes of formulae, 
  and focus instead on the right to left direction.

  Consider a \LTLnountil formula~$\phi$ that separates $A$ and
  $B$. We construct a deduction tree of size $\size(\phi)$.  
  We proceed by structural induction on $\phi$.  

  \proofitem{base case: $\phi$ literal} 
    The deduction tree consists of a single rule~\ruleAtomic. 

  \proofitem{induction step, case: $\phi = \phi_1 \lor \phi_2$} 
    Define $A_1 \coloneqq \{ a \in A : a \models \phi_1
    \}$ and $A_2 \coloneqq A \setminus A_1$. From $A
    \models \phi$ and $B \pperp \phi$ we get $A_i \models \phi_i$ and $B \pperp
    \phi_i$  for both $i \in \{1,2\}$. 
    By induction hypothesis $\separ{A_i,B}$ has
    a deduction tree of size $\size(\phi_i)$. 
    By applying the rule \ruleAnd, we obtain a
    deduction tree for $\separ{A,B}$ having size
    $\size(\phi_1) + \size(\phi_2) + 1 = \size(\phi)$.

  \proofitem{induction step, case: $\phi = \ltl{X} \psi$} 
    Since $A \models \ltl{X} \psi$, for every $\sigma
    \in A$ we have $\abs{\sigma} \geq 2$ and  $(\sigma,1)
    \models \psi$. By definition of $A^{\ltl{X}}$,
    $\abs{A^{\ltl{X}}} = \abs{A}$ and $A^{\ltl{X}} \models \psi$.
    From $B \pperp \ltl{X} \psi$, for every
    $\sigma' \in B$, if $\abs{\sigma'} \geq 2$ then
    $(\sigma',1) \not\models \psi$. By definition
    of~$B^{\ltl{X}}$, we have $B^{\ltl{X}} \pperp \psi$.
    By induction hypothesis, $\separ{A^{\ltl{X}},B^{\ltl{X}}}$ has a deduction tree 
    of size $\size(\psi)$. We apply the rule~\ruleNext 
    to obtain a deduction tree of $\separ{A,B}$ of size $\size(\psi)+1=\size(\phi)$.

  \proofitem{induction step: $\phi = \ltl{F} \psi$} 
    Since $A \models \ltl{F} \psi$, for every $\sigma \in A$ 
    there is $j_\sigma \in \position(\sigma)$
    such that $(\sigma,j_\sigma) \models \psi$. Let $f \in
    \FP_A$ be the map given by $f(\sigma) =
    j_{\sigma}$ for every $\sigma \in A$. 
    We have $A^f \models \psi$. We show
    that $B^{\ltl{G}} \pperp \psi$. 
    \emph{Ad absurdum},
    suppose there is $\sigma_1 \in
    B^{\ltl{G}}$ such that $\sigma_1 \models \psi$. By
    definition of $B^{\ltl{G}}$ there is $\sigma_2 \in B$ 
    such that $\sigma_1 \sqsubseteq \sigma_2$. 
    Then, $(\sigma_2,0) \models \eventually \psi$. 
    However, this contradicts the fact that
    $B \pperp \ltl{F} \psi$. 
    Therefore,  $B^{\ltl{G}} \pperp \psi$. 
    By induction hypothesis, $\separ{A^f, B^{\ltl{G}}}$ has
    a deduction tree of
    size $\size(\psi)$. 
    By applying the rule~\ruleFuture, 
    we obtain a deduction tree for $\separ{A,B}$ 
    of size $\size(\psi)+1 = \size(\phi)$. 

  \proofitem{induction step, cases $\phi = \phi_1 \land \phi_2$, $\phi = \weaktomorrow \psi$ and $\phi = \always \psi$} 
    The cases for $\phi = \phi_1 \land \phi_2$, $\phi = \weaktomorrow \psi$ and $\phi = \always \psi$ are analogous to the cases $\phi = \phi_1 \lor \phi_2$, $\phi = \tomorrow \psi$ and $\phi = \eventually \psi$, respectively.
\end{proof}

The right to left direction of
\Cref{th:ps} implies the following corollary that highlights how our proof system is used for formulae sizes lower bounds.

\begin{corollary}
  \label{corr:cc}
  Consider a formula $\phi$ in~\LTLnountil. Suppose that
  \begin{enumerate*}[label=(\roman*)]
    \item there are $A,B \subseteq \Sigma^+$ such that $\phi$ separates $A$ from $B$, and 
    \item every deduction tree of $\separ{A,B}$ has size at least~$k$.  
  \end{enumerate*}
  Then, $\size(\phi) \geq k$.
\end{corollary}

\subsection{Connections with the Adler--Immerman games}
\label{sec:connections-wit-Ad-Imm-games}

As outlined in~\Cref{sec:introduction},
our proof system can be seen as an adaptation of the games for \CTL introduced by Adler and Immerman in~\cite{adler2003n}.
We now illustrate this connection. 
Readers that are mostly interested in seeing our proof system in action may want to skip to~\Cref{sec:lbltlxf}.

The \emph{Adler--Immerman
games} extend the classical
Ehrenfeucht--Fraïssé games in order to bound 
the~\emph{sizes} of the formulae that separate two (sets of) structures, instead of their quantifier depths. 
As in the Ehrenfeucht--Fraïssé games, the {Adler--Immerman
games} are two-player games between a \emph{spoiler} and a
\emph{duplicator}. The game arena is a pair of sets of
structures~$(A,B)$, and at each round of the game the spoiler 
choses a rule~$r$ to play (there is one rule for each Boolean connective and operator of the logic) 
and plays on one set of structures accordingly to what~$r$ dictates. The duplicator replies on the other set, again accordingly to~$r$.
The goal of the spoiler is to separate $A$ from $B$ (i.e.,~to
show $\separ{A,B}$ in the context of~\CTL) in fewer rounds as
possible, whereas the duplicator must prolong the game as much
as she can. The length of the minimal game corresponds to
the size of the minimal formula separating $A$ from~$B$. The
main difference between an Adler--Immerman game and
an Ehrenfeucht--Fraïssé game is that, in the former, in each
round the duplicator is allowed to make copies of the structures
in the set she is playing on, and to play differently in each of these copies. This extra power given to the duplicator
is why the games end up capturing the notion of size of a
formula.

In the setting of the Adler--Immerman games, the rule for the operator~$\eventually$ in \LTL would be spelled as follows: 
\textit{``For each structure $\sigma \in A$, the spoiler moves to a future position of $\sigma$ (i.e., $\sigma\sffx{j}$ for some $j \in \position(\sigma)$). The duplicator answers by first making as many copies of elements in $B$ as she wants, and then selects a future position for each of these copies''}.
Because she can make copies, the duplicator 
has a trivial optimal strategy: at each round, copy the structures in $B$ as much as possible, choosing a different position in each copy. The rule for~$\eventually$ the simplifies to
\textit{``For each structure $\sigma \in A$, the spoiler moves to a future position of $\sigma$. The duplicator answers with $B^{\ltl{G}}$''}, which corresponds to our rule~\ruleFuture. 

While Adler and Immerman discuss the fact that the duplicator has a trivial optimal strategy, they do not restate the games with only one player (mainly to not lose the similarity with the Ehrenfeucht--Fraïssé games). Our work shows that removing the duplicator yields a natural one-player game 
based on building proofs within a proof system.
We think that this proof-system view has a few merits over the games. 
When proving lower bounds, it reduces the clumsiness of discussing the various moves of the spoiler and the replies of the duplicator.
The combinatorics is of course still there, but not the players, and this substantially simplifies the exposition.
Second, 
the proof system resembles 
the way in which one reasons
about the \emph{algorithmic} problem of separating $A$ from $B$. 
For instance, the algorithm presented in~\cite{Neider18} 
uses decision trees for solving this problem. These decision trees, when they encode a formula from~\LTLnountil, 
can be easily translated into proofs in our proof system.
We discuss more this line of work connected with \LTL~formulae learning and explainable planning in~\Cref{sec:related-and-future-works}.%

%% file: sections/lowerbound-monolitico.tex
  \section{The exponential lower bound for $\Phi_n$}
  \label{sec:lbltlxf}

  In this section, we show that, for every $n \geq 1$, all
  formulae of \LTLnountil characterising the~$\ltl{F}(\LTLPFinv)$
  formula~$\Phi_n$ defined in~\Cref{sec:difficult-property}
  have size at least~$2^n$. 
  According to the definition of $\Phi_n$, we consider a set of $2n+2$ distinct atomic propositions $\AP \coloneqq \{\widetilde{p},\widetilde{q}\} \cup P \cup Q$, with $P \coloneqq \{p_1,\dots,p_n\}$ and $Q \coloneqq \{q_1,\dots,q_n\}$; 
  and $\Sigma \coloneqq 2^{\AP}$.
  Throughout the section, let
  $\alpha(n) \coloneqq  2^{n+1} (n+2)^2$, i.e., the upper bound
  given in~\Cref{lemma:a-formula-for-Phi-n} for one of these
  formulae.

  Following~\Cref{corr:cc}, to prove the exponential lower bound we
  \begin{enumerate}
    \item define $\AA,\BB \subseteq \Sigma^+$ such that
    $\Phi_n$ separates $\AA$ from $\BB$ (\Cref{sec:lb:A-and-B}),
    and
    \item prove that every deduction tree for $\separ{\AA,\BB}$ has
    size at least $2^n$ (\Cref{sec:lb:exponential-proof}).%
  \end{enumerate}

  \subsection{Setting up the sets of traces $\AA$ and $\BB$}
  \label{sec:lb:A-and-B}

  We define the sets of \emph{types} $T_P \coloneqq \{ \tau \in \Sigma
  : \widetilde{p} \in \tau \text{ and } \tau \subseteq P \cup
  \{\widetilde{p}\} \}$ and $T_Q \coloneqq \{ \tau \in \Sigma
  : \widetilde{q} \in \tau \text{ and } \tau \subseteq Q \cup
  \{\widetilde{q}\} \}$.  Similarly to what done in the proof
  of~\Cref{lemma:a-formula-for-Phi-n}, we write $\overline{(\cdot)}$ for
  the involution on $T_P \cup T_Q$ sending a type $\tau \in T_Q$ into the
  (only) type~${\overline{\tau} \in T_P}$ with $q_i \in \tau$ if and only
  if~$p_i \in \overline{\tau}$, for every $i \in [1,n]$.

  Throughout the section, we fix a (arbitrary) strict total order~$\prec$ on the elements of~$T_Q$.
  Then, we denote by~$\Enum \in (\emptyset^{\alpha(n)} \cdot T_Q)^{2^n}
  \cdot \emptyset^{\alpha(n)}$ the (only) finite word enumerating
  all elements in $T_Q$, with respect to the order~$\prec$. 
  Note that, in~$\Enum$, between any two subsequent elements of $T_Q$
  there are $\alpha(n)$ positions of type~$\emptyset$. 
  This ``padding'' added to the enumeration is required to handle the rules~\ruleNext and~\ruleWeakNext.
  Given $\tau \in T_Q$, we write $\Enum|_{-\tau}$ for the
  trace obtained from~$\Enum$ by eliminating
  the only position of type $\tau$, together with the
  $\alpha(n)$ positions of type $\emptyset$ preceding it. 
  So, $\Enum_{-\tau}$ belongs to ${(\emptyset^{\alpha(n)} \cdot T_Q)^{2^n-1} \cdot
  \emptyset^{\alpha(n)}}$.

  For instance, consider the case of $n = 2$, so $Q = \{q_1,q_2\}$ and $\alpha(n) = 128$. 
  Suppose $\{\widetilde{q}\} \prec \{\widetilde{q},q_1\} \prec \{\widetilde{q},q_2\} \prec \{\widetilde{q},q_1,q_2\}$ to be the strict order on $T_Q$.
  Then,
  \begin{align*}
    \Enum &= \emptyset^{128} \cdot \{\widetilde{q}\} \cdot \emptyset^{128} \cdot \{\widetilde{q},q_1\} \cdot
      \emptyset^{128} \cdot \{\widetilde{q},q_2\} \cdot
      \emptyset^{128} \cdot \{\widetilde{q},q_1,q_2\} \cdot
      \emptyset^{128},\\
    \Enum|_{-\{\widetilde{q},q_2\}} &=
      \emptyset^{128} \cdot \{\widetilde{q}\} \cdot \emptyset^{128} \cdot \{\widetilde{q},q_1\} 
      \cdot
      \emptyset^{128} \cdot \{\widetilde{q},q_1,q_2\} \cdot
      \emptyset^{128}.
  \end{align*}

  For the rest of the paper, we denote with $\AA$ and $\BB$ the sets:
  \begin{center}
  $\AA \coloneqq \{ \emptyset^j \cdot \overline{\tau} \cdot \Enum : j \in \N, \tau \in T_Q\}$, \qquad $\BB \coloneqq \{ \emptyset^j \cdot \overline{\tau} \cdot (\Enum|_{-\tau}) : j \in \N, \tau \in T_Q \}$. 
  \end{center}

  \begin{lemma}
    \label{lemma:phi-n-separates}
    The formula~$\Phi_n$ separates $\AA$ from $\BB$.
  \end{lemma}

  \begin{proof}
    Let $j \in \N$ and $\tau \in T_Q$.
    In a nutshell, the fact that $\emptyset^j \cdot \overline{\tau} \cdot \Enum \models \Phi_n$ follows from the fact that $\tau$ occurs in $\Enum$, and from the position corresponding to $\tau$ one can refer back to $\overline{\tau}$ and find in this way a position satisfying $p_i$ if and only if $q_i \in \tau$, for every $i \in [1,n]$. 
    However, since $\tau$ is removed from $\Enum|_{-\tau}$, 
    we see that $b \coloneqq \emptyset^j \cdot \overline{\tau} \cdot (\Enum|_{-\tau}) \not \models \Phi_n$: 
    indeed, $b[j] = \overline{\tau}$ corresponds to the only position in $b$ satisfying~$\widetilde{p}$, but 
    $\tau$ does not appear in $b$ (since it does not appear in~$\Enum|_{-\tau}$).
    Therefore, $\AA \models \Phi_n$ and $\BB \pperp \Phi_n$. 
  \end{proof}%
\subsection{Separating $\AA$ from $\BB$ requires an exponential proof}
\label{sec:lb:exponential-proof}

We now show that every deduction tree for $\separ{\AA,\BB}$ has
size at least $2^n$. To do so, we use a relation~$\approx$ that, roughly speaking, states what elements $(a,b) \in \AA^{\ltl{G}} \times \BB^{\ltl{G}}$ are similar enough to require a non-trivial proof in order to be separated using the proof system. 
Formally, for $a,b \in \Sigma^+$,
we write $a \approx b$ whenever:
\begin{center}
  $a$ and $b$ are in the language $\emptyset^u \cdot \rho \cdot \emptyset^{\alpha(n)} \cdot \Sigma^*$, for some $u \in \N$ and $\rho \in T_Q \cup T_P$.
\end{center}
The central issue in the proof of the lower bound is 
counting how many of these pairs $a \approx b$ are preserved when applying the rules of the proof system. This count is done inductively on the size of the deduction tree, and allows us to derive the following lemma.

\begin{restatable}{lemma}{LemmaLowerBoundSpoonCase}
  \label{lemma:lower-bound:spoon-case-with-globally}
  Let $r_1,t_1,\dots,r_m,t_m \in \N$
  and let ${\tau_1,\dots,\tau_m \in T_Q}$ be pairwise distinct sets. Consider $A \subseteq \AA^{\ltl{G}}$, $B
  \coloneqq \{(\emptyset^{t_i} \cdot \overline{\tau_i} \cdot \Enum|_{-\tau_i})\sffx{r_i} : i \in [1,m]\}$, 
  and~$C \coloneqq \{ (a,b) \in A \times B : {a \approx b} \}$.
  Every deduction tree for $\separ{A,B}$ has size at least $\abs{C}+1$.
\end{restatable}

\begin{proof}
  Below, suppose that $\separ{A,B}$ has a deduction tree (else the statement is trivially true).
  In particular, 
  let~$\Tree$ be a minimal deduction tree for $\separ{A,B}$,
  and assume it has size $s$. 
  Note that the hypothesis that $\tau_1,\dots,\tau_m$ are distinct implies $\abs{B} \leq 2^n$, which in turn implies $\abs{C} < 2^n$ (by definition of $\approx$, for every $b \in B$ there is at most one $a \in \AA^{\ltl{G}}$ such that $a \approx b$).
  Then, w.l.o.g.~we can assume $s \leq \alpha(n)$; otherwise the lemma follows trivially.

  During the proof, we write $\prec$ for the strict total order on elements of $T_Q$ used to construct the trace $\Enum$ enumerating $T_Q$.
  Before continuing the proof of the lemma, we highlight a useful property of the elements of $C$.
  \begin{restatable}{clm}{ClaimRhoPrecRhok}
    \label{claim-rho-prec-rhok}
    Let $(a,b) \in C$ and $i \in [1,m]$ with $b = (\emptyset^{t_i} \cdot \overline{\tau_i} \cdot \Enum|_{-\tau_i})\sffx{r_i}$. 
    Then, $b = \emptyset^u \cdot \rho \cdot \emptyset^{\alpha(n)} \cdot \sigma$, 
    for some $u \in \N$, $\rho \in \{ \overline{\tau_i} \} \cup \{\tau \in T_Q : \tau \prec \tau_i \}$ and $\sigma \in \Sigma^*$.
  \end{restatable}
  \noindent
  In a nutshell, this claim 
  tells us that for every~$(a,b) \in C$ we have $b \not\sqsubseteq \Enum$.

  Let us go back to the proof 
  of~\Cref{lemma:lower-bound:spoon-case-with-globally}.
  If $A = \emptyset$ or $m = 0$ then $C = \emptyset$ and the lemma follows trivially. Below, let us assume $A \neq \emptyset$ and $m \geq 1$. We prove the statement by induction on the size $s$ of $\Tree$. 
  
  In the base case $s = 1$, $\Tree$ is a simple application of the rule~\ruleAtomic. This means that for every $a \in A$ and $b \in B$ we have $a[0] \neq b[0]$. By definition of~$\approx$, this implies $C = \emptyset$, and therefore $s \geq \abs{C}+1$.

  Let us then consider the induction step $s \geq 2$.
  Note that if $\abs{C} \leq 1$ then the statement follows trivially. Hence, below, we assume $\abs{C} \geq 2$.
  We split the proof depending on the rule applied to
  the root $\separ{A,B}$ of~$\Tree$. Since $s \geq 2$, this rule
  cannot be~\ruleAtomic. 
  We omit the cases for~\ruleOr and~\ruleAnd, as they  
  simply follow the induction hypothesis, 
  and focus on the rules related to temporal operators.

  \proofitem{$\bullet$ case: rules~{\rm{\ruleNext}} 
      and~{\rm{\ruleWeakNext}}} 
      \!\!We consider~\ruleNext and~\ruleWeakNext
      together, as both require $\separ{A^{\ltl{X}},B^{\ltl{X}}}$.
      Perhaps surprisingly, this case is non-trivial.
      The main difficulty stems from 
      the fact that $C' \coloneqq \{ (a,b) \in A^{\ltl{X}} \times B^{\ltl{X}} : {a \approx b} \}$ might in principle even be empty, and thus applying the induction hypothesis on $\separ{A^{\ltl{X}},B^{\ltl{X}}}$ is unhelpful for concluding that $s \geq \abs{C}+1$.
      We now show how to circumvent this issue.
      The minimal deduction tree for
      $\separ{A^{\ltl{X}},B^{\ltl{X}}}$ has size $s-1$. 
      Within this deduction tree, consider the maximal partial deduction tree~$\Tree'$ 
      rooted at $\separ{A^{\ltl{X}},B^{\ltl{X}}}$ and made solely 
      of applications of the rules~\ruleAnd,~\ruleOr,~\ruleNext, and~\ruleWeakNext.
      Let $\separ{A_1,B_1},\dots,\separ{A_q,B_q}$ be the leafs of such a tree. 
      Let ${j \in [1,q]}$.
      In the tree~$\Tree$, to $\separ{A_j,B_j}$ it is applied a rule among~\ruleAtomic, \ruleFuture and~\ruleGlobally.
      Let~${\xi_j \geq 1}$ be the number of \ruleNext and \ruleWeakNext rules used in the path of $\Tree$ from $\separ{A,B}$ to $\separ{A_j,B_j}$. 
      Note that, from $s \leq \alpha(n)$, we have $\xi_j \leq \alpha(n)$.
      We define the following two sets $C_j$ and $N_j$, whose role is essentially to ``track'' the evolution of pairs in $C$ with respect to $A_j \times B_j$:
      \begin{align*}
        C_j &\coloneqq \{(a\sffx{\xi_j},b\sffx{\xi_j}) \in A_j \times B_j : {(a,b) \in {C}},\, a\sffx{\xi_j} \approx b\sffx{\xi_j} \},\\
        N_j &\coloneqq \{(a\sffx{\xi_j},b\sffx{\xi_j}) \in A_j \times B_j : {(a,b) \in C},\,
        a\sffx{\xi_j} \not\approx b\sffx{\xi_j} \}.
      \end{align*}
      The minimal 
      deduction tree for $\separ{A_j,B_j}$ 
      has size ${s_j \geq \abs{C_j}+1}$; by induction hypothesis. \Cref{proof-lb1:clm2,proof-lb1:clm3,proof-lb1:clm4} below highlight a series of properties  on the sets $C_j$ and $N_j$ from which we derive $s \geq \abs{C}+1$.

      \begin{restatable}{clm}{ClmProofLBOneClmTwo}
        \label{proof-lb1:clm2}
        For every $j \in [1,q]$, if $C_j \cup N_j \neq \emptyset$ then the rule applied to $\separ{A_j,B_j}$ in $\Tree$ is either~\ruleFuture or~\ruleGlobally.
      \end{restatable}

      As already said, the rule applied to $\separ{A_j,B_j}$ 
      is among the rules~\ruleAtomic, \ruleFuture and~\ruleGlobally. 
      Then, showing that $a[0] = b[0]$ for every $(a,b) \in C_j \cup N_j$ excludes the rule~\ruleAtomic.

      \begin{restatable}{clm}{ClmProofLBOneClmThree}
        \label{proof-lb1:clm3}
        For every $j \in [1,q]$, $\abs{N_j} \leq 1$.
      \end{restatable}

      The proof of this claim is by contradiction, 
      assuming the existence of distinct $(a_1,b_1),(a_2,b_2) \in N_j$. In the proof,
      we analyse structural properties of the traces~$a_1$, $a_2$, $b_1$ and $b_2$, 
      and consider several cases depending on such properties (for instance, one case split depends on whether $a_1 \sqsubseteq a_2$).
      In all these cases,
      we reach   
      a contradiction with either $(a_1,b_1) \neq (a_2,b_2)$ 
      or~\Cref{proof-lb1:clm2}.

      \begin{restatable}{clm}{ClmProofLBOneClmFour}
        \label{proof-lb1:clm4}
        $\abs{C} \leq \sum_{j=1}^q \abs{C_j \cup N_j}$.
      \end{restatable}
      The claim follows as soon as one proves the 
      following two statements: 
        \begin{enumerate}
          \item\label{clm4-it1-body} for every $(a,b) \in C$ there is $j \in [1,q]$ such that $(a\sffx{\xi_j},b\sffx{\xi_j}) \in C_j \cup N_j$,
          \item\label{clm4-it2-body} for all distinct $(a_1,b_1),(a_2,b_2) \in C$, we have $(a_1\sffx{\ell},b_1\sffx{\ell}) \neq (a_2\sffx{\ell},b_2\sffx{\ell})$ for every 
          $\ell \leq \alpha(n)$ (recall that $\xi_j \leq \alpha(n)$, for every $j \in [1,q]$).
        \end{enumerate}
      \Cref{clm4-it1-body} is by induction on the size of $\Tree'$. Similarly to~\Cref{proof-lb1:clm3}, the proof of
      \Cref{clm4-it2-body} again requires to consider many cases, and uses properties of $\approx$, $\Enum$ and~$\Enum|_{-\tau_i}$.

      Thanks to~\Cref{proof-lb1:clm3,proof-lb1:clm4}, 
      one can then prove $s \geq \abs{C}+1$, concluding the proof for the rules~\ruleNext and~\ruleWeakNext:
      {\allowdisplaybreaks
      \begin{align*}
        s &\geq 1 + \textstyle\sum_{j=1}^q s_j 
          &\text{by definition~of $\Tree$ and $\Tree'$}\\
          &\geq 1+\textstyle\sum_{j=1}^q (\abs{C_j}+1) 
          &\text{by $s_j \geq \abs{C_j}+1$ (induction hypothesis)}\\
          &\geq 1+\textstyle\sum_{j=1}^q (\abs{C_j \cup N_j})
          &\text{by $\abs{N_j} \leq 1$ (\Cref{proof-lb1:clm3})}\\
          &\geq \abs{C}+1
          &\text{by $\textstyle\abs{C} \leq \textstyle\sum_{j=1}^q \abs{C_j \cup N_j}$ (\Cref{proof-lb1:clm4}).}
      \end{align*}
      }

    \proofitem{$\bullet$ case: rule~{\rm{\ruleFuture}}}
      Let $f \in \FP_A$ be the future point used when applying this rule. Define $C' \coloneqq \{(a',b') \in A^f \times B^{\ltl{G}} : a' \approx b' \}$. 
      The minimal
      deduction tree for $\separ{A^f,B^{\ltl{G}}}$ has size
      $s-1$. 
      By induction hypothesis, $s-1 \geq \abs{C'} + 1$, i.e., $s \geq \abs{C'} + 2$. We divide the proof 
      into two cases.

      \vspace{5pt}
      \textit{\underline{Case~1:} for every $a' \in A^f$, $a' \not\sqsubseteq \Enum$.}
      By definition of~$\approx$, every $(a,b) \in C$ is such that 
      $a$ and $b$ belong to the language 
      $\emptyset^u \cdot \overline{\tau_i} \cdot \emptyset^{\alpha(n)} \cdot \Sigma^*$
      for some~$u \in \N$, and $i \in [1,m]$.
      Since $a^f \not\sqsubseteq \Enum$, we must have $f(a) \leq u+1$. 
      Then, $a^f \approx b\sffx{f(a)}$. 
      Note that distinct $(a,b) \in C$ concern distinct $\overline{\tau_i}$ with $i \in [1,m]$, and therefore,
      together with $b\sffx{f(a)} \in B^G$, one concludes that $\abs{C'} \geq \abs{C}$; and so $s \geq \abs{C}+2$. 

      \vspace{5pt}
      \textit{\underline{Case~2:} there is $a' \in A^f$ such that $a' \sqsubseteq \Enum$.}
      Let us denote with $\widetilde{a}$ the element in $A^f$ such that $\widetilde{a} \sqsubseteq a$ for every $a \in A^f$. The existence of such an element 
      follows directly from the fact that $a' \sqsubseteq \Enum$ for some $a' \in A^f$. 

      Let $I \subseteq [1,m]$ be the subset of those indices $i \in [1,m]$ for which there is a pair $(a',b') \in C$ such that $b' = (\emptyset^{t_i} \cdot \overline{\tau_i} \cdot \Enum|_{\tau_i})\sffx{r_i}$. 
      Without loss of generality, suppose $I = \{1,\dots,q\}$ for some $q \leq m$,
      and that $\tau_{1} \prec \tau_{2} \prec \dots \prec \tau_{q}$; recall that all these types are pairwise distinct.
      By definition of $\approx$, for every $b' \in B$ there is at most one $a' \in \AA^{\ltl{G}}$ such that $a' \approx b'$,
      and therefore $q = \abs{C}$.
      To conclude the proof it suffices to show $\abs{C'} \geq q - 1$.
      We do so by establishing a series of claims.
      Recall that we are assuming $\abs{C}\geq 2$, so in particular $C$ and $I$ are non-empty.

      \begin{restatable}{clm}{ClmProofLBFClmOneNew}
        \label{proof-lbF-clm1-new}
        There are $u \in \N$, $\rho \in T_Q$ and $\sigma \in (2^{Q})^*$ s.t.~${\widetilde{a} = \emptyset^{u} \cdot \rho \cdot \emptyset^{\alpha(n)} \cdot \sigma}$.
        Moreover, $\rho \preceq \tau_i$ for every $i \in I$.
      \end{restatable}

      The first statement of this claim is established from the 
      definition of $\widetilde{a}$. 
      The second statement is proven by contradiction. 
      In particular, assuming that there is $i \in I$ such that $\tau_i \prec \rho$ yields a contradiction with~\Cref{claim-rho-prec-rhok}.

      Below, we write $u,\rho$ and $\sigma$ for the objects 
      appearing in~\Cref{proof-lbF-clm1-new}.
      Note that, from $\tau_1 \prec \dots \prec \tau_q$, the second statement of~\Cref{proof-lbF-clm1-new} implies 
      $\rho \prec \tau_2 \dots \prec \tau_q$. 
      For $i \in [2,q]$, let $(a_i',b_i')$ denote the pair in $C$ such that $b_i' = (\emptyset^{t_i} \cdot \overline{\tau_i} \cdot \Enum|_{\rho_i})\sffx{r_i}$.

      \begin{restatable}{clm}{ClmProofLBFClmTwoNew}
        \label{proof-lbF-clm2-new}
        For each $i \in [2,q]$ there is $\ell \in \N$
        such that $\widetilde{a} \approx b_i''$ with $b_i'' \coloneqq b_i'\sffx{\ell}$.
        Moreover, every type in $\{\tau_2, \dots, \tau_q\} \setminus \{\tau_i\}$ appears in some position of $b_i''$. 
      \end{restatable}

      This claim is proven using~\Cref{claim-rho-prec-rhok,proof-lbF-clm1-new} 
      and properties of~$\Enum|_{-\tau_i}$.

      Since all types $\tau_2, \dots, \tau_q$ are pairwise distinct, from the second statement in~\Cref{proof-lbF-clm2-new}
      we conclude that $b_i'' \neq b_j''$ for every two distinct $i,j \in I \setminus \{i_1\}$.
      Then, the first statement in~\Cref{proof-lbF-clm2-new}
      entails $\abs{C'} \geq q-1$. 

    \proofitem{$\bullet$ case: rule~{\rm{\ruleGlobally}}}
      Let $f \in \FP_A$ be the future point used when applying this rule. 
      The minimal deduction tree for $\separ{A^{\ltl{G}},B^f}$ has size $s-1$. 
      We define ${C' \coloneqq \{(a',b') \in A^{\ltl{G}} \times B^{f} : a' \approx b' \}}$. By induction hypothesis, $s-1 \geq \abs{C'} + 1$, i.e., $s \geq \abs{C'} + 2$. To conclude the proof it suffices to show that $\abs{C'} \geq \abs{C}-1$ (in fact, we prove $\abs{C'} \geq \abs{C}$). Let $\{(a_1,b_1),\dots,(a_{\abs{C}},b_{\abs{C}})\} = C$.

      \begin{restatable}{clm}{ClmProofLBGClmOne}
        \label{proof-lbG-clm1}
        For every $j \in [1,\abs{C}]$, 
        $b_j^f$ is not a suffix of $\Enum$. 
        More precisely, given $i \in [1,m]$ such that $b_j = (\emptyset^{t_i} \cdot \overline{\tau_i} \cdot \Enum|_{-\tau_i})\sffx{r_i}$, 
        we have $b_j^f = \emptyset^u \cdot \rho \cdot \emptyset^{\alpha(n)} \cdot \sigma$, 
        for some $u \in \N$, $\rho \in \{ \overline{\tau_i} \} \cup \{\tau \in T_P : \tau \prec \tau_i \}$ and $\sigma \in \Sigma^*$.
      \end{restatable}

      See the similarities between this claim and~\Cref{claim-rho-prec-rhok}. The first statement is proven by contradiction, deriving an absurdum 
      with the existence of~$\Tree$. 
      The second statement follows from the definition of~$\Enum|_{-\tau_i}$.

      Starting from \Cref{proof-lbG-clm1}, we conclude that 
      \begin{enumerate*}[label=(\roman*)]
        \item\label{gcc-i1} for every $j \neq k \in [1,\abs{C}]$, $b_j^f \neq b_k^f$, and 
        \item for every $j \in [1,\abs{C}]$ there is $\ell \in \N$ 
        such that $a_j\sffx{\ell} \approx b_j^f$.
      \end{enumerate*}
      This directly implies~$\abs{C'} \geq \abs{C}$.
      This concludes both the proof of the case~\ruleGlobally 
      and the proof of the lemma.
\end{proof}

Together,~\Cref{lemma:phi-n-separates,lemma:lower-bound:spoon-case-with-globally} 
yield an exponential lower bound for all formulae of \LTLnountil characterising $\Phi_n$ (\Cref{th:ltlnountilexponential}), which in turn implies~\Cref{theorem:LTLO-more-succint}.

\begin{lemma}
\label{th:ltlnountilexponential}
  Let $\Psi_n \in \LTLnountil$. 
  If $\Psi_n \equiv \Phi_n$ then $\size(\Psi_n) \geq 2^n$.
\end{lemma}

\begin{proof}
  We define the sets $A = \{ \overline{\tau} \cdot \Enum : \tau \in T_Q\}$ and $B = \{ \overline{\tau} \cdot \Enum_{-\tau} : \tau \in T_Q \}$. 
  Observe that $A \subseteq \AA \subseteq \AA^{\ltl{G}}$ and $B \subseteq \BB$.
  Let $C = \{(a,b) \in A \times B : a \approx b \}$. 
  From the definition of $\approx$, $\abs{C} = 2^n$. 
  We apply~\Cref{lemma:lower-bound:spoon-case-with-globally},
  and conclude that 
  the minimal deduction tree for $\separ{A,B}$ has size at least $2^n$ (in fact, $2^n+1$). 
  Since $A \subseteq \AA$ and $B \subseteq \BB$, 
  the same holds for the minimal deduction tree for $\separ{\AA,\BB}$.
  Then, the theorem follows from~\Cref{corr:cc} and~\Cref{lemma:phi-n-separates}.
\end{proof}

While we do not
prove it formally, we claim
that~\Cref{theorem:LTLO-more-succint} also holds for
infinite traces. It is in fact quite simple to see this: all traces in
$\AA$ and $\BB$, have a suffix of the form
$\emptyset^{\alpha(n)}$. Roughly speaking, these suitably long suffixes are added to make the far-end of the traces in $\AA$
and $\BB$ indistinguishable at the level of formulae, 
so that they cannot be used in deduction
trees to separate $\AA$ from $\BB$. Then, to
prove~\Cref{theorem:LTLO-more-succint} for infinite traces,
it suffices to update the proof system to handle these
structures and consider an infinite suffix
$\emptyset^{\omega}$ instead. The proof of~\Cref{lemma:lower-bound:spoon-case-with-globally} goes through with no significant change.

A second observation: traces in $\AA$ and $\BB$ are closed under taking arbitrary long prefixes of the form $\emptyset^j$.
This feature is not used to prove~\Cref{th:ltlnountilexponential} (see the definition of~$A$ and $B$ in the proof). However, these prefixes play a role in the next section, when studying the succinctness of~$\eventually(\LTLPnountil)$ on infinite~traces.

%% file: sections/pastification-lowerbounds.tex
\section{\cref{theorem:LTLF-more-succint}: a $2^{n}$ lower bound for $\LTL[\eventually]$ pastification}
\label{sec:past-lb}

In this section, we rely on~\cref{th:ltlnountilexponential} to prove 
\cref{theorem:LTLF-more-succint} and~\cref{corollary:optimal-pastification}.

\cref{theorem:LTLF-more-succint} is proven by relying on a ``future--past duality'' between future and past fragments of \LTL.
Given a trace $\sigma \in
\Sigma^+$ we define the \emph{reverse of
$\sigma$}, written $\sigma^{-}$, as the trace satisfying $\sigma^{-}[i]
= \sigma[\abs{\sigma}-i]$ for every $i \in \position(\sigma)$. 
The \emph{reverse of a language} $\lang \subseteq \Sigma^+$
is defined as the language $\lang^{-} \coloneqq \{ \sigma^{-} : \sigma \in \lang \}$. 
Clearly, $(\lang^-)^- = \lang$.
Given a set of temporal operators $S \subseteq \{\tomorrow,\weaktomorrow,\eventually,\always\}$, we write $S^-$ for the set of temporal operators among $\{\yesterday, \weakyesterday, \once, \historically\}$ such that 
$S^-$ contains~$\yesterday$ \mbox{(resp.~$\weakyesterday$; $\once$; $\historically$)}
if and only if $S$ contains $\tomorrow$ (resp.~$\weaktomorrow$; $\eventually$; $\always$).
For finite traces, the following lemma,
proves that if there is a family of languages $(\lang_n)_{n\ge 1}$ that can
be compactly defined in $\eventually(\LTLPFinv)$ but explodes in
$\LTLnountil$, then the family 
$(\lang_n^{-})_{n\ge 1}$ can be compactly defined in $\LTLonlyF$ but
explodes in $\eventually(\LTLPnountil)$.

\begin{restatable}{lemma}{LemmaReverseTwo}
  \label{lemma:reverseTwo}
  Let $\lang \subseteq \Sigma^+$, 
  $S \subseteq \{\tomorrow,\weaktomorrow,\eventually,\always\}$,
  and $\phi$ be a formula in $\eventually(\LTL[S^-])$. 
  There is a formula $\psi$ in $\eventually(\LTL[S])$
  such that $\lang(\psi) = \lang(\phi)^-$ \!and ${\size(\psi) = \size(\phi)}$.
\end{restatable}

\cref{theorem:LTLF-more-succint} follows by applying~\Cref{lemma:reverseTwo}
on the family of formulae $(\Phi_n)_{n \ge 1}$ defined
in~\cref{sec:difficult-property}, and by relying on
the exponential lower bounds of~\Cref{th:ltlnountilexponential}.
The sequence of languages 
showing that~$\LTL[\eventually]$ can be exponentially more succinct than 
$\eventually(\LTLPnountil)$ is given by~$(\lang(\Phi_n)^-)_{n \geq 1}$.

Next, we extend~\cref{theorem:LTLF-more-succint} to the case of
infinite traces.
As usual, let $\Sigma^\omega$ be the set of all infinite traces 
over the finite alphabet $\Sigma$.
We denote with $\lang^\omega(\phi)$ the language of $\phi$,
when $\phi$ is interpreted over infinite traces (we refer the reader to, e.g.,~\cite{DBLP:conf/aaai/ArtaleGGMM23} for the semantics of~$\LTL$ on infinite traces).  

\begin{restatable}{lemma}{lemmareverseinfinite}
\label{lemma:reverseinfinite}
  The family of languages of infinite traces $(\lang(\Phi_n)^{-} \cdot
  \Sigma^\omega)_{n \ge 1}$ is such that, for every $n\ge 1$,
  \begin{enumerate*}[label=(\roman*)]
    \item\label{item:revinf-1}
      there is a formula $\phi$ of $\LTLonlyF$ 
      such that $\size(\phi) \in \O(n)$ and $\lang^\omega(\phi) = \lang(\Phi_n)^{-}\cdot\Sigma^\omega$, and
    \item\label{item:revinf-2} for every formula $\psi$ in $\eventually(\LTLPnountil)$,
      if~$\lang^\omega(\psi) = \lang(\Phi_n)^{-}\cdot\Sigma^\omega$
      then $\size(\psi) \geq 2^{n}$.
  \end{enumerate*}
\end{restatable}

\noindent
\Cref{item:revinf-1} in the lemma above follows by applying~\Cref{lemma:reverseTwo}
and exploiting the fact that formulae~$\phi$ in $\LTLonlyF$ satisfy $\lang^{\omega}(\phi)
= \lang(\phi) \cdot \Sigma^\omega$ and ${\lang(\phi) = \lang(\phi) \cdot \Sigma^*}$ (cf.~\cite[Definition 5 and Lemma 5]{DBLP:conf/aaai/ArtaleGGMM23}). The proof of \Cref{item:revinf-2} is instead quite subtle.
One would like to use the hypothesis~$\lang^\omega(\psi) = \lang(\Phi_n)^{-}\cdot\Sigma^\omega$ and that $\lang(\psi)$ is a  cosafety language to derive $\lang(\psi) = \lang(\Phi_n)^-$. However, note that nothing prevents $\lang(\psi)$ to be equal to~$\lang(\Phi_n)^- \cdot \Sigma$, and as it stands we do not 
have bounds for characterising this language. We apply instead the following strategy. We consider the family of structures $A' \coloneqq \{ a^- \cdot \emptyset^\omega : a \in \AA \}$ 
and 
$B' \coloneqq \{ b^- \cdot \emptyset^\omega : b \in \BB \}$. Note that $A' \subseteq \lang^{\omega}(\psi)$ and $B' \cap \lang^{\omega}(\psi) = \emptyset$. 
Since $\psi$ is of the form $\eventually(\alpha)$ 
with $\alpha \in \LTLPnountil$, 
we can, roughly speaking, study the effects of applying to $A'$ and $B'$ a variant of the rule~\ruleFuture for infinite words and that does not ``forget the past'', and then reverse all traces in the resulting sets $(A')^f$ and $(B')^{\ltl{G}}$. 
In this way, we obtain two sets of finite traces $\widetilde{A} \subseteq \AA$ and $\widetilde{B} \subseteq \BB$ (this is where the prefixes~$\emptyset^j$ discussed at the end of~\Cref{sec:lbltlxf} play a role). We show that the hypotheses of~\Cref{lemma:lower-bound:spoon-case-with-globally} apply to $\widetilde{A}$ and a set $\widehat{B} \subseteq \widetilde{B}$ for which the set $\{(a,b) \in \widetilde{A} \times \widehat{B} : a \approx b \}$ has size at least $2^n-1$. 
By~\Cref{corr:cc}, we get that
$\alpha$, and thus~$\psi$, is of size at least $2^n$.

\Cref{lemma:reverseinfinite} shows that~\cref{theorem:LTLF-more-succint} holds over infinite
traces as well. 
Together with the~$2^{\O(n)}$ upper bound for the pastification problem for~$\LTLxf$ 
into $\eventually(\LTL[\ltl{Y,wY,O}])$ established\footnote{To be more precise, in~\cite{ArtaleGGMM23} the authors only provide a~$2^{\O(n^2)}$ upper bound for their algorithm. Their analysis can however be easily improved to~$2^{\O(n)}$.} 
in~\cite{ArtaleGGMM23}, 
this 
entails~\cref{corollary:optimal-pastification}.

%% file: sections/automata-struggles.tex
\section{The automata method does not work for $\eventually(\LTL[\once])$}
\label{sec:automata-struggles}

In this section we show that the classical method introduced by Markey in~\cite{markey2003temporal} to prove ``future against past'' succinctness discrepancies in fragments of \LTL cannot be used to prove the results
in~\Cref{sec:lbltlxf}, namely that $\eventually(\LTLPFinv)$
can be exponentially more succinct than $\LTLnountil$. 
Due to space constraints, we assume a basic familiarity with
\emph{non-deterministic B\"uchi automata} (\NBA{s}) (and
\emph{deterministic B\"uchi automata}, \DBA{s}), which are
central tools in~\cite{markey2003temporal}. 
Moreover, we work on~\LTL on infinite traces, as done in~\cite{markey2003temporal}, 
and write $\phi \equiv_{\omega} \psi$ whenever $\lang^{\omega}(\phi) = \lang^{\omega}(\psi)$.
We write $\lang^{\omega}(A)$ for the language of an \NBA $A$.

\Cref{proposition:markey} below summarises the method in~\cite{markey2003temporal}, 
which is parametric on a fixed prefix
$\Pi$ of operators among $\tomorrow$, $\eventually$ and $\always$.
Given two fragments $F,F'$ of $\LTL$, with $F'$ pure future,
to apply the method one has to provide the two families of formulae ${(\Phi_n)_{n \geq 1} \!\in\! F}$ and 
${(\Phi_n')_{n \geq 1} \!\in\! F'}$,
as well as the family of minimal \NBA{s}
${(A_n)_{n \geq 1}}$, satisfying the hypotheses of~\Cref{proposition:markey}. 
In~\cite{markey2003temporal}, this is done for ${F = \LTL}$ and $F'$ set as
the pure future fragment of \LTL, using the prefix $\Pi = \always$.

\begin{proposition}[{{\rm Markey's method \cite{markey2003temporal}}}]
  \label{proposition:markey}
  Let $F,F'$ be fragments of $\LTL$, with $F'$ pure future. Consider two
  families of formulae ${(\Phi_n)_{n \geq 1} \!\in\! F}$, 
  ${(\Phi_n')_{n \geq 1} \!\in\! F'}$,
  and a family of minimal \NBA{s}
  ${(A_n)_{n \geq 1}}$, 
  such that%
  \begin{center}%
    \hfill
    $\size(A_n) \in 2^{2^{\Omega(n)}}$\!\!,
    \hfill
    $\size(\Phi_n) \in \poly(n)$,
    \hfill
    $\Phi_n \equiv_{\omega} \Phi_n'$,
    \hfill
    $\lang^{\omega}(\Pi(\Phi_n')) = \lang^{\omega}(A_n)$.
    \hfill\,
  \end{center}
  Then, $\size(\Phi_n') \in 2^{\size(\Phi_n)^{\Omega(1)}}$\!\! 
  and $F$ can be exponentially more succinct than $F'$.
\end{proposition}

\noindent
The consequence $\size(\Phi_n') \in 2^{\size(\Phi_n)^{\Omega(1)}}$ obtained in~\Cref{proposition:markey} follows directly from the fact that, from every pure future \LTL formula $\phi$, one can build an \NBA $A$ of size $2^{\O(\size(\phi))}$ such that~$\lang^{\omega}(A) = \lang^{\omega}(\phi)$~\cite{vardi1986automata}.
Then, the hypotheses ${\size(A_n) \in 2^{2^{\Omega(n)}}}$ and $\lang^{\omega}(\Pi(\Phi_n')) = \lang^{\omega}(A_n)$
imply $\size(\Phi_n') \in 2^{\Omega(n)}$.

To show that~\Cref{proposition:markey} cannot be used to
derive that $F \coloneqq \eventually(\LTLPFinv)$ can be
exponentially more succinct than $F' \coloneqq
\LTLnountil$, it suffices to show that no families
${(\Phi_n)_{n \geq 1} \!\in\! F}$, ${(\Phi_n')_{n \geq 1}
\!\in\! F'}$ and ${(A_n)_{n \geq 1}}$ achieve the hypotheses
required by~\Cref{proposition:markey}, no matter the temporal prefix $\Pi$. We do so by establishing
that whenever $\size(\Phi_n) \in \poly(n)$ and $\Phi_n
\equiv_{\omega} \Phi_n'$, the minimal \emph{deterministic} B\"uchi automaton
for~$\lang^{\omega}(\Pi(\Phi_n'))$ has size in
$2^{\O(\poly(n))}$. 
Therefore, no family of minimal \NBA{s} $(A_n)_{n \geq 1}$ such that $\size(A_n) \in 2^{2^{\Omega(n)}}$ can also satisfy the hypothesis $\lang^{\omega}(\Pi(\Phi_n')) = \lang^{\omega}(A_n)$. Here is the formal statement:

\begin{restatable}{theorem}{TheoremWhyAutomataMethodFails}
  \label{theorem:why-automata-method-fails}
  Let $\Pi$ be a prefix of $k$ temporal operators among $\tomorrow$, $\eventually$ and $\always$. 
  Let~$\phi$ be a formula of $\eventually(\LTL[\once])$, 
  and~$\psi$ be a formula of $\LTLnountil$, with $\phi \equiv_{\omega} \psi$.
  The minimal \DBA for $\lang^{\omega}(\Pi(\psi))$ is of size
  in $(k+1) \cdot 2^{\O(\size(\phi))}$. 
\end{restatable}

\noindent
The proof of this theorem is divided into three steps. 

As a first step, one shows that $\psi \equiv_{\omega} \eventually \psi$;
which essentially follows from the fact that $\psi \equiv_{\omega} \phi$ with $\phi \in \eventually(\LTLPFinv)$. 
Together with tautologies of $\LTL$ such as ${\eventually \always \eventually \psi' \equiv_{\omega} \always \eventually \psi'}$, $\eventually \tomorrow \psi' \equiv_{\omega} \tomorrow \eventually \psi'$ 
and $\always \tomorrow \psi' \equiv_{\omega} \tomorrow \always \psi'$, the equivalence
$\psi \equiv_{\omega} \eventually\psi$ 
let us rearrange $\Pi$ into a prefix of the form either $\tomorrow^{j} \always \eventually$ or $\tomorrow^j \eventually$, for some $j \leq k$. 
Let us focus on the former (more challenging) case of $\Pi = \tomorrow^{j} \always \eventually$.

The second step required for the proof is to bound the size of the
minimal \DBA~$A$ recognising $\lang^{\omega}(\eventually\psi)$. Thanks
to the equivalences $\phi \equiv_{\omega} \psi
\equiv_{\omega} \eventually \psi$, such a \DBA has size
exponential in $\size(\phi)$ by the following lemma.

\begin{restatable}{lemma}{LemmaUpperBoundDBACosafety}
  \label{lemma:upper-bound-dba-cosafety}
  Let $\phi$ in $\eventually(\LTLPFinv)$.
  There is a \DBA for $\lang^{\omega}(\phi)$ of size $2^{\O(\size(\phi))}$.
\end{restatable}

Starting from $A$, the third and last step of the proof requires constructing a \DBA for $\lang^{\omega}(\tomorrow^j \always \eventually \psi)$ of size in $(j+1) \cdot 2^{\O(\size(\phi))}$.
The treatment for the prefix~$\tomorrow^j$ is simple, 
so this step is mostly dedicated to constructing a \DBA for 
$\lang^{\omega}(\always \eventually \psi)$. 
In the case of $\LTL$, it is known that closing an \NBA under the globally operator $\always$ might lead to a further exponential blow-up (in fact, this is one of the reasons Markey's method is possible in the first place). However, because $\phi$ is in $\eventually(\LTLPFinv)$, and it is thus a cosafety language (and so $\psi$ is a cosafety language too),
it turns out that the size of the minimal \DBA for $\lang^{\omega}(\always \eventually \psi)$ is in $\O(\size(A))$.

\begin{restatable}{lemma}{LemmaMinimalDBACosafety}
  \label{lemma:minimal-dba-cosafety}
  Let $\psi$ be in $\LTL$, such that $\lang^{\omega}(\psi)$ is a cosafety language. Let $A$ be a \DBA for $\lang^{\omega}(\eventually \psi)$. The minimal \DBA for $\lang^{\omega}(\always\eventually \psi)$ has size in $\O(\size(A))$.
\end{restatable}

Thanks to~\Cref{lemma:minimal-dba-cosafety}, 
the proof of~\Cref{theorem:why-automata-method-fails} can be easily completed. 
To prove this lemma, one modifies $A$ by redirecting all transitions exiting a final state so that they mimic the transitions exiting the initial state of the automaton. The reason why the obtained automaton recognises $\lang^{\omega}(\always\eventually \psi)$ uses in a crucial way the fact that $\psi$ and $\eventually\psi$ are cosafety languages.

%% file: sections/related-works.tex

\section{Related and Future Work}
\label{sec:related-and-future-works}

The proof systems we use in this work to establish~\Cref{theorem:LTLF-more-succint} and~\Cref{theorem:LTLO-more-succint}
are strongly related to recent
work in two seemingly disconnected areas of computer
science: 
\begin{enumerate*}[label=(\roman*)]
  \item combinatorial games for formulae lower bounds of first-order logics
    and 
  \item learning of \LTL formulae in explainable planning and program
    synthesis.
\end{enumerate*}
\paragraph{Combinatorial games.}
We have already discussed the connections between our proof system and 
the \CTLplus games by Adler and Immerman~\cite{adler2003n}. Recently, 
Fagin and coauthors~\cite{FaginLR021,FaginLVW22} have looked at combinatorial games 
that allow to count the number of quantifiers required to express a property in first-order logic. While these games simplify Adler--Immerman games, they come with a drawback: by design, they implicitly look at how to express properties with first-order formulae in \emph{prenex normal form}, and they are not able to give any bound on the quantifier-free part of such formulae. It seems then not possible to use these types of games in the context of \LTL. One could in principle consider translations of \LTL formulas into a prenex fragment of \SoneS. However, since \SoneS is both more expressive and more succinct than \LTL~\cite{StockmeyerM02}, concluding that \LTLonlyF can be exponentially more succinct than
$\eventually(\LTLPnountil)$ will ultimately require analysing structural properties of the quantifier-free part of the \SoneS formulae.

Closer to the case of \LTL are the games on linear orders (implicitly) used
by Grohe and Schweikartdt in~\cite{GroheS05}.  These are
formula-size games for a fixed number of variables of first-order logic.
Using our notation, the method used to derive lower bounds in~\cite{GroheS05} relies on
defining a function $\omega$ from terms of the form $\separ{A,B}$ to $\N$
that acts as a \emph{sub-additive measure} with respect to the rules of the
proof system. For instance, according to the rule~\ruleOr,
$\omega$ should satisfy
    ${\omega(\separ{A,B}) \leq \omega(\separ{A_1,B})
    + \omega(\separ{A_2,B})}$, whenever $A = A_1 \uplus A_2$.
One can use a sub-additive measure~$\omega$ to conclude that the minimal deduction tree for $\separ{A,B}$, if it exists, has size at least $\omega(\separ{A,B})$. 
When restricted to the objects in~\Cref{lemma:lower-bound:spoon-case-with-globally}, 
one can show that the function $\omega(\separ{A,B}) \coloneqq \abs{\{ (a,b) \in A \times B : {a \approx b} \}}+1$ is a sub-additive measure with respect to the rules~\ruleAtomic, \ruleOr, \ruleAnd, \ruleFuture and~\ruleGlobally 
(this is implicit in the proof of~\Cref{lemma:lower-bound:spoon-case-with-globally}). 
However, it is not a sub-additive measure for the rules~\ruleNext and~\ruleWeakNext: as stressed in the proof, we might have $\omega(\separ{A^{\ltl{X}},B^{\ltl{X}}}) = 1$ even for $\omega(\separ{A,B})$ arbitrary large. 
This partially explains why the proof of~\Cref{lemma:lower-bound:spoon-case-with-globally} turned out to be quite involved. 

In view of the optimality of the algorithm in~\cite{ArtaleGGMM23}
(\Cref{corollary:optimal-pastification}), the main open problem regarding
pastification is the optimality of the 
triply-exponential time procedure given
in~\cite{de2021pure} for the pastification of $\LTL[\tomorrow,\until]$ into~$\eventually(\pLTL)$. As far as we are aware, no super-polynomial lower bound for this problem is known. Our proof
system, properly extended with rules for the until and release operators, might be
able to tackle this issue.
Obtaining a matching triply-exponential lower bound might however be impossible:
when restricted to propositional logic, 
our proof system
is equivalent to the communication games introduced by Karchmer and Wigderson~\cite{Karchmer89}.
It is well-known that these games cannot prove 
super-quadratic lower bounds for formulae sizes,
and one should expect similar limitations for temporal logics, albeit with respect to some
function that is at least exponential.



%
\paragraph{\LTL formulae learning.} 
Motivated by the practical issue of understanding a complex
system starting from its execution traces, several recent
works study the algorithmic problem of finding an \LTL
formula separating two finite sets of traces,
see
e.g.~\cite{Neider18,CamachoM19,RahaRFN22,FortinKRSWZ22,FortinKRSWZ23}.
In light of~\Cref{th:ps}, this problem is equivalent to
finding a proof in (possibly variations of) our combinatorial proof
system. We believe that this simple observation will turn
out to be quite fruitful for both the ``combinatorial
games'' and the ``formula learning'' communities. From our
experience, tools such as the one developed
in~\cite{Neider18,CamachoM19,RahaRFN22} are quite useful
when studying combinatorial lower bounds, as they can be used
to empirically test whether families of structures are
difficult to separate, before attempting a formal proof. In
our case, we have used the tool in~\cite{Neider18} while
searching for the structures and formulae we ended up using
in~\Cref{sec:lbltlxf}, and discarded several other
candidates thanks to the evidences the tool gave us. On the
other side of the coin, combinatorial proof systems can be
seen as a common foundational framework for all these
formulae-learning procedures. With this in mind, we believe
that the techniques developed for proving lower bounds in
works such as~\cite{GroheS05} might be of help for improving these
procedures, for example using the minimization of a sub-additive measure
as a heuristic.





%% file: sections/appendix.tex

\section{Proofs of~\Cref{sec:difficult-property}}
\label{appendix:section-3}

\LemmaAFormulaForPhiN*
\begin{proof}
  For
  conciseness, given every $\tau \in 2^P$, we write $\overline{\tau}$ for the
  element of $2^Q$ such that $p_i \in \tau$ if and only if $q_i \in
  \overline{\tau}$, for every $i \in [1,n]$.  
  We define the formula $\Phi_n'$ as follows:
  \[ 
      \Phi_n' \coloneqq \bigvee_{\tau \,\in\, 2^{P}}
      \Big( 
        \bigwedge_{p \,\in\, \tau} \eventually \big(\widetilde{p} \land p \land \eventually (\widetilde{q} \land \psi_{\overline{\tau}})\big)
        \land 
        \bigwedge_{p \,\in\, P \setminus \tau} \eventually \big(\widetilde{p} \land \lnot p \land \eventually (\widetilde{q} \land \psi_{\overline{\tau}})\big)
      \Big),
  \]
  where $\psi_{\overline{\tau}} \coloneqq (\bigwedge_{q \,\in\,
  \overline{\tau}} q \land \bigwedge_{q \,\in\, Q \setminus
  \overline{\tau}} \lnot q)$, which has size~$2 n-1$. We now consider the
  size of $\Phi_n'$. Scanning the formula reveals that the disjunct
  corresponding to $\tau \in 2^{P}$ has size $|\tau| \cdot (2n + 8) + |P
  \setminus \tau| \cdot (2n+8) - 1$. Therefore, $\Phi_n'$ has size 
  {\allowdisplaybreaks
  \begin{align*}
    &
      |2^{P}| - 1 + \sum_{\tau \,\in\, 2^{P}} (|\tau| \cdot (2n + 8) + |P \setminus \tau| \cdot
      (2n+8) - 1)\\
    =\ & 
      2^n - 1 +  \sum_{\tau \,\in\, 2^{P}} (n \cdot (2n + 8) - 1) \\
    =\ & 
      2^n - 1 + 2^n \cdot (n \cdot (2n + 8) - 1)\\
    =\ & 
      2^n \cdot n \cdot (2n+8)-1 < 2^{n+1} \cdot (n+2)^2 .
  \end{align*}
  }
  Let us now show that $\Phi_n \equiv \Phi_n'$. 
  For the entailment $\Phi_n \models \Phi_n'$, consider a trace $\sigma \in \Sigma^+$
  such that $\sigma \models \Phi_n$. Then, there is $j \in \position(\sigma)$ 
  such that $\widetilde{q} \in \sigma[j]$ and for every $i \in [1,n]$ there is $k_i \in [0,j]$ such that $\widetilde{p} \in \sigma[k_i]$, and $q_i \in \sigma[j]$ if and only if $p_i \in \sigma[k_i]$.
  Let $\tau \in 2^P$ such that $\overline{\tau} = \sigma[j]$. We show that the disjunct of $\Phi_n'$ corresponding to the type $\tau$ is satisfied. Note that $\sigma,j \models \widetilde{q} \land \psi_{\overline{\tau}}$. 
  For every $i \in [1,n]$ such that $p_i \in \tau$, from $k_i \leq j$ we conclude $\sigma,k_i \models \widetilde{p} \land p_i \land \eventually(\widetilde{q} \land \psi_{\overline{\tau}})$. 
  Similarly, for every $i \in [1,n]$ such that $p_i \not\in \tau$, we have $\sigma,k_i \models \widetilde{p} \land \lnot p_i \land \eventually(\widetilde{q} \land \psi_{\overline{\tau}})$. 
  Then, $\sigma \models \bigwedge_{p \,\in\, \tau} \eventually (\widetilde{p} \land p \land \eventually (\widetilde{q} \land \psi_{\overline{\tau}}))
  \land 
  \bigwedge_{p \,\in\, P \setminus \tau} \eventually (\widetilde{p} \land \lnot p \land \eventually (\widetilde{q} \land \psi_{\overline{\tau}}))$.

  For the other direction, consider $\sigma \in \Sigma^*$ such that $\sigma \models \Phi_n'$. Then, there is $\tau \in 2^P$ such that  $\sigma \models \bigwedge_{p \,\in\, \tau} \eventually (\widetilde{p} \land p \land \eventually (\widetilde{q} \land \psi_{\overline{\tau}}))
  \land 
  \bigwedge_{p \,\in\, P \setminus \tau} \eventually (\widetilde{p} \land \lnot p \land \eventually (\widetilde{q} \land \psi_{\overline{\tau}}))$.
  This means that, for every $p \in \tau$, there are two positions $k_p \leq j_p$ such that 
  $\widetilde{p},p \in \sigma[k_p]$ and $\sigma,j_p \models \widetilde{q} \land \psi_{\overline{\tau}}$. 
  Similarly, for every $p \in P \setminus \tau$ there are two positions $k_p \leq j_p$ such that $\widetilde{p} \in \sigma[k_p]$, $p \not \in \sigma[k_p]$ and $\sigma,j_p \models \widetilde{q} \land \psi_{\overline{\tau}}$. We let $j \coloneqq \max\{j_p : p \in P \}$.
  From $\sigma,j \models \widetilde{q} \land \psi_{\overline{\tau}}$ we conclude that $\widetilde{q} \in \sigma[j]$ and for every $i \in [1,n]$, $q_i \in \sigma[j]$ if and only if $p_i \in \sigma[k_{p_i}]$.
  Moreover, $k_p \leq j$ for every~${p \in P}$. Therefore, $\sigma, j \models \widetilde{q} \land \bigwedge_{i=1}^n 
  ((q_i \land \once( \widetilde{p} \land p_i)) \lor
  (\lnot q_i \land \once (\widetilde{p} \land \lnot p_i))
  )$, and thus we conclude that $\sigma \models \Phi_n$.
\end{proof}

\section{Proofs of~\Cref{sec:psltlxf}}
\label{appendix:section-4}

\lemmaTrivialX*
\begin{proof}
  For the first statement, reasoning contrapositively, if there is a word $\sigma$ in both $A$ and $B$, then clearly no formula can separate $A$ from $B$. 

  Let us prove the second statement.
  Suppose $A \cap B = \emptyset$.
  To each word $\sigma \in \Sigma^*$ we associate
  the formula~$\phi(\sigma)$ of~$\LTL[\tomorrow]$ that is defined inductively as follows: 
  \[ 
    \phi(\sigma) = 
    \begin{cases}
      \bot &\text{ if } \sigma \text{ empty word}\\ 
      (\bigwedge_{p \in \sigma[0]} p \land \bigwedge_{q \in \AP \setminus \sigma[0]} \lnot q) \land \tomorrow \phi(\sigma\sffx{1}) &\text{ otherwise},
    \end{cases}
  \]
  where $\bot$ is short for $p \land \lnot p$, for some fixed $p \in \AP$.
  Let $\Phi \coloneqq \bigvee_{\sigma \in A} \phi(\sigma)$.
  It is simple to show that for every~$\sigma \in \Sigma^+$, $\sigma \in A$ if and only if $\sigma,0 \models \Phi$. 
  Then, by $A \cap B = \emptyset$, we conclude that $A \models \Phi$ and $B \pperp \Phi$.
\end{proof}

\MainTheoremProofSystem*
\begin{proof}
  We remark that the proof is constructive, i.e., from a
  deduction tree it is possible to construct a formula, and
  vice versa.

  \bigskip

  \noindent
  ($\Rightarrow$): Given a deduction tree of size $k$ for $\separ{A,B}$, we
  construct a formula $\phi$ of \LTLnountil having size $k$ and separating $A$ from~$B$. In a nutshell, the deduction tree can be seen as the parse tree of
  the formula $\phi$.
  We proceed by induction on $k$.

  \proofitem{base case: $k = 1$} In this case the deduction
    tree is a simple instantiation of the rule~\ruleAtomic,
    and the proof is straightforward.
    
  \proofitem{induction step: $k \geq 2$} 
    We proceed
    by cases depending on the rule applied to the root~$\separ{A,B}$ of the
    deduction tree. 
    \begin{itemize}[align=left]
      \item[\textbf{rule {\rm\ruleOr}.}] 
        Let $\separ{A_1,B}$ and $\separ{A_2,B}$ be the two hypotheses used to apply the rule~\ruleOr, 
        with $A = A_1 \uplus A_2$.
        Let $k_1$ and $k_2$ be the sizes of the deduction trees rooted in $\separ{A_1,B}$ and
        $\separ{A_2,B}$, respectively. Since $k = k_1+k_2$, by induction hypothesis, there
        are two formulae $\phi_1$ and $\phi_2$ of \LTLnountil such that $A_1 \models \phi_1$ and $B \pperp \phi_1$, $A_2 \models \phi_2$ and $B \pperp \phi_2$, 
        $\size(\phi_1) = k_1$ and $\size(\phi_2) = k_2$.
        Then, $A \models \phi_1 \lor \phi_2$, $B
        \pperp (\phi_2 \lor \phi_2)$ and $\size(\phi_2 \lor \phi_2) = k$.
      \item[\textbf{rule {\rm\ruleNext}.}] 
        By induction hypothesis, there is a \LTLnountil formula $\phi$ such that
        $A^{\ltl{X}} \models \phi$, $B^{\ltl{X}} \pperp \phi$ and $\size(\phi) = k-1$.  Since
        $|A^{\ltl{X}}| = |A|$, by definition of $A^{\ltl{X}}$, for every
        $\sigma \in A$ we have $\abs{\sigma} \geq 2$ and $(\sigma,1)
        \models \phi$, i.e., $A \models \ltl{X} \phi$.  We now show
        that $B \pperp \ltl{X} \phi$. \emph{Ad absurdum}, 
        suppose that there is $\sigma' \in B$ such that $(\sigma',0) \models \ltl{X}
        \phi$. This implies $\abs{\sigma} \geq 2$. Then, $\sigma'\sffx{1} \in B^{\ltl{X}}$ and $(\sigma'\sffx{1}, 0) \models \phi$. However, this contradicts $B^{\ltl{X}} \pperp \phi$. 
        Hence, $B \pperp \ltl{X} \phi$.
      \item[\textbf{rule {\rm\ruleFuture}.}] 
        By induction hypothesis, there is a \LTLnountil formula $\phi$ such that 
        ${A^f \models \phi}$, $B^{\ltl{G}} \pperp \phi$ and $\size(\phi) = k-1$. 
        Since $f \in \FP_A$,
        we have $A \models \ltl{F} \phi$. Let us show that $B \pperp
        \ltl{F} \phi$. \emph{Ad absurdum}, suppose that there is
        $\sigma \in B$ such that $(\sigma,0) \models \ltl{F} \phi$.
        Then, there is $j \in \position(\sigma)$ such that $(\sigma,j) \models
        \phi$. We get $(\sigma\sffx{j},0) \models \phi$ and, by definition of $B^{\ltl{G}}$, $\sigma\sffx{j} \in
        B^{\ltl{G}}$. However, this contradicts $B^{\ltl{F}} \pperp \phi$. 
        Hence, $B \pperp \ltl{F} \phi$.
      \item[\textbf{rules {\rm\ruleAnd}, {\rm\ruleWeakNext}, {\rm\ruleGlobally}.}]
        The proofs for \ruleAnd, \ruleWeakNext and \ruleGlobally are analogous 
        to the proofs of \ruleOr, \ruleNext and \ruleFuture, respectively. 
        Indeed, note that the former rules are obtained from the latter rules
        by applying the standard dualities
        \[
          \phi \land \psi \equiv \lnot(\lnot \phi \lor \lnot \psi),
          \qquad
          \weaktomorrow \phi \equiv \lnot \tomorrow \lnot \phi,
          \qquad 
          \always \phi \equiv \lnot \eventually \lnot \phi, 
        \] 
        where $\lnot \phi$ corresponds, in terms of combinatorial proof systems, to a rule of the form $\frac{\separ{B,A}}{\separ{A,B}}$. We leave further details to the reader.

    \end{itemize} 

  \noindent
  ($\Leftarrow$): Given a \LTLnountil formula~$\phi$ that separates $A$ and
  $B$, we construct a deduction tree of size $\size(\phi)$.  
  We proceed by structural induction on $\phi$.  

  \proofitem{base case: $\phi$ literal} 
    In this case, the deduction tree is a simple
    application of the rule~\ruleAtomic. 

  \proofitem{induction step, case: $\phi = \phi_1 \lor \phi_2$} 
    We define $A_1 \coloneqq \{ a \in A : a \models \phi_1
    \}$ and $A_2 \coloneqq A \setminus A_1$. From $A
    \models \phi$ we get $A_1 \models \phi_1$ and $A_2
    \models \phi_2$. From $B \pperp \phi$ we get $B \pperp
    \phi_1$ and $B \pperp \phi_2$. Since $A_i \models
    \phi_i$ and $B \pperp \phi_i$ for both $i \in
    \{1,2\}$, by induction hypothesis $\separ{A_i,B}$ has
    a deduction tree of size $\size(\phi_i)$. With an
    application of the rule \ruleAnd, we obtain a
    deduction tree for $\separ{A,B}$ having size
    $\size(\phi_1) + \size(\phi_2) + 1 = \size(\phi)$.

  \proofitem{induction step, case: $\phi = \ltl{X} \psi$} 
    Since $A \models \ltl{X} \psi$, for every $\sigma
    \in A$ we have $\abs{\sigma} \geq 2$ and  $(\sigma,1)
    \models \psi$. Then, by definition of $A^{\ltl{X}}$,
    $\abs{A^{\ltl{X}}} = \abs{A}$ and $A^{\ltl{X}} \models \psi$.
    Similarly, from $B \pperp \ltl{X} \psi$, for every
    $\sigma' \in B$, if $\abs{\sigma'} \geq 2$ then
    $(\sigma',1) \not\models \psi$. By definition
    of~$B^{\ltl{X}}$, we have $B^{\ltl{X}} \pperp \psi$.
    By induction hypothesis, $\separ{A^{\ltl{X}},B^{\ltl{X}}}$ has a deduction tree 
    of size $\size{\psi}$. With an application to the rule~\ruleNext, 
    we obtain a deduction tree of $\separ{A,B}$ of size $\size(\psi)+1=\size(\phi)$.

  \proofitem{induction step: $\phi = \ltl{F} \psi$} 
    Since $A \models \ltl{F} \psi$, for every $\sigma \in A$ 
    there is $j_\sigma \in \position(\sigma)$
    such that $(\sigma,j_\sigma) \models \psi$. Let $f \in
    \FP_A$ be the map given by $f(\sigma) =
    j_{\sigma}$ for every $\sigma \in A$. 
    We have $A^f \models \psi$. We show
    that $B^{\ltl{G}} \pperp \psi$. 
    \emph{Ad absurdum},
    suppose there is $\sigma_1 \in
    B^{\ltl{G}}$ such that $\sigma_1 \models \psi$. By
    definition of $B^{\ltl{G}}$ there is $\sigma_2 \in B$ 
    such that $\sigma_1 \sqsubseteq \sigma_2$. 
    Then, $(\sigma_2,0) \models \eventually \psi$. 
    However, this contradicts the fact that
    $B \pperp \ltl{F} \psi$. 
    Therefore,  $B^{\ltl{G}} \pperp \psi$. 
    By induction hypothesis, $\separ{A^f, B^{\ltl{G}}}$ has
    a deduction tree of
    size $\size(\psi)$. 
    With an application to the rule~\ruleFuture, 
    we obtain a deduction tree for $\separ{A,B}$ 
    of size $\size(\psi)+1 = \size(\phi)$. 

  \proofitem{induction step, cases $\phi = \phi_1 \land \phi_2$, $\phi = \weaktomorrow \psi$ and $\phi = \always \psi$} 
    As in the previous direction of the theorem, 
    the cases for $\phi = \phi_1 \land \phi_2$, $\phi = \weaktomorrow \psi$ and $\phi = \always \psi$ are analogous to the cases $\phi = \phi_1 \lor \phi_2$, $\phi = \tomorrow \psi$ and $\phi = \eventually \psi$, respectively.
\end{proof}

\section{A list of useful lemmas on deduction trees}

In this appendix we prove a series of helpful lemmas on the sizes of deduction trees.
As usual, below $\Sigma \coloneqq
2^{\AP}$ for some finite set of atomic propositions $\AP$.

\begin{restatable}{lemma}{LemmaUsefulOne}
  \label{lemma:weakening-sets}
  Let $A,B \subseteq \Sigma^+$.
  Consider $A' \subseteq A$ and $B' \subseteq B$. 
  If $\separ{A,B}$ has a deduction tree of size $k$, 
  then $\separ{A',B'}$ has a deduction tree of size $k$.
\end{restatable}

\begin{proof}
  By definition, every formula separating $A$ from $B$ also separates $A'$ from $B'$. 
  Then, the statement follows by applying~\Cref{th:ps}.
\end{proof}


\begin{restatable}{lemma}{LemmaUsefulTwo}
  \label{lemma:prefix-length-proof}
  Let $u \in \Sigma$, $\sigma \in \Sigma^+$ and $i,j \in \N$, with~$u \neq \sigma[0]$.
  The minimal deduction tree~$\Tree$ of 
  $\separ{\{u^i \cdot \sigma\}, \{u^j \cdot \sigma\}}$, if it exists,
  has size $s \geq \min(i,j)+1$. 
\end{restatable}

\begin{proof}
  Note that if $i = j$ then, by~\Cref{lemma:unclosable-tree-condition},
  $\separ{\{u^i \cdot \sigma\}, \{u^j \cdot \sigma\}}$ does not hold, and thus the tree~$\Tree$ does not exist. Below, let us assume $i \neq j$.

  Let $a \coloneqq u^i \cdot \sigma$ and $b \coloneqq u^j \cdot \sigma$.
  The proof is by induction on the size $s$. 

  \proofitem{base case: $s = 1$} 
  In this case, $\Tree$ is a simple application of the rule~\ruleAtomic. It must therefore be the case that $a[0] \neq b[0]$, and so by definition of $a$ and $b$, either $i = 0$ or $j = 0$. Then, $s \geq \min(i,j)+1$.

  \proofitem{induction step: $s \geq 2$} We split the
  proof depending on the rule applied to the root
  $\separ{\{a\}, \{b\}}$. Since $s \geq 2$, this rule cannot be~\ruleAtomic. Note that this implies $i,j \geq 1$, as otherwise by $u \neq \sigma[0]$, which in turn implies $a[0] \neq b[0]$, enabling us to use the rule~\ruleAtomic to derive $\separ{\{a\}, \{b\}}$.

  \begin{itemize}[align=left]
    \item[\textbf{case: rule~{\rm{\ruleNext} and~{\rm{\ruleWeakNext}}}.}] We consider the~\ruleNext and~\ruleWeakNext rules together. 
    We have $\{a\}^{\ltl{X}} = \{u^{i-1} \cdot \sigma\}$ and $\{b\}^{\ltl{X}} = \{u^{j-1} \cdot \sigma\}$. 
    The minimal deduction tree of $\separ{\{a\}^{\ltl{X}}, \{b\}^{\ltl{X}}}$ has size $s-1$, and 
    by induction hypothesis, $s-1 \geq \min(i-1,j-1) + 1$; 
    i.e., $s \geq \min(i,j)+1$.
    \item[\textbf{case: rule~{\rm{\ruleFuture}}.}] Let $f \in \FP_{\{a\}}$ be the future point used when applying this rule. 
    The minimal deduction tree of $(\{a\}^f,\{b\}^{\ltl{G}})$ has size $s-1$.
    Let us distinguish two cases: 
    \begin{itemize}
      \item $i - f(a) \leq j$. In this case, $a^f \in \{b\}^{\ltl{G}}$. 
      This means that the pair $(\{a\}^f,\{b\}^{\ltl{G}})$ does not belong to $\separ{\cdot,\cdot}$.
      It is thus not possible to rely on this pair to 
      build $\Tree$, i.e.,~this case cannot occur.
      \item $i - f(a) > j$. Note that this implies $i > f(a)$. 
      Then, $(u^i \cdot \sigma)^f = u^{i-f(a)} \cdot \sigma$. 
      Since $b \in \{b\}^{\ltl{G}}$, by~\Cref{lemma:weakening-sets}, 
      the minimal deduction tree of $\separ{\{a\}^f,\{b\}}$ has size $k \leq s-1$. By induction hypothesis, $k \geq \min(i-f(a),j) + 1$. We have $\min(i-f(a),j) = \min(i,j)$, 
      and so $s \geq s-1 \geq k \geq \min(i,j)+1$.
    \end{itemize}
    \item[\textbf{case: rule~{\rm{\ruleGlobally}}.}] This case is similar to the one for the rule~\ruleFuture; by symmetry it suffices to swap the roles of $\{a\}$ and $\{b\}$.
    \qedhere
  \end{itemize}
\end{proof}

\begin{restatable}{lemma}{LemmaUsefulThree}
  \label{lemma:future-globally-disabled}
  Consider $A,B \subseteq \Sigma^+$ such that one of the following holds:
  \begin{compactenum}
    \item\label{suffix-lemma-c1} there are $a \in A$ and $b_1,b_2 \in B$ such that $b_1 \sqsubseteq a$ and $a \sqsubseteq b_2$, or 
    \item\label{suffix-lemma-c2} there are $a_1,a_2 \in A$ and $b \in B$ such that $a_1 \sqsubseteq b$ and $b \sqsubseteq a_2$.
  \end{compactenum}
  In every deduction tree~$\Tree$ of~$\separ{A,B}$, the rule applied to the root of $\Tree$ is neither the rule~\ruleFuture nor the rule~\ruleGlobally.
\end{restatable}

\begin{proof}
  Due to the symmetry between the rules~\ruleFuture and~\ruleGlobally, the two cases~\eqref{suffix-lemma-c1} and~\eqref{suffix-lemma-c2} are analogous. Here, we only show the lemma assuming~\eqref{suffix-lemma-c1} as a hypothesis. The statement follows directly from the fact that $\separ{A^f,B^{\ltl{G}}}$ and $\separ{A^{\ltl{G}},B^f}$ have no deduction tree:

  \proofitem{case: rule {\rm\ruleFuture}} 
  Let $f \in \FP_A$. 
  As $a \sqsubseteq b_2$, we have $a^f \sqsubseteq b_2$. 
  So, $a^f \in A^f \cap B^{\ltl{G}}$. By~\Cref{lemma:unclosable-tree-condition} (first statement taken contrapositively) and~\Cref{th:ps}, no formula separates $A^f$ from $B^{\ltl{G}}$. $\separ{A^f,B^{\ltl{G}}}$ does not have any deduction tree.
  
  \proofitem{case: rule {\rm\ruleGlobally}}
  Let $f$ be a future point of $B$. 
  Since $b_1 \sqsubseteq a$, we have $b_1^f \sqsubseteq a$.
  Then, $b_1^f \in A^{\ltl{G}}$ and by~\Cref{lemma:unclosable-tree-condition} (first statement) and~\Cref{th:ps} $\separ{A^{\ltl{G}},B^f}$ does not have any deduction tree.
\end{proof}

\begin{restatable}{lemma}{LemmaUsefulFour}
  \label{lemma:all-pairs-are-considered}
  Consider a partial deduction tree~$\mathcal{T}$ with root $\separ{A,B}$ 
  and only using the rules~\ruleOr,~\ruleAnd,~\ruleNext,~\ruleWeakNext. 
  For every ${(a, b) \in A \times B}$ there is
  a leaf $\separ{A',B'}$ of~$\mathcal{T}$ such that $(a\sffx{\xi}, b\sffx{\xi}) \in A' \times B'$, 
  where $\xi \in \N$ is the number of rules~\ruleNext and~\ruleWeakNext used in the path of $\Tree$ from $\separ{A,B}$ to $\separ{A',B'}$.
\end{restatable}

\begin{proof}
  The proof is by induction on the size $s$ of $\Tree$ and by cases on the rule applied to the root $\separ{A,B}$.

  \proofitem{base case: $s = 1$} In this case $\Tree$ is obtained with a single application of one of the rules in the statement of the lemma. We show the base case for the rules~\ruleOr and~\ruleNext. The cases for~\ruleAnd and~\ruleWeakNext are analogous.
  \begin{itemize}[align=left]
    \item[\textbf{case: rule~{\rm\ruleOr}.}] In this case $\Tree$ is of the form 
    \begin{mathpar}
      \inferrule[\ruleOr]{\separ{A_1,B} \and \separ{A_2,B}}{\separ{A_1 \uplus A_2, B}}
    \end{mathpar}
    where $A = A_1 \uplus A_2$. Then, given $(a,b) \in A \times B$ we either have $(a,b) \in A_1 \times B$ or $(a,b) \in A_2 \times B$, as required by the lemma.
    \item[\textbf{case: rule~{\rm\ruleNext}.}] In this case $\Tree$ is of the form 
    \begin{mathpar}
      \inferrule[\ruleNext]{\separ{A^{\ltl{X}},B^{\ltl{X}}} \and |A^{\ltl{X}}| = |A|}{\separ{A,B}}
    \end{mathpar}
    and so for every $(a,b) \in A \times B$ we have $(a\sffx{1},b\sffx{1}) \in A^{\ltl{X}} \times B^{\ltl{X}}$.
  \end{itemize}
  \proofitem{induction step: $s \geq 2$} The induction step is essentially equivalent to the base case, the sole difference being that the induction hypothesis must be used on the terms $\separ{A',B'}$ required to apply the rule to the root $\separ{A,B}$. We show below the case for the rule~\ruleNext. The cases of the other rules are similar. 
  
  Consider $\separ{A^{\ltl{X}},B^{\ltl{X}}}$. Since $\Tree$ has size $s$, its subtree~$\Tree'$ rooted in~$\separ{A^{\ltl{X}},B^{\ltl{X}}}$ has size~$s-1$. Given $(a,b) \in A \times B$, we have $(a\sffx{1},b\sffx{1}) \in A^{\ltl{X}} \times B^{\ltl{X}}$.
  By applying the induction hypothesis on $\Tree'$, we conclude that there is a leaf $\separ{A',B'}$ of $\Tree'$ such that $(a\sffx{\xi+1},b\sffx{\xi+1}) \in A' \times B'$, where $\xi$ is the number of rules~\ruleNext and~\ruleWeakNext used in the path of $\Tree'$ from $\separ{A^{\ltl{X}},B^{\ltl{X}}}$ to $\separ{A',B'}$.
  Taking into account the rule~\ruleNext applied to the root $\separ{A,B}$ of $\Tree$,
  we conclude that $\xi+1$ is the number of rules~\ruleNext and~\ruleWeakNext used in the path of $\Tree$ from $\separ{A,B}$ to $\separ{A',B'}$. 
\end{proof}

\section{Proof of~\Cref{sec:lbltlxf}}
\label{appendix:proof-section-lower-bound-monolitico}

\LemmaLowerBoundSpoonCase*

\begin{proof}
  First, if $\separ{A,B}$ has no deduction tree then the 
  statement is trivially true. Below we assume that $\separ{A,B}$ has at least one deduction tree. In particular, 
  let~$\Tree$ be the minimal deduction tree for $\separ{A,B}$
  and assume it has size $s$. 
  Without loss of generality, we assume $s \leq \alpha(n)$; we recall that $\alpha(n)$ is the upper bound on the size of the formula of 
  \LTLnountil expressing $\Phi_n$. 
  Indeed, note that the hypothesis that the sets $\tau_1,\dots,\tau_m$ are distinct implies $\abs{B} \leq 2^n$. 
  Then, by definition of $\approx$, 
  for every $b \in B$ there is at most one $a \in A$ such that $a \approx b$. This implies
  $\abs{C} < 2^n$. Therefore, if $s > \alpha(n)$ the statement trivially holds.

  During the proof, we write $\prec$ for the strict total order on elements of $T_Q$ used to construct the trace $\Enum$ enumerating $T_Q$.
  Before continuing the proof of the lemma, we show a useful property of the elements of $C$.
  \ClaimRhoPrecRhok*
  \begin{proofclaim}
    Since $b \in B$, we have $b = (\emptyset^{t_i} \cdot \overline{\tau_i} \cdot \Enum|_{-\tau_i})\sffx{r_i}$.
    From $(a,b) \in C$ and $\approx$,
    \[ 
      a = \emptyset^u \cdot \rho \cdot \emptyset^{\alpha(n)} \cdot \sigma_1 
      \qquad 
      b = \emptyset^u \cdot \rho \cdot \emptyset^{\alpha(n)} \cdot \sigma_2,
    \]
    where $u \in \N$, $\rho \in T_Q \cup T_P$, and $\sigma_1,\sigma_2 \in \Sigma^*$. 
    If $\rho \in T_P$, then $\rho = \overline{\tau_i}$ and the claim follows.
    Therefore, let us consider the case of~$\rho \in T_Q$.
    \emph{Ad absurdum}, suppose $\tau_i \preceq \rho$. 
    Since $\rho$ occurs in $b$, by definition of $\Enum|_{-\tau_i}$ we have $\rho \neq \tau_i$, and thus $\tau_i \prec \rho$. Again by definition of $\Enum|_{-\tau_i}$, 
    this implies that $\sigma$ is a suffix of $\Enum$. By definition of~$\Enum$ and from the two equations above, this means $\sigma_1 = \sigma_2$ 
    and so $a = b$. However, this contradicts the existence of $\Tree$ by~\Cref{lemma:unclosable-tree-condition} and~\Cref{th:ps}.
    Hence, $\rho \in \{\tau \in T_Q : \tau \prec \tau_i\}$.
  \end{proofclaim}

  \vspace{5pt}
  \noindent
  In a nutshell,~\Cref{claim-rho-prec-rhok} 
  tells us that for no element~$(a,b) \in C$
  it can be the case that $b$ is a suffix of $\Enum$. 
  This claim is used later to tackle the case of the rule~\ruleFuture, but it shows a property that is useful to keep in mind throughout the proof.

  Let us go back to the proof 
  of~\Cref{lemma:lower-bound:spoon-case-with-globally}.
  If $A = \emptyset$ or $m = 0$ then $C = \emptyset$ and the lemma follows trivially. Below, let us assume $A \neq \emptyset$ and $m \geq 1$. We prove the statement by induction on the size $s$ of $\Tree$. 

  \proofitem{base case: $s = 1$} 
  In this case, $\Tree$ is a simple application of the rule~\ruleAtomic. This means that for every $a \in A$ and $b \in B$ we have $a[0] \neq b[0]$. By definition of~$\approx$, this implies $C = \emptyset$, and therefore $s \geq \abs{C}+1$.

  \proofitem{induction step: $s \geq 2$}
  Note that if $\abs{C} \leq 1$ then the statement follows trivially. Hence, below, we assume $\abs{C} \geq 2$.
  We split the proof depending on the rule applied to
  the root $\separ{A,B}$ of~$\Tree$. Since $s \geq 2$, this rule
  cannot be~\ruleAtomic.
  \begin{itemize}[align=left]
    \setlength{\itemsep}{3pt}
    \item[\textbf{case: rule~{\rm{\ruleOr}}.}] 
      Consider $A_1,A_2$ such that $A_1 \uplus A_2 = A$.
      The cases where $A_1 = A$ or $A_2 = A$ yield a
      contradiction with the minimality of $\Tree$, 
      as one of the hypotheses used to apply the rule~\ruleOr is $\separ{A,B}$ itself. 
      Hence,
      assume $A_1$ and $A_2$ non-empty.
      For $i \in \{1,2\}$, 
      let $s_i$ be the size of the minimal deduction 
      tree of $\separ{A_i,B}$. 
      Let $C_i \coloneqq \{(a,b) \in A_i \times B :
      {a \approx b} \}$.
      By induction hypothesis, $s_1 \geq |C_1|+1$ 
      and $s_2 \geq |C_2|+1$.
      By minimality of $\Tree$,
      we have $s = s_1 + s_2 + 1 \geq |C_1|+|C_2|+3$.
      Since $A_1$ and $A_2$
      partition~$A$, given $(a,b) \in C$ 
      we have $(a,b) \in C_1$ or $(a,b) \in C_2$.
      Then, 
      $|C| \leq |C_1| + |C_2|$; 
      and $s \geq |C| + 2$.
    \item[\textbf{case: rule~{\rm{\ruleAnd}}.}] 
      This case is analogous to the previous one. 
      Consider $B_1,B_2$ such that $B_1 \uplus B_2 = B$.
      Assume $B_1 \neq \emptyset$ and $B_2 \neq \emptyset$; 
      else we have a contradiction with the
      minimality of $\Tree$.
      For $i \in \{1,2\}$, 
      let $s_i$ be the size of the minimal deduction 
      tree of $\separ{A,B_i}$, and define 
      $C_i \coloneqq \{(a,b) \in A \times B_i :
      {a \approx b} \}$.
      By induction hypothesis, $s_1 \geq |C_1|+1$ 
      and ${s_2 \geq |C_2|+1}$.
      From the minimality of $\Tree$,
      $s = s_1 + s_2 + 1 \geq |C_1|+|C_2|+3$.
      Since $B_1$ and $B_2$
      partition~$B$,
      we derive
      $|C| \leq |C_1| + |C_2|$. 
      Then, $s \geq |C| + 2$.
    \item[\textbf{case: rule~{\rm{\ruleNext}} 
      and~{\rm{\ruleWeakNext}}.}] 
      \!\!We consider the~\ruleNext and~\ruleWeakNext rules
      together, as both rely on deriving $\separ{A^{\ltl{X}},B^{\ltl{X}}}$.
      Perhaps surprisingly, their treatment is non-trivial.
      The main difficulty is given by 
      the fact that the set $C' \coloneqq \{ (a,b) \in A^{\ltl{X}} \times B^{\ltl{X}} : {a \approx b} \}$ might in principle even be empty, and thus relying on the induction hypothesis to derive that $\separ{A^{\ltl{X}},B^{\ltl{X}}}$ has a minimal deduction tree of size at least $\abs{C'}+1$ is unhelpful for concluding that $s \geq \abs{C}+1$.
      Below, we show how to circumvent this issue.

      The minimal deduction tree for
      $\separ{A^{\ltl{X}},B^{\ltl{X}}}$ has size $s-1$. 
      Within this deduction tree, consider the maximal partial deduction tree~$\Tree'$ 
      rooted at $\separ{A^{\ltl{X}},B^{\ltl{X}}}$ and made solely 
      of applications of the rules~\ruleAnd,~\ruleOr,~\ruleNext, and~\ruleWeakNext.
      Let $\separ{A_1,B_1},\dots,\separ{A_q,B_q}$ be the leafs of such a tree. 
      Let $j \in [1,q]$.
      In the original tree~$\Tree$, to $\separ{A_j,B_j}$ it is applied a rule different from~\ruleAnd,~\ruleOr,~\ruleNext and~\ruleWeakNext.
      Let~${\xi_j \geq 1}$ be the number of rules \ruleNext and \ruleWeakNext used in the path of $\Tree$ from $\separ{A,B}$ to $\separ{A_j,B_j}$. 
      Note that, from $s \leq \alpha(n)$, we have $\xi_j \leq \alpha(n)$.
      We define the following two sets $C_j$ and $N_j$, whose role is essentially to ``track'' the evolution of pairs in $C$ with respect to $A_j \times B_j$:
      \begin{align*}
        C_j &\coloneqq \{(a\sffx{\xi_j},b\sffx{\xi_j}) \in A_j \times B_j : {(a,b) \in C},\, a\sffx{\xi_j} \approx b\sffx{\xi_j} \},\\
        N_j &\coloneqq \{(a\sffx{\xi_j},b\sffx{\xi_j}) \in A_j \times B_j : {(a,b) \in C},\,
        a\sffx{\xi_j} \not\approx b\sffx{\xi_j} \}.
      \end{align*}
      By induction hypothesis, the minimal 
      deduction tree for $\separ{A_j,B_j}$ 
      has size $s_j \geq \abs{C_j}+1$.
      We now show series of properties on the sets $C_j$ and $N_j$, from which we derive $s \geq \abs{C}+1$.

      \begin{restatable}{clm}{ClmProofLBOneClmOne}
        \label{proof-lb1:clm1}
        Let $j \in [1,q]$ and $(a,b) \in N_j$. There 
        are $(a',b') \in C$, $u \in [0,\alpha(n)]$, $i \in [1,m]$, $\rho \in \{ \overline{\tau_i} \} \cup \{\tau \in T_Q : \tau \prec \tau_i \}$ and $\sigma \sqsubseteq \Enum$ satisfying:
        \begin{align*}
          a' &\ =\ \emptyset^u \cdot \rho \cdot \emptyset^{\alpha(n)-v} \cdot a,
          &
          a  &\ =\ \emptyset^v \cdot \tau_i \cdot \emptyset^{\alpha(n)-v} \cdot b,
          \\
          b' &\ =\ \emptyset^u \cdot \rho \cdot \emptyset^{\alpha(n)-v} \cdot b,
          &
          b &\ =\ \emptyset^v \cdot \sigma,\\
          b' & \ =\ (\emptyset^{t_i} \cdot \overline{\tau_i} \cdot \Enum|_{-\tau_i})\sffx{r_i},
        \end{align*}
        where $v \coloneqq u+1+\alpha(n)-\xi_j$, 
        and if $\rho = \overline{\tau_i}$ then $\tau_i$ is the smallest 
        element with respect to the strict order~$\prec$.
      \end{restatable}

      \begin{proofclaim}
        Since $(a,b) \in N_j$, there is $(a',b') \in C$ such that $a'\sffx{\xi_j} = a$ and
        $b'\sffx{\xi_j} = b$.
        By~\Cref{claim-rho-prec-rhok},
        there is $i \in [1,m]$, 
        $u \in \N$, 
        $\rho \in \{ \overline{\tau_i} \} \cup \{\tau \in T_Q : \tau \prec \tau_i \}$ and $\sigma_2 \in \Sigma^*$ 
        such that 
        \begin{center}
        $b' = (\emptyset^{t_i} \cdot \overline{\tau_i} \cdot \Enum|_{-\tau_i})\sffx{r_i}$ \
        and \ $b' = \emptyset^u \cdot \rho \cdot \emptyset^{\alpha(n)} \cdot \sigma_2$.
        \end{center}
        From $(a',b') \in C$ we have $a' \approx b'$, and thus there is $\sigma_1 \in \Sigma^*$ such that 
        $a' = \emptyset^u \cdot \rho \cdot \emptyset^{\alpha(n)} \cdot \sigma_1$.
        Since $\xi_j \leq \alpha(n)$, $a'\sffx{\xi_j} = a$ and
        $b'\sffx{\xi_j} = b$, one of the following cases holds: 

        \begin{enumerate}[label=(\Alph*)]
          \item\label{cc10-i1} 
            $a = \emptyset^{t} \cdot \rho \cdot \emptyset^{\alpha(n)} \cdot \sigma_1$ and $b = \emptyset^{t} \cdot \rho \cdot \emptyset^{\alpha(n)} \cdot \sigma_2$, with $t \coloneqq u - \xi_j \geq 0$, or 
          \item\label{cc10-i2} 
            $a = \emptyset^{v} \cdot \sigma_1$ and $b = \emptyset^{v} \cdot \sigma_2$ with $v$ such that $\xi_j =u+1+\alpha(n)-v$.
        \end{enumerate}

        \emph{Ad absurdum}, assume that the case~\ref{cc10-i1} holds. 
        Then, we derive $a \approx b$,
        which however contradicts $(a,b) \in N_j$. Hence, case~\ref{cc10-i2} holds.
        Note that, then, from $\xi_j \leq \alpha(n)$ we get $u \in [0,\alpha(n)]$.
        Since $a' \approx b'$ and $a \not\approx b$, by definition of $\Enum$ and $\Enum|_{-\tau_i}$ we conclude that 
        $\sigma_1 = \tau_i \cdot \emptyset^{\alpha(n)} \cdot \sigma_2$.
        Note that then $\sigma_2 \sqsubseteq a \sqsubseteq \Enum$. 
        Moreover, note that if $\rho = \overline{\tau_i}$, then it must be the case that $\tau_i$ is the smallest element with respect to the strict order~$\prec$. 
        Indeed, if $\tau_i$ is any other element, $\sigma_1$ and $\sigma_2$ would start with that smallest element of~$\prec$, and thus we would have $a \approx b$.
        Putting together the equations we derived, i.e.,  
        \begin{align*}
          &a' = \emptyset^u \cdot \rho \cdot \emptyset^{\alpha(n)} \cdot \sigma_1,
          &&b' = \emptyset^u \cdot \rho \cdot \emptyset^{\alpha(n)} \cdot \sigma_2,
          &&b' = (\emptyset^{t_i} \cdot \overline{\tau_i} \cdot \Enum|_{-\tau_i})\sffx{r_i},\\
          &a = \emptyset^{v} \cdot \sigma_1,
          &&b = \emptyset^{v} \cdot \sigma_2,
          &&\sigma_1 = \tau_i \cdot \emptyset^{\alpha(n)} \cdot \sigma_2,
        \end{align*}
        the claim follows (w.r.t.~the statement, $\sigma = \sigma'$).
      \end{proofclaim}

      \ClmProofLBOneClmTwo*

      \begin{proofclaim}
        By definition of $\Tree'$, the only rules that can be applied to $\separ{A_j,B_j}$ are~\ruleAtomic,~\ruleFuture and~\ruleGlobally. Suppose  $C_j \cup N_j \neq \emptyset$. We consider two cases, depending on the non-emptiness of $C_j$ and $N_j$. 
        \begin{itemize}
          \item If there is a pair $(a,b) \in C_j$, then $a \approx b$ and $(a,b) \in A_j \times B_j$.
          By definition of $\approx$, $a[0] = b[0]$, and so the rule~\ruleAtomic cannot be applied to $\separ{A_j,B_j}$.
          \item If there is $(a,b) \in N_j$ then, by \Cref{proof-lb1:clm1}, ${a,b \in (\emptyset^{v} \cdot \Sigma^*)}$ for some $v \geq 1$.
          Hence $a[0] = b[0]$, so the rule~\ruleAtomic cannot be applied to $\separ{A_j,B_j}$.
        \end{itemize}
        Since $C_j \cup N_j \neq \emptyset$, from the two cases above we conclude that only 
        the rules~\ruleFuture and~\ruleGlobally can be 
        applied to $\separ{A_j,B_j}$.
      \end{proofclaim}

      \ClmProofLBOneClmThree*
      
      \begin{proofclaim}
        \emph{Ad absurdum}, suppose $\abs{N_j} \geq 2$. Pick distinct $(a_1,b_1),(a_2,b_2) \in N_j$. 
        From~\Cref{proof-lb1:clm1}, the following equations hold:
        \begin{align*}
          a_1' &\ =\ \emptyset^{u_1} \cdot \rho_1 \cdot \emptyset^{\alpha(n)-v_1} \cdot a_1,
          &
          a_1  &\ =\ \emptyset^{v_1} \cdot \tau_i \cdot \emptyset^{\alpha(n)-v_1} \cdot b_1,
          \\
          b_1' &\ =\ \emptyset^{u_1} \cdot \rho_1 \cdot \emptyset^{\alpha(n)-v_1} \cdot b_1,
          &
          b_1 &\ =\ \emptyset^{v_1} \cdot \sigma_1,\\
          b_1' & \ =\ (\emptyset^{t_i} \cdot \overline{\tau_i} \cdot \Enum|_{-\tau_i})\sffx{r_i},
        \end{align*}
        and 
        \begin{align*}
          a_2' &\ =\ \emptyset^{u_2} \cdot \rho_2 \cdot \emptyset^{\alpha(n)-v_2} \cdot a_2,
          &
          a_2  &\ =\ \emptyset^{v_2} \cdot \tau_k \cdot \emptyset^{\alpha(n)-v_2} \cdot b_2,
          \\
          b_2' &\ =\ \emptyset^{u_2} \cdot \rho_2 \cdot \emptyset^{\alpha(n)-v_2} \cdot b_2,
          &
          b_2 &\ =\ \emptyset^{v_2} \cdot \sigma_2,\\
          b_2' & \ =\ (\emptyset^{t_k} \cdot \overline{\tau_k} \cdot \Enum|_{-\tau_k})\sffx{r_k},
        \end{align*}
        for some $(a_1',b_1'), (a_2',b_2') \in C$, $u_1,u_2 \in [0,\alpha(n)]$, 
        $i,k \in [1,m]$, 
        $\rho_1 \in \{ \overline{\tau_i} \} \cup \{\tau \in T_Q : \tau \prec \tau_i \}$, 
        $\rho_2 \in \{ \overline{\tau_i} \} \cup \{\tau \in T_Q : \tau \prec \tau_k \}$,
        $\sigma_1,\sigma_2 \sqsubseteq \Enum$, 
        and where $v_1 = u_1+1+\alpha(n)-\xi_j$ and $v_2 = u_2+1+\alpha(n)-\xi_j$. Moreover, if $\rho_1 = \overline{\tau_i}$ (resp.~$\rho_2 = \overline{\tau_k}$) then $\tau_i$ (resp.~$\tau_k$) is the smallest element with respect to $\prec$.
        
        Note that $\xi_j = u_2+1+\alpha(n)-v_2 = u_1+1+\alpha(n)-v_1$. Then,
        since $(a_1,b_1) \neq (a_2,b_2)$ we conclude that $(a_1',b_1') \neq (a_2',b_2')$.

        \emph{Ad absurdum}, suppose $\rho_1 = \overline{\tau_i}$ and $\rho_2 = \overline{\tau_k}$. 
        Hence, both $\tau_i$ and $\tau_k$ must be the smallest element with respect to~$\prec$,
        meaning that $\tau_i = \tau_k$. Since the elements $\tau_1,\dots,\tau_m$ are pairwise distinct, this implies $i = k$.
        Therefore, $b_1' = b_2'$ and, from $a_1' \approx b_1'$ and $a_2' \approx b_2'$, we conclude that $a_1'=a_2'$. 
        However, this contradicts $(a_1',b_1') \neq (a_2',b_2')$, 
        hence at least one among $\rho_1$ and $\rho_2$ belong to $T_Q$. By definition of $A$, this implies that at least one among $a_1'$ and $a_2'$ is a suffix of $\Enum$, and therefore either $a_1' \sqsubseteq a_2'$ or $a_2' \sqsubseteq a_1'$. Without loss of generality, below we assume $a_2' \sqsubseteq a_1'$. The case of $a_1' \sqsubseteq a_2'$ is analogous. 
        Note that $a_2' \sqsubseteq a_1'$ implies $\rho_2 \in \{\tau \in T_Q : \tau \prec \tau_k\}$.

        We divide the proof in three cases $\rho_2 \prec \tau_i$, $\rho_2 = \tau_i$,  and $\tau_i \prec \rho_2$.
        In all cases, we derive an absurdum, and conclude that then $\abs{N_j} \leq 1$.
        
        \vspace{5pt}
        \textit{\underline{Case~1:} $\rho_2 \prec \tau_i$.} In this case, $\tau_i$ is not the smallest element with respect to $\prec$, and therefore $\rho_1  \in \{\tau \in T_Q : \tau \prec \tau_i\}$. 
        From the equation $a_1' \ =\ \emptyset^{u_1} \cdot \rho_1 \cdot \emptyset^{\alpha(n)} \cdot \tau_i \cdot \emptyset^{\alpha(n)-v_1} \cdot b_1$ 
        and the fact that $a_2' \sqsubseteq a_1'$, we conclude that $\rho_1 = \rho_2$. From the equations at the beginning of the proof, we conclude that $\tau_i = \tau_k$. Since the elements $\tau_1,\dots,\tau_m$ are pairwise distinct, this implies $i = k$. Then, $b_1' = b_2'$. 
        As already discussed above, $b_1' = b_2'$ implies $a_1' = a_2'$. 
        However, this contradicts $(a_1',b_1') \neq (a_2',b_2')$. 
        Hence, $\abs{N_j} \leq 1$.

        \vspace{5pt}
        \textit{\underline{Case~2:} $\rho_2 = \tau_i$.} In this case, 
        since every element of $T_Q$ only appears once in $\Enum$, 
        and $a_2' \sqsubseteq a_1'$,
        from the equations at the beginning of the proof we get: 
        \begin{align*}
          a_1' =&& \emptyset^{u_1} \cdot \rho_1 \cdot \emptyset^{\alpha(n)-v_1} \cdot{} & \tau_i \cdot \emptyset^{\alpha(n)-v_1} \cdot \emptyset^{v_1} \cdot \sigma,\\
           a_2' =&& \emptyset^{u_2} \cdot{} & \tau_i \cdot \emptyset^{\alpha(n)-v_2} \cdot \emptyset^{v_2} \cdot \sigma,
        \end{align*}
        for some $\sigma \in \Sigma^*$. Then, we get $\sigma_1 = \sigma_2 = \sigma$, 
        and $b_1 = \emptyset^{v_1} \cdot \sigma$ and $a_2 = \emptyset^{v_2} \cdot \sigma$. 
        Since $(a_2,b_1) \in A_j \times B_j$, by~\Cref{lemma:prefix-length-proof,lemma:weakening-sets}, 
        the minimal deduction tree for $\separ{A_j,B_j}$ 
        has size $\min(v_1,v_2)+1$. Then, taking into account the $\xi_j$ applications of the rules \ruleNext and \ruleWeakNext, we conclude that 
        \begin{align*}
          s &{}\geq \xi_j + \min(v_1,v_2)+1\\
            &{}=    \xi_j + \min(u_1+1+\alpha(n)-\xi_j,u_2+1+\alpha(n)-\xi_j)+1\\
            &{}= \alpha(n)+2 + \min(u_1,u_2) > \alpha(n).
        \end{align*}
        However, this contradicts the hypothesis $s \leq \alpha(n)$. 
        Hence, $\abs{N_j} \leq 1$.

        \vspace{5pt}
        \textit{\underline{Case~3:} $\tau_i \prec \rho_2$.}
        From $a_2' \sqsubseteq a_1'$ and the equations at the beginning of the proof, and by definition of $\Enum$, we conclude that $\rho_2$ occurs in $\sigma_1$.
        This implies $a_2 \sqsubseteq b_1$. 
        Since moreover $b_1 \sqsubseteq a_1$,
        by~\Cref{lemma:future-globally-disabled} we conclude that the rule applied in $\Tree$ in order to derive $\separ{A_j,B_j}$ is neither~\ruleFuture 
        nor~\ruleGlobally. However, this contradicts~\Cref{proof-lb1:clm2}. 
        Hence $\abs{N_j} \leq 1$.
      \end{proofclaim}

      \ClmProofLBOneClmFour*

      \begin{proofclaim}
        The claim follows as soon as we prove the following 
        two statements: 
        \begin{enumerate}
          \item\label{clm4-it1} for every $(a,b) \in C$ there is $j \in [1,q]$ such that $(a\sffx{\xi_j},b\sffx{\xi_j}) \in C_j \cup N_j$,
          \item\label{clm4-it2} for all distinct $(a_1,b_1),(a_2,b_2) \in C$, we have $(a_1\sffx{\ell},b_1\sffx{\ell}) \neq (a_2\sffx{\ell},b_2\sffx{\ell})$ for every 
          $\ell \leq \alpha(n)$ (recall that $\xi_j \leq \alpha(n)$, for every $j \in [1,q]$).
        \end{enumerate}
        
        Let us prove~\Cref{clm4-it1}.
        By~\Cref{lemma:all-pairs-are-considered}, 
        for every $(a,b) \in C$ there is $j \in [1,q]$ 
        such that ${(a\sffx{\xi_j},b\sffx{\xi_j}) \in A_j \times B_j}$. By definition of $C_j$ and $N_j$, 
        if $a\sffx{\xi_j} \approx b\sffx{\xi_j}$ 
        then $(a\sffx{\xi_j},b\sffx{\xi_j}) \in C_j$, 
        and otherwise $(a\sffx{\xi_j},b\sffx{\xi_j}) \in N_j$.
        
        We move to~\Cref{clm4-it2}.
        Pick two distinct pairs $(a_1,b_1),(a_2,b_2) \in C$.
        We have $a_1 \approx
        b_1$ and $a_2 \approx b_2$. Note that, by definition of
        $\approx$ and~$\Enum$, this means that the length of
        the traces $a_1,a_2,b_1,b_2$ is at least
        $\alpha(n)+1$. 
        
        \emph{Ad absurdum},
        suppose that $b_1 = b_2$.
        Since $b_1,b_2 \in B$, there is 
        $i \in [1,m]$ such that $b_1 = b_2 = (\overline{\tau_i} \cdot \Enum|_{-\tau})\sffx{r_i}$. 
        By $a_1 \approx b_1 $ and $a_2 \approx b_2$, this implies $a_1 = a_2$. 
        However, this contradicts $(a_1,b_1) \neq (a_2,b_2)$. 
        Therefore, we have $b_1 \neq b_2$.
        By~\Cref{claim-rho-prec-rhok} and $a_1 \approx b_1$ and $a_2 \approx b_2$,
        \begin{align*}
          a_1 &\ =\ \emptyset^{u_1} \cdot \rho_1 \cdot \emptyset^{\alpha(n)} \cdot \sigma_1,
          &
          a_2  &\ =\ \emptyset^{u_2} \cdot \rho_2 \cdot \emptyset^{\alpha(n)} \cdot \sigma_2,
          \\
          b_1 &\ =\ \emptyset^{u_1} \cdot \rho_1 \cdot \emptyset^{\alpha(n)} \cdot \sigma_1',
          &
          b_2 &\ =\ \emptyset^{u_2} \cdot \rho_2 \cdot \emptyset^{\alpha(n)} \cdot \sigma_2',\\ 
          b_1 &\ =\ (\emptyset^{t_i} \cdot \overline{\tau_i} \cdot \Enum|_{-\tau_i})\sffx{r_i},
          & 
          b_2 &\ =\ (\emptyset^{t_k} \cdot \overline{\tau_k} \cdot \Enum|_{-\tau_i})\sffx{r_k},
        \end{align*}
        for some $i,k \in [1,m]$,
        $u_1,u_2 \in \N$,
        $\rho_1 \in \{ \overline{\tau_i} \} \cup \{\tau \in T_Q : \tau \prec \tau_i \}$,
        $\rho_2 \in \{ \overline{\tau_k} \} \cup \{\tau \in T_Q : \tau \prec \tau_k \}$,
        and $\sigma_1,\sigma_2,\sigma_1',\sigma_2' \in \Sigma^*$.
        Since $b_1 \neq b_2$, we have $i \neq k$. We divide the proof into three cases. 

        \vspace{5pt}
        \textit{\underline{Case~1:} $\rho_1 = \rho_2$.} 
        In this case, from $i \neq k$ we get $\rho_1 = \rho_2 \in \{\tau \in T_Q : \tau \prec \tau_i \text{ and } \tau \prec \tau_k\}$. 
        By definition of $\Enum|_{-\tau_i}$ and $\Enum|_{-\tau_k}$, 
        we conclude that $\sigma_1'$ has a position of type $\tau_k$ but not a position of type $\tau_i$, whereas $\sigma_2'$ has a position of type $\tau_i$ but not a position of type $\tau_k$. Therefore, 
        we conclude that for every $\ell \in [0,\alpha(n)]$, $b_1\sffx{\ell} \neq b_2\sffx{\ell}$; which in turn implies $(a_1\sffx{\ell},b_1\sffx{\ell}) \neq (a_2\sffx{\ell},b_2\sffx{\ell})$.

        \vspace{5pt}
        \textit{\underline{Case~2:} $\rho_1 \neq \rho_2$ and at least one among $\rho_1$ and $\rho_2$ belong to $T_Q$.} In this case, $a_1 \neq a_2$. 
        Moreover, since one among $\rho_1$ and $\rho_2$ belong to $T_Q$, we conclude by definition of $\AA$ and $\Enum$
        that either $a_1 \sqsubseteq a_2$ or $a_2 \sqsubseteq a_1$. 
        Then, for every $\ell \in [0,\alpha(n)]$, $a_1\sffx{\ell} \neq a_2\sffx{\ell}$; which in turn implies $(a_1\sffx{\ell},b_1\sffx{\ell}) \neq (a_2\sffx{\ell},b_2\sffx{\ell})$.

        \vspace{5pt}
        \textit{\underline{Case~3:} $\rho_1 \neq \rho_2$ and $\rho_1,\rho_2 \not\in T_Q$.} We have $\rho_1 = \overline{\tau_i}$ and $\rho_2 = \overline{\tau_k}$. 
        As in the first case, 
        from $\tau_i \neq \tau_k$
        and by definition of $\Enum|_{-\tau_i}$ and $\Enum|_{-\tau_k}$, 
        we conclude that $\sigma_1'$ has a position of type $\tau_k$ but not a position of type $\tau_i$, and $\sigma_2'$ has a position of type $\tau_i$ but not a position of type $\tau_k$. Therefore, for every $\ell \in [0,\alpha(n)]$, $b_1\sffx{\ell} \neq b_2\sffx{\ell}$; which in turn implies $(a_1\sffx{\ell},b_1\sffx{\ell}) \neq (a_2\sffx{\ell},b_2\sffx{\ell})$. 
      \end{proofclaim}

      \vspace{5pt}
      We are now ready to prove that $s \geq \abs{C}+1$, concluding the proof for~\ruleNext and~\ruleWeakNext:
      {\allowdisplaybreaks
      \begin{align*}
        s &\geq 1 + \textstyle\sum_{j=1}^q s_j 
          &\text{by definition~of $\Tree$ and $\Tree'$}\\
          &\geq 1+\textstyle\sum_{j=1}^q (\abs{C_j}+1) 
          &\text{by $s_j \geq \abs{C_j}+1$ (induction hypothesis)}\\
          &\geq 1+\textstyle\sum_{j=1}^q (\abs{C_j \cup N_j})
          &\text{by $\abs{N_j} \leq 1$ (\Cref{proof-lb1:clm3})}\\
          &\geq \abs{C}+1
          &\text{by $\textstyle\abs{C} \leq \textstyle\sum_{j=1}^q \abs{C_j \cup N_j}$ (\Cref{proof-lb1:clm4}).}
      \end{align*}
      }

    \item[\textbf{case: rule~{\rm{\ruleFuture}}.}]
      Let $f \in \FP_A$ be the future point used when applying this rule. Define $C' \coloneqq \{(a',b') \in A^f \times B^{\ltl{G}} : a' \approx b' \}$. 
      The minimal
      deduction tree for $\separ{A^f,B^{\ltl{G}}}$ has size
      $s-1$. 
      By induction hypothesis, $s-1 \geq \abs{C'} + 1$, i.e., $s \geq \abs{C'} + 2$. We divide the proof 
      into two cases.

      \vspace{5pt}
      \textit{\underline{Case~1:} for every $a' \in A^f$, $a' \not\sqsubseteq \Enum$.}
      In this case, by definition of~$\approx$, every $(a,b) \in C$ is such that 
      $a$ and $b$ belong to the language 
      $\emptyset^u \cdot \overline{\tau_i} \cdot \emptyset^{\alpha(n)} \cdot \Sigma^*$
      for some~$u \in \N$, and $i \in [1,m]$.
      Since $a^f \not\sqsubseteq \Enum$, we must have $f(a) \leq u+1$. 
      Then, $a^f \approx b\sffx{f(a)}$. 
      Note that distinct $(a,b) \in C$ concern distinct $\overline{\tau_i}$ with $i \in [1,m]$, and therefore,
      together with $b\sffx{f(a)} \in B^G$, one concludes that $\abs{C'} \geq \abs{C}$; and so $s \geq \abs{C}+2$. 

      \vspace{5pt}
      \textit{\underline{Case~2:} there is $a' \in A^f$ such that $a' \sqsubseteq \Enum$.}
      Let us denote with $\widetilde{a}$ the element in $A^f$ such that $\widetilde{a} \sqsubseteq a$ for every $a \in A^f$. The existence of such an element 
      follows directly from the fact that $a' \sqsubseteq \Enum$ for some $a' \in A^f$. 

      Let $I \subseteq [1,m]$ be the subset of those indices $i \in [1,m]$ for which there is a pair $(a',b') \in C$ such that $b' = (\emptyset^{t_i} \cdot \overline{\tau_i} \cdot \Enum|_{\tau_i})\sffx{r_i}$. 
      W.l.o.g., suppose $I = \{1,\dots,q\}$ for some $q \leq m$,
      and that $\tau_{1} \prec \tau_{2} \prec \dots \prec \tau_{q}$; recall that all these types are pairwise distinct.
      By definition of $\approx$, $A$ and $\Enum$, 
      one has $a_1 = a_2$
      for every $(a_1,b'),(a_2,b') \in C$.
      Therefore, $q = \abs{C}$.
      To conclude the proof it suffices to show that $\abs{C'} \geq q - 1$.
      We do so by establishing a series of claims.
      Below, recall that we are assuming $\abs{C}\geq 2$, so in particular $C \neq \emptyset$.

      \ClmProofLBFClmOneNew*
      \begin{proofclaim}
        Let us focus on the first part of the claim.
        We have $\widetilde{a} \sqsubseteq \Enum$, and so
        either (1) $\widetilde{a} = \emptyset^v$ for some $v \in [1,\alpha(n)]$, or (2) there are $u \in \N$, $\rho \in T_Q$ and $\sigma \in (2^{Q})^*$ such that $\widetilde{a} = \emptyset^{u} \cdot \rho \cdot \emptyset^{\alpha(n)} \cdot \sigma$.

        \emph{Ad absurdum}, suppose that we are in the case (1).
        Since $C \neq \emptyset$, there is $(a',b') \in C$. Then, by definition of $\approx$, 
        we have
        $\emptyset^{\alpha(n)} \sqsubseteq b'$.
        Therefore, $\emptyset^v \in B^{\ltl{G}}$. 
        However, this contradicts the existence of $\Tree$ by~\Cref{lemma:unclosable-tree-condition} and~\Cref{th:ps}.
        Hence, we are in case (2), which shows the first part of the claim.

        For the second part of the claim, 
        \emph{ad absurdum} suppose there is $i \in I$ such that $\tau_i \prec \rho$. 
        We show that this implies again $\widetilde{a} \in B^{\ltl{G}}$; thus contradicting the existence of $\Tree$.
        By definition of $I$,
        there is $(a',b') \in C$ such that 
        $b' =  (\emptyset^{t_i} \cdot \overline{\tau_i} \cdot \Enum|_{-\tau_i})\sffx{r_i}$. 
        By~\Cref{claim-rho-prec-rhok},
        $b' = \emptyset^u \cdot \rho \cdot \emptyset^{\alpha(n)} \cdot \sigma$, 
        for some $u \in \N$, $\rho \in \{ \overline{\tau_i} \} \cup \{\tau \in T_P : \tau \prec \tau_i \}$ and $\sigma \in \Sigma^*$. 
        By definition of
        $\Enum|_{-\tau_i}$, 
        and since $\tau_i \prec \rho$, 
        we conclude that the suffix~$\gamma$ of $\Enum$ 
        beginning with $\emptyset^{\alpha(n)} \cdot \rho$ is also a suffix of $b'$.
        We have $\widetilde{a} \sqsubseteq \gamma \sqsubseteq b'$, which implies $\widetilde{a} \in B^{\ltl{G}}$.
      \end{proofclaim}

      \vspace{5pt}
      Below, we write $u,\rho$ and $\sigma$ for the objects 
      appearing in~\Cref{proof-lbF-clm1-new}.
      Note that, from $\tau_1 \prec \dots \prec \tau_q$, the second statement of~\Cref{proof-lbF-clm1-new} implies 
      $\rho \prec \tau_2 \dots \prec \tau_q$. 
      For $i \in [2,q]$, let $(a_i',b_i')$ denote the pair in $C$ such that $b_i' = (\emptyset^{t_i} \cdot \overline{\tau_i} \cdot \Enum|_{\rho_i})\sffx{r_i}$.

      \ClmProofLBFClmTwoNew*
      \begin{proofclaim}
        By~\Cref{claim-rho-prec-rhok},
        $b_i' = \emptyset^v \cdot \rho' \cdot \emptyset^{\alpha(n)} \cdot \sigma'$, 
        for some $v \in \N$, $\rho' \in \{ \overline{\tau_i} \} \cup \{{\tau \in T_P} : \tau \prec \tau_i \}$ and $\sigma' \in \Sigma^*$. 
        Since $i \geq 2$, by definition of $\Enum|_{-\tau_i}$, and since $\rho \prec \tau_i$,
        we conclude that there is a suffix $b_i''$ of $b_i'$ having the form $b_i'' = \emptyset^u \cdot \rho \cdot \emptyset^{\alpha(n)} \cdot \sigma''$, for some $\sigma'' \in \Sigma^*$.
        Following the first statement of~\Cref{proof-lbF-clm1-new}, we have $\widetilde{a} \approx b_i''$; i.e., the first statement of the claim. The second statement follows from
        $\widetilde{a} \approx b_i''$, the fact that $\rho \prec \tau_2 \dots \prec \tau_q$, and by definition of $\Enum|_{-\tau_i}$.
      \end{proofclaim}

      \vspace{5pt}
      Since all types $\tau_2, \dots, \tau_q$ are pairwise distinct, from the second statement in~\Cref{proof-lbF-clm2-new}
      we conclude that $b_i'' \neq b_j''$ for every two distinct $i,j \in I \setminus \{i_1\}$.
      Then, the first statement in~\Cref{proof-lbF-clm2-new}
      entails $\abs{C'} \geq q-1$. 
      This concludes the proof of the case~\ruleFuture.

    \item[\textbf{case: rule~{\rm{\ruleGlobally}}.}] 
      Let $f \in \FP_A$ be the future point used when applied this rule. 
      The minimal deduction tree for $\separ{A^{\ltl{G}},B^f}$ has size $s-1$. 
      We define ${C' \coloneqq \{(a',b') \in A^{\ltl{G}} \times B^{f} : a' \approx b' \}}$. By induction hypothesis, $s-1 \geq \abs{C'} + 1$, i.e., $s \geq \abs{C'} + 2$. To conclude the proof it suffices to show that $\abs{C'} \geq \abs{C}-1$ (in fact, we prove $\abs{C'} \geq \abs{C}$). Let $\{(a_1,b_1),\dots,(a_{\abs{C}},b_{\abs{C}})\} = C$.

      \ClmProofLBGClmOne*
      \begin{proofclaim}
        For the first statement,
        \emph{ad absurdum}, suppose that $b_j^f$ is a suffix of~$\Enum$.
        By definition of $\Enum$ and since $a_j \approx b_j$,
        we have $b_j^f \sqsubseteq a_j$.
        Since $a_j \in A$, by definition of $A^{\ltl{G}}$ 
        we get $b_j^f \in A^{\ltl{G}}$. However, this contradicts the existence of~$\Tree$ by~\Cref{lemma:unclosable-tree-condition} and~\Cref{th:ps}. Hence, $b_j^f$ is not a suffix of $\Enum$. 

        By definition of $b_j$, the fact that $b_j^f \not\sqsubseteq \Enum$ and $b_j^f \sqsubseteq b_j$, we conclude that $b_j^f = \emptyset^u \cdot \rho \cdot \alpha^{\alpha(n)} \cdot \sigma$ 
        for some $\rho \in \{\overline{\tau_i}\} \cup \{\tau \in T_P : \tau \prec \tau_i\}$ and $\sigma \in \Sigma^*$; 
        i.e., the second statement of the claim also holds.
      \end{proofclaim}

      \begin{restatable}{clm}{ClmProofLBGClmTwo}
        \label{proof-lbG-clm2}
        For every $j \neq k \in [1,\abs{C}]$, $b_j^f \neq b_k^f$.
      \end{restatable}
      \begin{proofclaim}
        \emph{Ad absurdum}, suppose $b_j^f = b_k^f$. 
        By definition of $B$, $b_j = (\emptyset^{t_i} \cdot \overline{\tau_i} \cdot \Enum|_{-\tau_i})\sffx{r_i}$ and 
        $b_k = (\emptyset^{t_\ell} \cdot \overline{\tau_\ell} \cdot \Enum|_{-\tau_{\ell}})\sffx{r_{\ell}}$ 
        for some $i \neq \ell \in [1,m]$.
        From the hypotheses of the lemma, $\tau_i \neq \tau_{\ell}$.
        Then, by definition of $\Enum|_{-\tau_i}$ and $\Enum|_{-\tau_\ell}$, 
        $b_j^f = b_k^f$ 
        implies $b_j^f \sqsubseteq \Enum$. 
        However, this contradicts~\Cref{proof-lbG-clm1}. 
        So, $b_j^f \neq b_k^f$.
      \end{proofclaim}

      \begin{restatable}{clm}{ClmProofLBGClmThree}
        \label{proof-lbG-clm3}
        For every $j \in [1,\abs{C}]$ there is $\ell \in \N$ 
        such that $a_j\sffx{\ell} \approx b_j^f$.
      \end{restatable}
      \begin{proofclaim}
        Following~\Cref{proof-lbG-clm1}, 
        given $i \in [1,m]$ such that $b_j = (\emptyset^{t_i} \cdot \overline{\tau_i} \cdot \Enum|_{-\tau_i})\sffx{r_i}$, 
        we have $b_j^f = \emptyset^u \cdot \rho \cdot \emptyset^{\alpha(n)} \cdot \sigma$, 
        for some $u \in \N$, $\rho \in \{ \overline{\tau_i} \} \cup \{\tau \in T_P : \tau \prec \tau_i \}$ and $\sigma \in \Sigma^*$.
        Since $a_j \approx b_j$, 
        and by definition of $\Enum$ and $\Enum|_{-\tau_i}$,
        we have that the prefix $\emptyset^u \cdot \rho$ 
        occurs in $a_j$. So, there is an $\ell \in \N$ such that $a_j\sffx{\ell} \approx b_j^f$.
      \end{proofclaim}

      \vspace{5pt}
      Together,~\Cref{proof-lbG-clm2,proof-lbG-clm3}
      directly imply $\abs{C'} \geq \abs{C}$.
      This concludes both the proof of the case~\ruleGlobally 
      and the proof of the lemma.
      \qedhere
  \end{itemize}
\end{proof}



\section{Proofs of~\Cref{sec:past-lb}}
\label{appendix:past-lb}

In this and the following appendices, given a (possibly infinite) trace $\sigma \in \Sigma^+ \cup \Sigma^\omega$ and natural numbers $j \leq k$, 
we write $\sigma[j,k)$ for the finite trace that is the prefix of $\sigma\sffx{j}$ having length $k-j$.
For instance, given $\sigma = w_0w_1w_2w_3w_4\dots$, 
$\sigma[1,3) = w_1w_2$. 
For finite traces~$\sigma$, we always make sure to satisfy $k \leq \abs{\sigma}+1$ when picking such a subtrace $\sigma[j,k)$.

We will need the following result: 

\begin{lemma}[{\cite[Lemma 5]{DBLP:conf/aaai/ArtaleGGMM23}}]
  \label{lemma:lemmafiveotherpaper}
  Let $\phi$ be a formula such that $\lang(\phi)$ is a cosafety language (e.g.~$\phi \in \LTL[\eventually]$ or $\phi \in \eventually(\LTLPnountil)$). 
  Then, $\lang^\omega(\phi) = \lang(\phi) \cdot \Sigma^\omega$
  and $\lang(\phi) = \lang(\phi) \cdot \Sigma^*$.
\end{lemma}

\begin{lemma}
  \label{lemma:reverse-helper}
  \label{clm:alpha-alpharev}
  Let $S \subseteq \{\tomorrow,\weaktomorrow,\eventually,\always\}$.
  Let $\alpha \in \LTL[S^-]$. Let $\alpha^-$ be the formula obtained from $\alpha$ by replacing every 
  operator~$\yesterday$ (resp.~$\weakyesterday$; $\once$; $\historically$)
  with~$\tomorrow$ \mbox{(resp.~$\weaktomorrow$; $\eventually$; $\always$)}. 
  For every $\sigma \in \Sigma^+$ and $i \in \position(\sigma)$, 
  $\sigma,i \models \alpha$ 
  if and only if $\sigma^-,\abs{\sigma}-i-1 \models \alpha^-$.
\end{lemma}

\begin{proof}
  The proof follows with a simple induction on the structure of $\alpha$.

  \proofitem{base case: $\alpha$ literal} 
    The statement follows from $\sigma[i] = \sigma^{-}[\abs{\sigma}-i-1]$.

  \proofitem{induction step, case: $\alpha = \alpha_1 \lor \alpha_2$} 
    This case is straightforward:
    \begin{align*}
      \sigma, i \models \alpha 
      &\text{ \ iff \ }
        \sigma,i \models \alpha_j \text{ for some } j \in \{1,2\}\\
      &\text{ \ iff \ }
        \sigma^-,\abs{\sigma}-i-1 \models \alpha_j^- \text{ for some } j \in \{1,2\}
        &\text{ (by induction hypothesis)}\\ 
      &\text{ \ iff \ }
        \sigma^-, \abs{\sigma}-i-1 \models \alpha^-
        &\text{ (by $\alpha^- = \alpha_1^- \lor \alpha_2^-$)}
    \end{align*}

  \proofitem{induction step, case: $\alpha = \alpha_1 \land \alpha_2$} 
    Analogous to the previous case. 

  \proofitem{induction step, case: $\alpha = \yesterday \alpha'$}
    We have,
    \begin{align*}
      \sigma, i \models \alpha 
      &\text{ \ iff \ }
        \sigma,i-1 \models \alpha'\\
      &\text{ \ iff \ }
        \sigma^-,\abs{\sigma}-(i-1)-1 \models (\alpha')^-
        &\text{ (by induction hypothesis)}\\ 
      &\text{ \ iff \ }
        \sigma^-, \abs{\sigma}-i-1 \models \alpha^-
        &\text{ (by $\alpha^- = \tomorrow (\alpha')^-$)}
    \end{align*}

  \proofitem{induction step, case: $\alpha = \weakyesterday \alpha'$} 
    Similar to the previous case. 

  \proofitem{induction step, case: $\alpha = \once \alpha'$}
    We have,
    \begin{align*}
      \sigma, i \models \alpha 
      &\text{ \ iff \ }
        \sigma,i-j \models \alpha' \text{ for some } j \in [0,i]\\
      &\text{ \ iff \ }
        \sigma^-,\abs{\sigma}-i+j-1 \models (\alpha')^-  \text{ for some } j \in [0,i]
        &\text{ (by ind. hyp.)}\\ 
      &\text{ \ iff \ }
        \sigma^-, \abs{\sigma}-i-1 \models \alpha^-
        &\text{ (by $\alpha^- = \eventually (\alpha')^-$)}
    \end{align*}

  \proofitem{induction step, case: $\alpha = \historically \alpha'$}
    Similar to the previous case.
\end{proof}

\LemmaReverseTwo*

\begin{proof}
  By definition, $\phi
  = \ltl{F}(\alpha)$, for some formula $\alpha \in \LTL[S^-]$. 
  Let $\alpha^-$ be the formula obtained from $\alpha$ by replacing every 
  operator~$\yesterday$ (resp.~$\weakyesterday$; $\once$; $\historically$)
  with~$\tomorrow$ \mbox{(resp.~$\weaktomorrow$; $\eventually$; $\always$)}.
  Define $\psi \coloneqq \eventually(\alpha^-)$, so that $\psi \in \eventually(\LTL[S])$ and $\size(\psi) = \size(\phi)$. 
  For every $\sigma \in \Sigma^+$ we have 
  \begin{align*}
    \sigma \models \phi 
     &\text{ \ iff \ }
      \sigma,i \models \alpha \text{ for some } i \in [0,\abs{\sigma}-1]\\
    &\text{ \ iff \ }
      \sigma^-,\abs{\sigma}-i-1 \models \alpha^- \text{ for some } i \in [0,\abs{\sigma}-1]
    &\text{ (by~\Cref{clm:alpha-alpharev})}\\
    &\text{ \ iff \ } 
      \sigma \models \psi.
    &&\qedhere
  \end{align*}
\end{proof}

\theoremLTLFsuccinct*

\begin{proof}
Consider the family of formulas $(\Phi_n)_{n \ge 1}$ defined
in~\cref{sec:difficult-property}. 
For every $n \geq 1$, $\Phi_n$ is a formula of $\eventually(\LTLPFinv)$ of size $\O(n)$. 
By~\Cref{lemma:reverseTwo} there is a formula $\Phi_n^-$ in $\eventually(\LTL[\eventually])$ (which is a subset of $\LTL[\eventually]$) 
satisfying $\lang(\Phi_n^-) = \lang(\Phi_n)^-$ and~$\size(\Phi_n^-) \in \O(n)$.  

Consider now a formula~$\Psi_n^-$ in $\eventually(\LTLPnountil)$
satisfying $\lang(\Psi_n^-) = \lang(\Phi_n)^-$. 
We apply~\Cref{lemma:reverseTwo}, and conclude that 
there is a formula $\Psi_n$ in $\eventually(\LTLnountil)$ such that $\size(\Psi_n) = \size(\Psi_n^-)$ and $\lang(\Psi_n) = \lang(\Phi_n)$.
Since $\eventually(\LTLnountil)$ is a fragment of $\LTLnountil$, by~\Cref{th:ltlnountilexponential} 
the formula $\Psi_n$ has size at least $2^n$.
Hence, $\size(\Psi_n^-) \geq 2^n$.

We conclude that the sequence of languages $(\lang(\Phi_n)^-)_{n \geq 1}$ 
witnesses the fact that~$\LTL[\eventually]$ can be exponentially more succinct than 
$\eventually(\LTLPnountil)$.
\end{proof}

Below, we extend the notion of \emph{separation} of $A$ from $B$ to sets of infinite traces. The extension is trivial:
A formula $\phi$ separates $A \subseteq \Sigma^\omega$ from $B \subseteq \Sigma^\omega$ whenever $A \subseteq \lang^{\omega}(\phi)$ and $B \cap \lang^{\omega}(\phi) = \emptyset$.

\begin{lemma}
  \label{lemma:flipping-games}
  Let $A,B \subseteq \Sigma^{\omega}$. 
  Suppose there is a formula $\phi \in \eventually(\LTLPnountil)$ such that $A \subseteq \lang^{\omega}(\phi)$ and $B \cap \lang^{\omega}(\phi) = \emptyset$. 
  Then, there are a map $f \colon A \to \N$ and a formula $\psi \in \LTLnountil$ such that $\psi$ separates 
  $A' \coloneqq \{a[0,f(a))^{-} : a \in A\}$ form ${B' \coloneqq \{b[0,j)^- : b \in B \text{ and } j \in \N \}}$, 
  and~$\size(\psi) = \size(\phi)-1$.
\end{lemma}

Above, note that $A'$ and $B'$ are sets of finite traces, so the notion of separation is the standard one used throughout the body of the paper, that is, $A \subseteq \lang(\psi)$ and $B \cap \lang(\psi) = \emptyset$ (equivalently, $A \models \psi$ and $B \pperp \psi$).

\begin{proof}
  Since $\phi \in \eventually(\LTLPnountil)$, 
  $\phi =
  \eventually(\alpha)$ for some $\alpha \in \LTLPnountil$. 
  We define the map $f$ following the semantics of $\phi$.
  Since $\phi$ separates $A$ form $B$, 
  \begin{itemize}
    \item for every $a \in A$ there is $j \in \N$ such that $a,j \models \alpha$. Set $f(a) \coloneqq j+1$, so that the prefix $a[0,f(a))$ end with the $j$th position in $a$.
    \item for every $b \in B$ and $j \in \N$, we have $b,j \not \models \alpha$.
  \end{itemize}
  Let us define 
    $A'' \coloneqq \{a[0,f(a)) : a \in A\}$ 
    and $B'' \coloneqq \{b[0,j) : b \in B \text{ and } j \in \N \}$. 

  Since $\alpha$ is a pure past formula, 
  from the two items above we conclude that 
  \begin{itemize}
    \item for every $a \in A''$, $a,\abs{a}-1 \models \alpha$,
    \item for every $b \in B''$ and $j \in \position(b)$, we have $b,j \not \models \alpha$.
  \end{itemize}
  Let $\alpha^-$ be the formula in~$\LTLnountil$ obtained by replacing every 
  operator~$\yesterday$ (resp.~$\weakyesterday$; $\once$; $\historically$)
  with~$\tomorrow$ \mbox{(resp.~$\weaktomorrow$; $\eventually$; $\always$)}.
  Then, directly from~\Cref{lemma:reverse-helper},
  \begin{itemize}
    \item for every $a \in A''$ $a^-,0 \models \alpha^-$,
    \item for every $b \in B''$ and $j \in \position(b)$, we have $b^-,j \not \models \alpha^-$.
  \end{itemize}
  By definition of $A''$ and $B''$,
  we conclude that
  $\alpha^-$ separates $A'$ form $B'$ (the sets defined as in the statement of the lemma). 
  Note that $\size(\alpha^-) = \size(\phi)-1$.
\end{proof}

\lemmareverseinfinite*

\begin{proof}
  Fix $n \geq 1$, and let $\Phi_n$ be the formula defined in~\Cref{sec:difficult-property}.

  We begin with~\Cref{item:revinf-1}. 
  Since $\Phi_n$ belongs to $\eventually(\LTLPFinv)$,
  by~\Cref{lemma:reverseTwo}, there is a formula~$\phi$ in $\LTL[\eventually]$ such that $\lang(\phi) = \lang(\Phi_n)^{-}$ 
  and $\size(\phi) = \size(\Phi_n) \in \O(n)$.
  Since $\lang(\phi)$ is cosafety,
  by~\Cref{lemma:lemmafiveotherpaper}, 
  $\lang^{\omega}(\phi) = \lang(\phi) \cdot \Sigma^\omega = \lang(\Phi_n)^{-} \cdot \Sigma^\omega$.
  This proves~\Cref{item:revinf-1}. We remark that, following the construction 
  in~\Cref{lemma:reverseTwo,clm:alpha-alpharev}, 
  \[ 
    \phi \coloneqq  \eventually \left( \widetilde{q} \land \bigwedge_{i=1}^n 
    \Big( 
      \big(q_i \land \eventually( \widetilde{p} \land p_i)\big) 
      \lor
      \big(\lnot q_i \land \eventually (\widetilde{p} \land \lnot p_i)\big)
    \Big) \right),
  \]
  i.e.,~it is simply obtained form $\Phi_n$ by replacing every occurrence of $\once$ with $\eventually$.

  We now prove~\Cref{item:revinf-2}. 
  For this item, we need to rely on objects and results from~\Cref{sec:lbltlxf}. 
  Consider~$\psi$ in $\eventually(\LTLPnountil)$ such that
  $\lang^\omega(\psi) = \lang(\Phi_n)^{-}\cdot\Sigma^\omega$. 
  Let $\AA = \{ \emptyset^j \cdot \overline{\tau} \cdot \Enum : j \in \N, \tau \in
  T_Q\}$ and $\BB = \{ \emptyset^j \cdot \overline{\tau} \cdot \Enum_{-\tau} :
  j \in \N, \tau \in T_Q \}$ 
  be the sets of finite traces defined in~\Cref{sec:lb:A-and-B}. 
  We recall that $\Enum$ is an enumeration of the types in $T_Q$, with respect to a strict order~$\prec$.
  We define 
  \begin{center}
    $A' \coloneqq \{ a^- \cdot \emptyset^\omega : a \in \AA \}$ 
    \ 
    and 
    \
    $B' \coloneqq \{ b^- \cdot \emptyset^\omega : b \in \BB \}$. 
  \end{center}
  Note that
  the formula $\phi$ defined during~\Cref{item:revinf-1} separates $A'$ from~$B'$ (the proof of this can be done directly from the definition of these sets and the definition of $\phi$, and it is analogous to the proof of~\Cref{lemma:phi-n-separates}). 
  Then, from $\lang^{\omega}(\phi) = \lang(\Phi_n)^{-}\cdot\Sigma^\omega = \lang^{\omega}(\psi)$, 
  we conclude that also $\psi$ separates $A'$ from $B'$.

  Since~$\psi \in \eventually(\LTLPnountil)$, 
  by~\Cref{lemma:flipping-games}, 
  there is a map $f : A' \to \N$ 
  and a formula $\widetilde{\psi} \in \LTLnountil$ such that
  $\widetilde{\psi}$ separates  
  $\widetilde{A} \coloneqq \{a[0,f(a))^{-} : a \in A'\}$ form 
  ${\widetilde{B} \coloneqq \{b[0,j)^- : b \in B' \text{ and } j \in \N \}}$. 
  Moreover, $\size(\widetilde{\psi}) = \size(\psi)-1$.

  By definition of $A'$ and $B'$, 
  we conclude that $\widetilde{A} \subseteq \AA^{\ltl{G}}$ and $\widetilde{B} = \BB^{\ltl{G}}$. 
  We now aim at applying~\Cref{lemma:lower-bound:spoon-case-with-globally} to $\widetilde{A}$ together with a suitable subset $\widehat{B} \subseteq \widetilde{B}$ such that 
  the set ${C \coloneqq \{ (a,b) \in \widetilde{A} \times \widehat{B} : a \approx b \}}$ has size at least $2^n-1$. This allows us (by~\Cref{lemma:lower-bound:spoon-case-with-globally}) 
  to conclude that the minimal deduction tree 
  for $\separ{\widetilde{A},\widehat{B}}$ has size at least $2^n$, and thus, by~\Cref{th:ps}, 
  $\size(\widetilde{\psi}) \geq 2^n$.
  By $\size(\widetilde{\psi}) = \size(\psi)-1$,
  this implies $\size(\psi) \geq 2^n$ (in fact, $\size(\psi) \geq 2^n+1$), concluding the proof. 

  To find the set $\widehat{B}$, 
  let us look at $\widetilde{A}$ and $\widetilde{B}$ more closely. 
  Let $\tau_0 \in T_Q$ be the smallest element with respect to the order~$\prec$ used to define $\Enum$.
  \emph{Ad absurdum}, suppose that $\tau_0$ does not appear in some $a \in \widetilde{A}$. 
  Then, by definition of $\AA$, 
  $a$ is a suffix of $\Enum|_{-\tau_0}$. By definition of $\BB$, we note that $\overline{\tau_0}\cdot\Enum|_{-\tau_0}$ and all its suffixes occur in $\widetilde{B}$. 
  This means that there is~$b \in \widetilde{B}$ 
  such that $a = b$. By~\Cref{lemma:unclosable-tree-condition} (taking the contrapositive of the first statement), 
  we conclude that $\separ{\widetilde{A},\widetilde{B}}$
  does not hold. However, this contradicts the fact that $\widetilde{\psi}$ separates $\widetilde{A}$ from $\widetilde{B}$. 
  Hence, $\tau_0$ occurs in every $a \in \widetilde{A}$.
  By definition of $\AA$, 
  this means that, 
  one of the following holds: 
  \begin{enumerate}[label=(\Roman*)]
    \item\label{ll10-c1} $a = \emptyset^j \cdot \overline{\tau} \cdot \Enum$ for some $j \in \N$ and $\tau \in T_Q$, 
    \item\label{ll10-c2} $a = \emptyset^j \cdot \tau_0 \cdot \Enum|_{-\tau_0}$ for some $j \in [0,\alpha(n)]$.
  \end{enumerate}
  We divide the proof in two cases.

  \proofitem{case: there is $a \in \widetilde{A}$
  satisfying~\ref{ll10-c2}} Let $\widetilde{a} \in
  \widetilde{A}$ and $j \in [0,\alpha(n)]$ such that
  $\widetilde{a} = \emptyset^j \cdot \tau_0 \cdot
  \Enum|_{-\tau_0}$. By definition of~$\BB$ and
  $\widetilde{B} = \BB^{\ltl{G}}$, for every $\tau \in T_Q$ there is
  $b_{(\tau)} \in \widetilde{B}$ such that $b_{(\tau)} =
  \Enum|_{-\tau}$. Let $\widehat{B} \coloneqq \{
  b_{(\tau)}\sffx{\alpha(n)-j} : \tau \in T_Q \setminus
  \{\tau_0\} \} \subseteq \widetilde{B}$. Note that every $b
  \in \widehat{B}$ belongs to the language $\emptyset^j
  \cdot \tau_0 \cdot \Sigma^+$, and that moreover
  $\abs{\widehat{B}} = 2^n-1$. By definition of $\approx$,
  the set $\{ (\widetilde{a},b) : b \in \widehat{B}, \widetilde{a}
  \approx b \}$ has size $2^n-1$, and therefore the set
  $C \coloneqq \{(a,b) \in \widetilde{A} \times \widehat{B} : a
  \approx b\}$ has size at least $2^n-1$.
  As discussed above, this implies $\size(\psi) \geq 2^n$. 

  \proofitem{case: no $a \in \widetilde{A}$
  satisfies~\ref{ll10-c2}} In this case, by definition of $\widetilde{A}$, we conclude that for every $\tau \in T_Q$ 
  there is $a_{(\tau)} \in \widetilde{A}$ and $j_{(\tau)} \in \N$ such that $a_{(\tau)} = \emptyset^{j_{(\tau)}} \overline{\tau} \cdot \Enum$. 
  By definition of $\BB$ and $\widetilde{B} = \BB^{\ltl{G}}$, 
  for every $\tau \in T_Q$ there is $b_{(\tau)} \in \widetilde{B}$ such that $b_{(\tau)} = \emptyset^{j_{(\tau)}} \cdot \overline{\tau} \cdot \Enum|_{-\tau}$.
  For every two distinct~$\tau_1,\tau_2 \in T_Q$ we have $b_{(\tau_1)} \neq b_{(\tau_2)}$. 
  Moreover, by definition of $\approx$, for every $\tau \in T_Q$, $a_{(\tau)} \approx b_{(\tau)}$.
  Define $\widehat{B} \coloneqq \{b_{(\tau)} : \tau \in T_Q\}$ and let $C \coloneqq \{(a,b) \in \widetilde{A} \times \widehat{B} : a
  \approx b\}$.
  We have $2^n = \abs{\widehat{B}} \leq \abs{C}$. 
  Again, this allows us to conclude $\size(\psi) \geq 2^n$.
\end{proof}

\section{Proofs of~\Cref{sec:automata-struggles}}
\label{appendix:automata-struggles}

We recall the notion of deterministic automaton --- and later consider the standard acceptance conditions of finite automaton (DFA) and B\"uchi automaton (DBA). 
A finite automaton is a tuple $A = (Q, \Sigma, \delta, q_0, F)$ consisting of the following components: 
\begin{itemize}[nosep]
  \item $Q$ is a finite set of \emph{states} whose elements are called the \emph{states} of $A$, 
  \item $\Sigma$ is a finite alphabet,  
  \item $\delta : Q \times \Sigma \to Q$ is a \emph{transition} function, 
  \item $q_0 \in Q$ is the \emph{initial state} of $A$, 
  \item $F \subseteq Q$ is a set of \emph{accepting states}.
\end{itemize}
Given a transition function $\delta$, as usual we write $\delta^* : Q \times \Sigma^+ \to Q$ for the extension of $\delta$ to finite traces. That is, for every $q \in Q$ and $\sigma \in \Sigma^+$,
\[ 
  \delta^*(q,\sigma) \coloneqq 
    \begin{cases}
    \delta(q,\sigma) 
      &\text{if $\abs{\sigma} = 1$}\\
    \delta^*(\delta(q,\sigma[0]),\sigma\sffx{1}) 
      &\text{otherwise}.
    \end{cases}
\]

\paragraph*{DFA acceptance condition.} Given a finite automaton $A = (Q, \Sigma, \delta, q_0, F)$, we write $\lang(A) \coloneqq \{\sigma \in \Sigma^* : \delta^*(q_0,\sigma) \in F \}$. 
When focusing on finite languages, we say that $A$ is a DFA and consider $\lang(A)$ as the language it recognises.

\paragraph*{DBA acceptance condition.}
When working on infinite traces, an automaton $A = (Q, \Sigma, \delta, q_0, F)$ is understood as a DBA.
An infinite trace $\sigma \in \Sigma^\omega$ is accepted by the DBA $A$ whenever, during its run on $\sigma$, the automaton reaches states in $F$ an infinite amount of times.
Formally, this means that there is an infinite increasing sequences of positions $(j_i)_{i \geq 1}$ and an infinite sequence of finite states $(q_i)_{i \geq 1} \in F$ 
such that $j_i = 0$ 
and for every $i \geq 1$, $\delta^*(q_{i-1},\sigma[j_i,j_{i+1})) = q_i$ (note: $q_0$ is the initial state).
W.l.o.g., in this decomposition one can also assume that 
for every $i \geq 1$ and $k \in [j_i,j_{i+1}-1]$, $\delta^*(q_{i-1},\sigma[j_{i},k)) \not \in F$; that is the sequence of $(j_i)_{i \geq 1}$ corresponds to exactly those positions for which $A$ reaches a final state.
We write $\lang^{\omega}(A)$ for the language accepted by a DBA (or an NBA) $A$.

In the next subsections, 
we need the following folklore result (for which however we failed to find an adequate reference). 

\begin{lemma} 
  \label{lemma:folklore-dba-cosafety}
  Let $\Sigma$ be a finite alphabet, and consider a regular language ${\lang \subseteq \Sigma^*}$. 
  There is a deterministic automaton $A = (Q, \Sigma, \delta, q_0, F)$ 
  such that 
  \begin{itemize}
    \item $\lang^{\omega}(A) = \lang \cdot \Sigma^\omega$ (i.e., when considering $A$ as a DBA), 
    \item $\lang(A) = \lang \cdot \Sigma^*$ (i.e., when considering $A$ as a DFA),
    \item ${\delta(q,a) = q}$ for every $q \in F$ and $a \in \Sigma$. 
  \end{itemize}
\end{lemma}

\begin{proof}
  Given a DFA $A' = (Q, \Sigma, \delta, q_0, F)$
  for $\lang$, one can produce the automaton~$A$ 
  by simply updating $\delta$ so that every final state becomes an~\emph{accepting trap states}.
  More precisely, we define the deterministic 
  automaton $A = (Q, \Sigma, \gamma, q_0, F)$ 
  where $\gamma$ defined from $\delta$ as follows: given $(q,a) \in Q \times \Sigma$: 
  \[ 
    \gamma(q,a) = 
      \begin{cases} 
        q &\text{if $q \in F$}\\
        \delta(q,a) &\text{otherwise}.
      \end{cases}
  \]
  It is easy to see that $\lang^{\omega}(A) = \lang \cdot \Sigma^\omega$. 
  For the inclusion $\lang^{\omega}(A) \subseteq \lang \cdot \Sigma^\omega$, 
  because the accepting trap states correspond to final states in $A'$, every word~$\sigma \in \lang^{\omega}(A)$ is of the form $\sigma = \sigma_1 \cdot \sigma_2$ with $\sigma_1 \in \lang$ and $\sigma_2 \in \Sigma^\omega$;
  and so $\sigma \in \lang \cdot \Sigma^\omega$.
  For the left to right inclusion, every word $\sigma \in \lang \cdot \Sigma^\omega$ is of the form $\sigma = \sigma_1 \cdot \sigma_2$ with $\sigma_1 \in \lang$ and $\sigma_2 \in \Sigma^\omega$. 
  By definition of $A$, we have $\gamma^*(q_0,\sigma_1) = q$ for some $q \in F$. Since $q$ is an accepting trap state, 
  we then conclude that $\sigma \in \lang^{\omega}(A)$.

  The proof that $\lang(A) = \lang \cdot \Sigma^*$ is analogous.
\end{proof}

\begin{restatable}{lemma}{LemmaFisForFree}
  \label{lemma:F-is-for-free}
  Consider two formulae $\phi$ and $\psi$ in $\eventually(\LTLPFinv)$ and $\LTLnountil$, respectively.
  If $\phi \equiv_{\omega} \psi$, then ${\psi \equiv_{\omega} {\eventually\psi}}$. 
\end{restatable}
\begin{proof} 
  We first prove two claims.

  \begin{clm}
    \label{clm:fforfree1}
    For every $\psi$ in $\LTLnountil$,
    $\lang^\omega(\ltl{F}\psi) = \Sigma^* \cdot \lang^\omega(\psi)$.
  \end{clm}
  \begin{proofclaim}
    This claim follows directly form the semantics of the $\ltl{F}$ operator and the
    fact that $\psi$ is a pure future formula.
  \end{proofclaim}

  \begin{clm}
    \label{clm:fforfree2}
    For any formula $\phi$ in $\ltl{F}(\LTL[\ltl{O}])$, it holds that
    $\lang^\omega(\phi) = \Sigma^* \cdot \lang^\omega(\phi)$.
  \end{clm}

  \begin{proofclaim}
    The left to right inclusion $\lang^\omega(\phi) \subseteq \Sigma^* \cdot \lang^\omega(\phi)$ is trivial. 
    For the other inclusion, consider 
    $\sigma \in \Sigma^* \cdot \lang^\omega(\phi)$.
    Therefore, there are $\sigma_1 \in \Sigma^*$ and $\sigma_2 \in \lang^{\omega}(\phi)$ such that $\sigma = \sigma_1 \cdot \sigma_2$. We show that $\sigma_1 \cdot \sigma_2 \in \lang^{\omega}(\phi)$. 

    Since $\phi$ is in $\ltl{F}(\LTL[\ltl{O}])$, 
    we have $\phi = \eventually(\alpha)$ for some $\alpha \in \LTL[\ltl{O}]$. The fact that $\sigma_1 \cdot \sigma_2 \in \lang^{\omega}(\phi)$ follows directly form the following statement:

    \begin{center}
    for every $\mu_1 \in \Sigma^*$, $\mu_2 \in \Sigma^{\omega}$ and $i \in \position(\mu_2)$,\\ 
    if $\mu_2,i \models \alpha$ then $\mu_1 \cdot \mu_2, i + \abs{\mu_1} \models \alpha$. 
    \end{center}
    Indeed, note that $\sigma_2 \models \phi$, so there is $i \in \position(\sigma_2)$ such that $\sigma_2,i \models \alpha$.
    Then, $\sigma_1 \cdot \sigma_2, i + \abs{\sigma_1} \models \alpha$, and so $\sigma_1 \cdot \sigma_2 \models \phi$. 

    To prove the statement above, we proceed by induction on the structure of~$\alpha$.

    \proofitem{base case: $\alpha$ literal}
      This case follows from $\mu[i] = (\mu'\cdot\mu)[i+\abs{\mu'}]$.

    \proofitem{induction step: $\alpha = \alpha_1 \lor \alpha_2$}
      We have 
      \begin{align*}
        \mu, i \models \alpha 
        &\text{ \ iff \ }
          \mu,i \models \alpha_j \text{ for some } j \in \{1,2\}\\
        &\text{ \ iff \ }
        \mu' \cdot \mu,i+\abs{\mu'} \models \alpha_j \text{ for some } j \in \{1,2\}
          &\text{ (by induction hypothesis)}\\ 
        &\text{ \ iff \ }
        \mu' \cdot \mu, i+\abs{\mu'} \models \alpha.
      \end{align*}

    \proofitem{induction step, case: $\alpha = \alpha_1 \land \alpha_2$}
      Analogous to the previous case. 

    \proofitem{induction step: $\alpha = \once(\alpha')$} We have 
      \begin{align*}
        \mu, i \models \alpha 
        &\text{ \ iff \ }
          \mu,j \models \alpha' \text{ for some } j \in [0,i]\\
        &\text{ \ iff \ }
        \mu' \cdot \mu,j+\abs{\mu'} \models \alpha' \text{ for some } j \in [0,i]
          &\text{ (by induction hypothesis)}\\ 
        &\text{ \ then \ }
        \mu' \cdot \mu, i+\abs{\mu'} \models \once(\alpha)
          &\text{ (by def. of~$\once$)}.
      \end{align*}
  \end{proofclaim}

  \vspace{5pt}
  \noindent
  Let us now prove~\Cref{lemma:F-is-for-free}.
  By~\Cref{clm:fforfree1}, $\lang^{\omega}(\eventually \psi) = \Sigma^* \cdot \lang^{\omega}(\psi)$.
  From the hypothesis $\phi \equiv_\omega \psi$, i.e., $\lang^\omega(\phi) = \lang^\omega(\psi)$, we have
  $\lang^\omega(\ltl{F}\psi) = \Sigma^* \cdot \lang^\omega(\phi)$.  By~\Cref{clm:fforfree2}, $\lang^\omega(\phi) = \Sigma^* \cdot \lang^\omega(\phi)$, and
  thus $\lang^\omega(\ltl{F}\psi) = \lang^\omega(\phi)$. 
  Again from $\phi
  \equiv_\omega \psi$, we conclude $\lang^\omega(\ltl{F}\psi)
  = \lang^\omega(\psi)$, i.e., $\psi \equiv_\omega \ltl{F}\psi$.
\end{proof}

\LemmaUpperBoundDBACosafety*
\begin{proof} 
  Let $\phi = \ltl{F}(\alpha)$ with $\alpha$ a formula of $\LTL[\ltl{O}]$.
  For any trace $\sigma \in \Sigma^+$, it holds that $\sigma,0 \models
  \ltl{F(\alpha)}$ iff $\sigma,|\sigma|-1 \models \ltl{O}(\alpha)$.
  By~\cite[Theorem 2]{de2021pure}, there exists a DFA $A$ recognizing
  $\lang(\ltl{O}(\alpha)) \coloneqq \set{\sigma \in \Sigma^+ : \sigma,
  |\sigma|-1 \models \ltl{O}(\alpha)}$ of size
  $2^{\O(\size(\ltl{O}(\alpha)))}$.  
  So, $\lang(A) = \lang(\phi)$.
  Below, we sometimes see $A$ as a DBA, hence considering the B\"uchi acceptance 
  condition on $A$, instead of the acceptance condition of DFAs. 
  Recall that $\lang(A)$ is the language recognised by $A$ under the 
  (standard) acceptance condition of DFAs, 
  whereas $\lang^{\omega}(A)$ is the language recognised by $A$ 
  under B\"uchi acceptance condition. 
  We show that $\lang^{\omega}(\phi) = \lang^{\omega}(A)$; concluding the proof. 
  
  Let $F$ be the set of finite states in $A$.
  Given~$q \in F$ of $A$, we write 
  $A_q$ as the automaton obtained from $A$ by changing its initial state to $q$. 
  From $\phi \in \eventually(\LTLPFinv)$, 
  $\lang(\phi)$ is a cosafety language, 
  and therefore for every $q \in F$, $\lang(A_q) = \Sigma^*$.
  This means that, in $A$, every state reachable from a $q \in F$ is an accepting state. 
  So, given $q \in F$, we have $\lang^{\omega}(A_q) = \Sigma^\omega$. 
  Let us write $A(\sigma) = q$ whenever the automaton~$A$ stops on the state $q$ when it runs on $\sigma \in \Sigma^*$.
  Then, 
  \begin{align*}
    \lang^{\omega}(A) 
      &= \{ \sigma \cdot \sigma' : \sigma \in \Sigma^* \text{ and there is } q \in F \text{ s.t. } \sigma' \in \lang^{\omega}(A_q) \text{ and } A(\sigma) = q \}\\
      &= \{ \sigma \cdot \sigma' : \sigma \in \Sigma^*, \sigma' \in \Sigma^\omega \text{ and there is } q \in F \text{ s.t. } A(\sigma) = q \}\\
      &= \lang(A) \cdot \Sigma^\omega\\ 
      &= \lang(\phi) \cdot \Sigma^{\omega} \qquad \text{(by $\lang(A) = \lang(\phi)$)}.
  \end{align*}
  By~\Cref{lemma:lemmafiveotherpaper}, $\lang^\omega(\phi) = \lang(\phi) \cdot \Sigma^\omega$, and therefore
  $\lang^{\omega}(\phi) = \lang^{\omega}(A)$.
\end{proof}

\LemmaMinimalDBACosafety*
\begin{proof}
  Since $\lang^\omega(\psi)$ is a cosafety language, the same holds for $\lang^{\omega}(\eventually \psi)$, i.e., 
  $\lang^{\omega}(\eventually \psi)$ is of the form $\lang \cdot \Sigma^{\omega}$ for some regular language $\lang$. 
  By~\Cref{lemma:folklore-dba-cosafety}
  There is a DBA $A = (Q,\Sigma,\delta,q_0,F)$
  such that $\lang^{\omega}(A) = \lang^{\omega}(\eventually\psi)$
  and that for every $q \in F$ and $a \in \Sigma$, 
  $\delta(q,a) = q$. To simplify the exposition, 
  let us furthermore assume that $q_0 \not \in F$. 
  This restriction is w.l.o.g.: if $q_0 \in F$ then $\lang^{\omega}(A) = \Sigma^\omega$, and one 
  derives $\lang^{\omega}(A) = \lang^{\omega}(\eventually \psi) = \lang^{\omega}(\always\eventually \psi) = \Sigma^\omega$, thus concluding the proof.

  We define the DBA $A' = (Q,\Sigma,\gamma,q_0,F)$ 
  where $\gamma$ defined from $\delta$ as follows: given $(q,a) \in Q \times \Sigma$: 
  \[ 
    \gamma(q,a) = 
      \begin{cases} 
        \delta(q_0,a) &\text{if $q \in F$}\\
        \delta(q,a) &\text{otherwise}.
      \end{cases}
  \]
  We show that $\lang^{\omega}(\always\eventually\psi) = \lang^{\omega}(A')$, and so the minimal DBA for $\lang^{\omega}(\always\eventually\psi)$ must have size in $\O(\size(A')) = \O(\size(A))$.

  ($\subseteq$): Let $\sigma \in \Sigma^+$ such that $\sigma \models \always\eventually\psi$. 
  For every $j \in \N$, $\sigma\sffx{j} \models \eventually\psi$, 
  and so $\sigma\sffx{j} \in \lang^{\omega}(A)$.
  In particular, for every $j \in \N$, there is $k > j$ such that $\delta^*(q_0,\sigma[j,k)) = q_f$, with $q_f \in F$. 
  Therefore, there is an infinite increasing sequence of positions $(j_i)_{i \geq 1}$ such that $j_1 = 0$ and 
  \begin{itemize}
    \item for every $i \geq 1$, $\delta^*(q_0,\sigma[j_i,j_{i+1})) \in F$, and 
    \item for every $i \geq 1$ and $k \in [j_i,j_{i+1}-1]$, $\delta^*(q_0,\sigma[j_i,k)) \not\in F$,
  \end{itemize}
  that is, the prefix of $\sigma\sffx{j_i}$ of length $j_{j+1}-j_i$ is the smallest prefix that lead to an accepting state when running $A$ on $\sigma\sffx{j_i}$.
  By definition of $\gamma$, we conclude that 
  \begin{itemize}
    \item $\gamma^*(q_0,\sigma[0,j_{2})) \in F$, and 
    \item for every $q \in F$ and $i \geq 2$, $\gamma^*(q,\sigma[j_{i},j_{i+1})) \in F$.
  \end{itemize}
  From the acceptance condition of DBAs, we conclude that $\sigma \in \lang^{\omega}(A')$.

  ($\supseteq$): Consider $\sigma \in \lang^{\omega}(A')$. 
  
  From the acceptance condition of DBAs, there is an
  infinite increasing sequence of positions $(j_i)_{i \geq
  1}$ and an infinite sequence of finite states $(q_i)_{i
  \geq 1} \in F$ such that $j_1 = 0$ and (recall: $q_0$ is
  the initial state)
  \begin{itemize}
    \item for every $i \geq 1$, $\gamma^*(q_{i-1},\sigma[j_{i},j_{i+1})) = q_{i}$, 
    \item for every $i \geq 1$ and $k \in [j_i,j_{i+1}-1]$, $\gamma^*(q_{i-1},\sigma[j_{i},k)) \not \in F$.
  \end{itemize}
  By definition of $\gamma$, we conclude that for every $i
  \geq 1$, ${\delta^*(q_{0},\sigma[j_{i},j_{i+1})) = q_{i}
  \in F}$. Since $\delta(q,a) = q$ for every $q \in F$ and $a
  \in \Sigma$, this means that, for every ${i \geq 1}$,
  $\sigma\sffx{j_i} \in \lang^{\omega}(A)$, and thus
  $\sigma\sffx{j_i} \models \eventually\psi$.
  Since $(j_i)_{i \geq 1}$ is an infinite increasing sequence,
  we obtain $\sigma \models \always\eventually \psi$.
\end{proof}

We are now ready to prove~\Cref{theorem:why-automata-method-fails}.

\TheoremWhyAutomataMethodFails*
\begin{proof}
  By~\Cref{lemma:F-is-for-free}, $\psi \equiv_{\omega} \eventually \psi$.
  Let $\Pi$ be a prefix of $k$ temporal operators among $\tomorrow$, $\eventually$ and~$\always$. 
  Note that, for every formula $\gamma$ of $\LTL$ without past operators,
  \begin{center}
    $\eventually \tomorrow \gamma \equiv_{\omega} \tomorrow \eventually \gamma$,
    \quad 
    $\always \tomorrow \gamma \equiv_{\omega} \tomorrow \always \gamma$,
    \quad
    $\eventually \always \eventually \gamma \equiv_{\omega} \always \eventually \gamma$, 
    \quad 
    $\eventually \eventually \gamma \equiv_{\omega} \eventually \gamma$ 
    \quad 
    $\always \always \gamma \equiv_{\omega} \always \gamma$.
  \end{center} 
  Since $\psi \equiv_{\omega} \eventually \psi$,
  can use these tautologies to rearrange $\Pi$ into 
  a prefix of the form either $\tomorrow^j \always \eventually$
  or $\tomorrow^j \eventually$, depending on whether $\Pi$ contains a $\always$, i.e.,
  there is $j \in [0,k]$ 
  such that either $\Pi(\psi) \equiv_{\omega} \tomorrow^j \always \eventually \psi$ or $\Pi(\psi) \equiv_{\omega} \tomorrow^j \eventually \psi$.
  Below we consider the (more interesting) case of  
  $\Pi(\psi) \equiv_{\omega} \tomorrow^j \always \eventually \psi$. The proof for $\Pi(\psi) \equiv_{\omega} \tomorrow^j \eventually \psi$ carries out in a similar way.

  By~\Cref{lemma:upper-bound-dba-cosafety} and $\phi \equiv_{\omega} \psi \equiv_{\omega} \eventually \psi$, we conclude that there minimal DBA for $\lang^{\omega}(\eventually \psi)$ has size $2^{\O(\size(\phi))}$.
  Then, by~\Cref{lemma:minimal-dba-cosafety}, 
  we can construct a DBA~$A'$ such that $\lang^{\omega}(A') = \lang^{\omega}(\always\eventually\psi)$ having size~$2^{\O(\size(\phi))}$.
  To conclude the proof, it now suffices to update $A'$ in order to produce a DBA~$A$ recognising $\lang^{\omega}(\tomorrow^j \always \eventually \psi)$. 
  Note that, $\always \eventually \psi$ is a pure future formula, and thus, given the alphabet $\Sigma$ 
  of $\lang^{\omega}(\always \eventually \psi) \subseteq \Sigma^\omega$,
  We have $\lang^{\omega}(\tomorrow^j \always \eventually \psi) = \Sigma^j \cdot \lang^{\omega}(\always \eventually \psi)$.
  Then, at the level of DBAs, $A$ can be intuitively constructed by adding a chain of $j$ states leading to the initial state of $A'$.
  For completeness, us formalise this intuition.
  If $j = 0$ then $A' \coloneqq A$ and the statement follows.
  Otherwise, suppose $j \geq 1$.
  Let $A' = (Q,\Sigma,\delta,q_0,F)$, 
  and consider a set $Q' \coloneqq \{q_1',\dots,q_{j}'\}$ of $j$ states not belonging to $Q$. 
  We define $A \coloneqq (Q \cup Q',\Sigma,\delta',q_1',F)$, 
  where $\delta'$ is defined as follows:
  \begin{itemize} 
    \item for every $q \in Q$ and $a \in \Sigma$, $\delta'(q,a) = \delta(q,a)$, 
    \item for every $k \in [1,j-1]$ and $a \in \Sigma$, $\delta'(q_k',a) = q_{k+1}'$, and 
    \item for every $a \in \Sigma$, $\delta'(q_{j},a) = q_0$.
  \end{itemize}
  The size of the automaton
  only grow by $\O(j \cdot \abs{\Sigma})$, so $\size{A} \leq (k+1) \cdot 2^{\O(\size(\phi))}$.
  With a simple induction, one can show that 
  for every $i \in [0,j-1]$, the automaton $A_i \coloneqq (Q \cup Q',\Sigma,\delta',q_{j-i}',F)$
  recognises the language $\lang^{\omega}(\tomorrow^{i+1} \always \eventually \psi)$. 
  Therefore,~$A$ recognises $\lang^{\omega}(\tomorrow^j \always \eventually \psi)$. 
\end{proof}

\section{On the pastification algorithm for $\LTL[\tomorrow,\eventually]$}
\label{sec:new-analysis-upper-bound}

In this appendix, we briefly discuss how to improve the
complexity analysis of the pastification algorithm
in~\cite{ArtaleGGMM23} to obtain a $2^{\O(n)}$ upper bound
for the problem, instead of the $2^{\O(n^2)}$ bound proved in
that paper. We do this to substantiate our claim that, in
view of our $2^{\Omega(n)}$ lower bound, the algorithm
in~\cite{ArtaleGGMM23} is an optimal procedure for the
pastification problem of~\LTLxf into
$\eventually(\LTLPnountil)$.

As explained in~\cite[Section 4.3]{ArtaleGGMM23}, the
bottleneck for the algorithm is given by a procedure
translating a formula of $\LTLxf$ into an equivalent formula
in disjunctive normal form, which in this context means a
formula of the form $\bigvee_{i = 1}^k \gamma_i$ where each
$\gamma_i$ is from the following grammar 
\begin{align}
  \label{grammar:conjunctive-reducts}
  \gamma \ \coloneqq \ p \ \mid \ \lnot p \ \mid \ \gamma \land \gamma \ \mid \ \tomorrow \gamma \ \mid \ \eventually \gamma,
\end{align}
with $p \in \AP$. All other parts of the procedure
in~\cite{ArtaleGGMM23} run in polynomial time, and so they
do not affect our upper bound (since $\poly(2^{\O(n)}) =
2^{\O(n)}$). Since $\land$, $\tomorrow$ and $\eventually$
distribute over disjunctions, the algorithm to put a formula
in disjunctive normal form is straightforward. Bottom-up,
starting from a formula of \LTLxf, it suffices to repeatedly
apply the following equivalences as rewrite rules (to be
applied from left to right) in order to push the
disjunctions upward:
\begin{align*}
  (\phi_1 \lor \phi_2) \land \psi &\ \equiv\ (\phi_1 \land \psi) \lor (\phi_2 \land \psi),\\
  \psi \land (\phi_1 \lor \phi_2) &\ \equiv\ (\phi_1 \land \psi) \lor (\phi_2 \land \psi),\\
  \tomorrow (\phi_1 \lor \phi_2) &\ \equiv\ (\tomorrow \phi_1) \lor (\tomorrow \phi_2),\\
  \eventually (\phi_1 \lor \phi_2) &\ \equiv\ (\eventually \phi_1) \lor (\eventually \phi_2).
\end{align*}
In~\cite{ArtaleGGMM23} a $2^{\O(n^2)}$ upper bound for the
size (and time) required to compute the resulting formula is
established. Below we improve this bound to~$2^{\O(n)}$.

Let $\phi$ be a formula in \LTLxf. A formula $\psi$ is said
to be a \emph{conjunctive reduct} of $\phi$ whenever $\psi$
can be obtained from $\phi$ by replacing (bottom-up) every
disjunction of the form $\phi_1 \lor \phi_2$ with either
$\phi_1$ or $\phi_2$. More precisely,
\begin{itemize}
  \item if $\phi$ is of the form $p$ or $\lnot p$, then
  $\psi = \phi$,
  \item if $\phi$ is of the form $\phi_1 \land \phi_2$, then
  $\psi = \psi_1 \land \psi_2$ where, for every $i \in
  \{1,2\}$, $\psi_i$ is a conjunctive reduct of $\phi_i$,
  \item if $\phi$ is of the form $\tomorrow \phi'$, then
  $\psi = \tomorrow \psi'$ where $\psi'$ is a conjunctive
  reduct of $\phi'$,
  \item if $\phi$ is of the form $\eventually \phi'$, then
  $\psi = \eventually \psi'$ where $\psi'$ is a conjunctive
  reduct of $\phi'$,
  \item if $\phi$ is of the form $\phi_1 \lor \phi_2$, then
  $\psi$ is a conjunctive reduct of either $\phi_1$ or
  $\phi_2$.
\end{itemize}
It is easy to see that every conjunctive reduct is a formula
from the grammar in~\Cref{grammar:conjunctive-reducts}. We
denote with $S(\phi)$ the set of all conjunctive reducts of
$\phi$. 

Since every $\psi \in S(\phi)$ is obtained by essentially removing
disjunctions from~$\phi$, $\size(\psi) \leq \size(\phi)$.
Moreover, $\abs{S(\phi)} \leq 2^{(\text{number of
disjunctions in $\phi$})} \leq 2^{\size(\phi)}$. With these
inequalities at hand,~\Cref{proposition-reducts-are-enough} below suffices to
show that the disjunctive normal form  
of every formula $\phi$ from \LTLxf has a size in
$\O(2^{\size(\phi)} \cdot \size(\phi)) \in
2^{\O(\size(\phi))}$. The time required to compute the
disjunctive normal form is also bounded in
$2^{\O(\size(\phi))}$, as it can be computed by simply
enumerating the various element in $S(\phi)$.
Note that this is then also the running time of the algorithm described in~\cite{ArtaleGGMM23}, as the above equivalences are essentially performing the computation of $S(\phi)$. 
For instance, given $(\phi_1 \lor \phi_2) \land \psi$ where $\phi_1,\phi_2$ and $\psi$ are conjunctive reducts, 
the formula $(\phi_1 \land \psi) \lor (\phi_2 \land \psi)$ is capturing the conjunctive reducts $\phi_1 \land \psi$ and $\phi_2 \land \psi$ obtained by reducing $\phi_1 \lor \phi_2$ to either $\phi_1$ or $\phi_2$.

\begin{proposition}
  \label{proposition-reducts-are-enough}
  Let $\phi$ be a formula in \LTLxf. Then, $\phi \equiv
  \bigvee_{\psi \in S(\phi)} \psi$.
\end{proposition}

\begin{proof}
The proof follows with a simple structural induction on~$\phi$.

\proofitem{base case: $\phi = p$ or $\phi = \lnot p$} 
  We have $S(\phi) = \{\phi\}$ and the proof is trivial.

\proofitem{induction step: $\phi = \phi_1 \lor \phi_2$}
  By induction hypothesis, $\phi_1 \equiv \bigvee_{\psi_1
  \in S(\phi_1)} \psi_1$ and $\phi_2 \equiv \bigvee_{\psi_2
  \in S(\phi_2)} \psi_2$. Then,
  \begin{align*}
    \phi 
      \ \equiv \ \Big(\bigvee_{\psi_1 \in S(\phi_1)} \psi_1\Big) \lor \Big(\bigvee_{\psi_2 \in S(\phi_2)} \psi_2\Big)
      \ \equiv \ \bigvee_{\psi \in S(\phi_1) \cup S(\phi_2)} \psi.
  \end{align*}
  From the definition of conjunctive reduct, $S(\phi)$ is
  the set of all formulae that belong to $S(\phi_1)$ or
  $S(\phi_2)$, i.e., $S(\phi) = S(\phi_1) \cup S(\phi_2)$,
  concluding the proof.

\proofitem{induction step: $\phi = \tomorrow \phi'$}
  By induction hypothesis, $\phi' \equiv \bigvee_{\psi' \in
  S(\phi')} \psi'$. Then,
  \begin{align*}
    \phi 
      \ \equiv \ \tomorrow\bigvee_{\psi' \in S(\phi')} \psi'
      \ \equiv \ \bigvee_{\psi' \in S(\phi')} \tomorrow \psi'.
  \end{align*}
  By definition of conjunctive reducts, $S(\phi)$ is the set
  of all formulae $\psi$ of the form $\tomorrow \psi'$ where
  $\psi' \in S(\phi')$. Therefore, $\bigvee_{\psi' \in
  S(\phi')} \tomorrow \psi'$ is syntactically equal (up to
  associativity and commutativity of disjunctions) to
  $\bigvee_{\psi \in S(\phi)} \psi$.

\proofitem{induction step: $\phi = \eventually \phi'$}
  This case is analogous to the previous one.

\proofitem{induction step: $\phi = \phi_1 \land \phi_2$}
  By induction hypothesis, $\phi_1 \equiv \bigvee_{\psi_1
  \in S(\phi_1)} \psi_1$ and $\phi_2 \equiv \bigvee_{\psi_2
  \in S(\phi_2)} \psi_2$. Then, 
  \begin{align*}
    \phi 
      \ \equiv \ \Big(\bigvee_{\psi_1 \in S(\phi_1)} \psi_1\Big) \land \Big(\bigvee_{\psi_2 \in S(\phi_2)} \psi_2\Big)
      \ \equiv \ \bigvee_{(\psi_1,\psi_2) \in S(\phi_1) \times S(\phi_2)} \psi_1 \land \psi_2.
  \end{align*}
  By definition of conjunctive reduct, $S(\phi)$ is the set
  of all formulae $\psi$ of the form $\psi_1 \land \psi_2$
  where, for every $i \in \{1,2\}$, $\psi_i$ is a
  conjunctive reduct of $\phi_i$. Therefore,
  $\bigvee_{(\psi_1,\psi_2) \in S(\phi_1) \times S(\phi_2)}
  \psi_1 \land \psi_2$ is syntactically equal (up to
  associativity and commutativity of disjunctions) to
  $\bigvee_{\psi \in S(\phi)} \psi$.
\end{proof}